\documentclass[a4paper,notitlepage,twocolumn,superscriptaddress,longbibliography,nofootinbib]{revtex4-1}
\pdfoutput=1
\usepackage[utf8]{inputenc}
\usepackage{floatrow}
\usepackage{amsmath,amsthm,amssymb,amsfonts}
\usepackage{adjustbox}
\usepackage{mathtools}
\usepackage{graphicx}
\usepackage{subfigure}
\usepackage[colorlinks = true,
            linkcolor = blue,
            urlcolor  = blue,
            citecolor = blue,
            anchorcolor = blue]{hyperref}
\usepackage{mathrsfs}
\usepackage{bbold}
\usepackage[]{units}
\usepackage{bm}
\usepackage[normalem]{ulem}
\usepackage{braket}
\usepackage{color}
\usepackage{makecell}
\usepackage{enumitem}
\usepackage{upgreek}
\usepackage{blindtext}
\usepackage{graphics}
\usepackage{verbatim} % for math
\usepackage{adjustbox}

\usepackage[dvipsnames]{xcolor}
\usepackage{bbm}
\usepackage{longtable}
\usepackage[bb=boondox]{mathalfa}
\usepackage{array}
\usepackage{multirow}
\usepackage{tabularx}
\usepackage{float}

\usepackage{tikz}
\usetikzlibrary{shapes,arrows,patterns}

\newcommand{\id}{\mathbf{1}}

\DeclareMathOperator{\arcsinh}{sinh^{-1}}

\begin{document}

\title{Probing quantum scars and weak ergodicity breaking through quantum complexity}
\author{Budhaditya Bhattacharjee}
\email{budhadityab@iisc.ac.in}
\affiliation{Centre for High Energy Physics, Indian Institute of Science, C.V. Raman Avenue, Bangalore-560012, India}
\author{Samudra Sur}
\email{samudrasur@iisc.ac.in}
\affiliation{Centre for High Energy Physics, Indian Institute of Science, C.V. Raman Avenue, Bangalore-560012, India}
\author{Pratik Nandy}
\email{pratik@yukawa.kyoto-u.ac.jp}
\affiliation{Centre for High Energy Physics, Indian Institute of Science, C.V. Raman Avenue, Bangalore-560012, India}
\affiliation{Center for Gravitational Physics and Quantum Information, Yukawa Institute for Theoretical Physics,\\ Kyoto University, Kitashirakawa Oiwakecho, Sakyo-ku, Kyoto 606-8502, Japan}

\begin{abstract}
Scar states are special many-body eigenstates that weakly violate the eigenstate thermalization hypothesis (ETH). Using the explicit formalism of the Lanczos algorithm, usually known as the forward scattering approximation in this context, we compute the Krylov state (spread) complexity of typical states generated by the time evolution of the PXP Hamiltonian, hosting such states. We show that the complexity for the N\'eel state revives in an approximate sense, while complexity for the generic ETH-obeying state always increases. This can be attributed to the approximate SU(2) structure of the corresponding generators of the Hamiltonian. We quantify such ``closeness'' by the $q$-deformed SU(2) algebra and provide an analytic expression of Lanczos coefficients for the N\'eel state within the approximate Krylov subspace. We intuitively explain the results in terms of a tight-binding model. We further consider a deformation of the PXP Hamiltonian and compute the corresponding Lanczos coefficients and the complexity. We find that complexity for the N\'eel state shows nearly perfect revival while the same does not hold for a generic ETH-obeying state.

\end{abstract}
\maketitle
%\tableofcontents
\section{Introduction}
%\emph{Introduction:} 

Thermalization is a well-known fact for a generic, isolated quantum system, where a statistical description emerges at late times in the thermodynamic limit \cite{Calabrese:2005in, Santos_2012, Deutsch_2013, Collura_2014, Kaufman_2016}. The system relaxes locally, with energy being the only conserved quantity \cite{Essler:2016ufo}. This is often attributed to the eigenstate thermalization hypothesis (ETH), which states that the highly excited eigenstates of generic many-body systems are thermal \cite{PhysRevA.43.2046, PhysRevE.50.888, PhysRevLett.98.050405, Rigol_2008}. However, integrable systems \cite{Caux:2010by} and many-body localized states \cite{Basko_2006, PhysRevLett.111.127201, PhysRevB.75.155111, PhysRevB.82.174411, PhysRevB.90.174202, nand_huse, RevModPhys.91.021001} are known to strongly violate ETH. Thus, for a quantum-chaotic systems, the local observables are expected to reach the thermal value irrespective of the choice of the initial state. %\sout{Thus, for any translationally invariant, non-integrable Hamiltonian, it is expected that all the time-evolved states obey ETH and should thermalize fast irrespective of the choice of the initial state.} \par
Recently, a weak violation of ETH has been observed for a few specific states in quantum many-body systems, although most states still follow ETH and show fast relaxation \cite{Turner_2018}. The full system is still non-integrable, as can be verified using the level statistics \cite{Turner_2018}. These weak ergodicity breaking states are commonly referred to as ``scar states'' \cite{PhysRevLett.53.1515}, which manifested themselves in the experimentally observed perfect revival of some special initial states in Rydberg atom chains \cite{Bernien_2017}, and optical lattices \cite{Scherg:2020mcp, PhysRevLett.124.160604}, and led to a flurry of theoretical studies in recent times \cite{Turner_2018, PhysRevLett.122.220603, PhysRevB.105.125123, PhysRevB.98.155134, PhysRevB.101.165139, PhysRevB.99.161101, PhysRevLett.122.173401, PhysRevLett.127.150601, Desaules:2021acx, PhysRevB.105.L060301, ODea:2020ooe, Chandran:2022jtd, tang2021multi, PhysRevLett.124.180604, Pakrouski:2020hym,  Pakrouski:2021jon, PhysRevB.101.024306, Dooley:2022kge, Desaules:2022ibp, Su:2022glk, Desaules:2022kse, Magnifico:2019kyj}. Time evolution of these specific initial states shows a finite overlap with themselves even after a sufficiently long time \cite{Serbyn:2020wys}. This weak breaking of ergodicity is also evident in the behavior of bipartite entanglement entropy, which fails to respect the volume-law scaling \cite{PhysRevB.98.235156, papic}. The slow thermalization is often attributed to the approximate Hilbert space division into thermalizing and non-thermalizing part $H \approx H_{\mathrm{nonth}} \bigoplus H_{\mathrm{th}}$ \cite{Moudgalya:2021ixk}. In this space, scar states are represented in terms of the Krylov basis vectors. The evolution focuses inside the Krylov space, usually termed as the ``Krylov-restricted thermalization'' \cite{moudgalya2022thermalization}.

In this paper, we aim to look at this weak ergodicity breaking in terms of quantum complexity. Complexity, primarily borrowed from computer science, is a fairly new concept that has emerged as a new tool for diagnosing quantum chaos \cite{Parker:2018yvk, Avdoshkin:2019trj, Dymarsky:2019elm, Bhattacharyya:2020art, Rabinovici:2021qqt} and scrambling \cite{Barbon:2019wsy, Bhattacharjee:2022vlt} in many-body systems. The term complexity refers to the cost of implementing any task at hand in the minimum number of steps. For our purpose, we consider the difficulty of spreading an initial state in the Hilbert space through the time evolution of a Hamiltonian. The difficulty is naturally understood in terms of complexity, dubbed as ``spread complexity'' \cite{Balasubramanian:2022tpr}. The definition is straightforward and suitable compared to the other theoretical cousins, namely, the circuit complexity \cite{Jefferson:2017sdb, Chapman:2017rqy}. Recently, it has been shown to detect topological and non-topological phases in many-body systems \cite{Caputa:2022eye}. The formulation is based upon the iterative process of the Lanczos algorithm \cite{viswanath1994recursion}, often known as the forward scattering approximation  \cite{PhysRevB.98.155134}. Although the formulation works in any generic case, symmetry greatly simplifies the problem, and Lanczos coefficients can be extracted analytically. Here we should mention that this formulation is conceptually different than studying the operator growth, where the behavior of Lanczos coefficients is speculated by the universal operator growth hypothesis \cite{Parker:2018yvk}.

We begin by studying the time evolution of the $\ket{\mathbf{Z}_{2}}$ (i.e., $\ket{1010101010101010}$ for $N = 16$ lattice size) state for a simple paramagnetic spin chain Hamiltonian $H_p = \sum_{n=1}^N \sigma^x_{n}$, where $\sigma^x_{n}$ denotes the Pauli $X$ matrices. This state is the lowest-weight state in the $j = N/2$ representation of the SU(2) symmetry of the Hamiltonian. By this, we mean that the Hamiltonian can be separated into two parts $H_{\pm}$, where $H_{\pm}$ and $H_z$ follows the SU(2) algebra ($H_z$ is obtained by the commutator $[H_{+}, H_{-}]$). Then we study the evolution of the $\ket{\mathbf{Z}_{2}}$ and a generic $\ket{0}$ state (i.e., an initial state without any \textbf{Z} symmetry which does not falls into the representation of the symmetry) for the PXP Hamiltonian. For numerical consistency (lattice size $N = 16$), we fix the choice $\ket{0} = \ket{0010100100100010}$ (in $\sigma^{z}$ basis). It is known that the paramagnetic Hamiltonian (more generally $H_{\pm}$ and $H_z$) satisfies the SU(2) symmetry relations. Therefore, the Lanczos coefficients and complexity can be computed analytically. On the other hand, the PXP Hamiltonian is known to break the SU(2) symmetry algebra. For this purpose, we implement the Lanczos algorithm numerically to compute the Lanczos coefficients, the Krylov wave functions, and complexity. We find that the complexity for the N\'eel state (i.e., the $\ket{\mathbf{Z}_{2}}$ state) demonstrates an oscillatory component while that of a generic ($\ket{0}$) state does not. Furthermore, the growth of the spread complexity for the N\'eel state appears to be slower than that of the generic state. We explain these observations in terms of the weak ergodicity breaking observed in the PXP Hamiltonian. 

We investigate the Lanczos coefficients of the PXP Hamiltonian in detail. This allows us to study the SU(2) algebra breaking since the underlying symmetry algebra controls the Lanczos coefficients of any system. We find that scaling the PXP Hamiltonian by a factor (which we determine numerically) gives rise to a Hamiltonian that approximately follows a $q$-deformed SU(2) algebra, denoted as $\mathrm{SU}_q(2)$. We study the system for sizes $N = 12$ to $30$, and find a system-size-dependent $q$ value. Extrapolation to $1/N \rightarrow 0$ gives us the $q$ value in the thermodynamic limit. We explicitly write the algebra and determine the algebra-breaking terms. 

We finally consider first-order perturbative correction to the PXP Hamiltonian. Here, we show that a scaled perturbed PXP Hamiltonian is very well approximated by an SU(2) algebra for the N\'eel state, with some explicit algebra-breaking terms. We then evaluate the Krylov basis wave functions and complexity for the N\'eel state and the $\ket{0}$ state. The N\'eel state is found to demonstrate strong revival with nearly full oscillatory behavior for the Krylov wave functions and complexity. The growth rate for the complexity is much slower [characteristic of SU(2) algebra] than the growth without perturbation. Therefore, adding first order perturbations moves the approximate algebra from $q < 1$ to $q\approx1$. Adding first-order perturbation initiates a stronger ergodicity breaking and a nearly exact division of the Hilbert space into the thermalizing and non-thermalizing parts.

The paper is organized as follows. In Section \ref{secII}, we provide a brief review of Krylov complexity for states (spread complexity) and describe the Lanczos algorithm. In Section \ref{secIII}, we briefly review $q$-deformed SU(2) algebra. We provide our analytical and numerical results in Section \ref{secIV} for the paramagnetic Hamiltonian and the PXP Hamiltonian. In Section \ref{secV}, we present the results for both PXP Hamiltonian and perturbed PXP Hamiltonian and describe a physical picture to understand the behavior of Lanczos coefficients and complexity. Section \ref{secVI} summarizes our results and proposes future directions.

\section{Krylov (spread) complexity for states}
\label{secII}
Here we introduce the ideas of Krylov complexity for quantum states \cite{Balasubramanian:2022tpr}. This is a somewhat different formalism as compared to the formalism of Krylov complexity for describing operator growth \cite{Parker:2018yvk}. This notion finds various applications in the study of chaos, scrambling and integrability in many-body quantum and semiclassical systems \cite{Avdoshkin:2019trj, Dymarsky:2019elm, Barbon:2019wsy, Jian:2020qpp, Rabinovici:2020ryf, PhysRevE.104.034112, Cao:2020zls, Rabinovici:2021qqt, Yates:2021asz, Kim:2021okd, Caputa:2021ori, Trigueros:2021rwj, Bhattacharjee:2022vlt, Caputa:2021sib, Patramanis:2021lkx, Hornedal:2022pkc, Heveling:2022hth, Heveling:2022orr, Bhattacharya:2022gbz, Banerjee:2022ime, Rabinovici:2022beu, Liu:2022god}.

To describe the notion of Krylov complexity for states, we consider the Hamiltonian evolution of a quantum state under a time-independent Hamiltonian
\begin{align}
\ket{\Psi(t)} = e^{-i H t} \ket{\Psi(0)}\,,
\end{align}
where $\ket{\Psi(0)}$ is not an eigenstate of the Hamiltonian $H$. This is a well-known quantum quench protocol \cite{RevModPhys.83.863}. This time evolution can be visualized (by writing a series expansion for $e^{-i H t}$) as the successive application of $H$ on the state $\ket{\Psi(0)}$. This generates a basis $\{H^n \ket{\Psi(0)}, \, n \in \mathbb{N}\}$ in which the time evolved state $\ket{\Psi(t)}$ can be expanded into, as a Taylor series in $t$. From a physical perspective, it quantifies the \textit{spreading} of the state $\ket{\Psi(0)}$ in the Hilbert space $\mathcal{H}$. Generally, there exist an infinite number of possible choices of such bases to describe the time evolution of $\ket{\Psi(0)}$. The question, then, is to find the most \textit{optimal} basis in the sense of minimizing some \textit{cost function}. The way to construct such an optimal basis is by orthonormalizing the above basis in a way similar to the Gram-Schmidt orthonormalization process. One performs a recursive algorithm known as the Lanczos algorithm \cite{viswanath1994recursion, Balasubramanian:2022tpr}
\begin{align}
    \ket{A_{n+1}} &= (H - a_n) \ket{\mathcal{K}_n} - b_n \ket{\mathcal{K}_{n-1}} \,, \nonumber \\ \ket{\mathcal{K}_n} &= b_n^{-1} \ket{A_{n}}\,. \label{lancalgo1}
\end{align}
where one starts from an initial state $\ket{\mathcal{K}_0}$ and recursively orthonormalizes the states $\ket{A_n}$. The constants $a_{n}$'s and $b_{n}$'s (with $b_0 = 0$) are known as Lanczos coefficients and usually obtained by the normalization factor from the orthonormalization procedure as
\begin{align}
    a_n = \braket{\mathcal{K}_n| H | \mathcal{K}_n}\,, ~~~~ b_n = \sqrt{\braket{A_n|A_n}}\,. \label{lancalgo2}
\end{align}
The basis thus generated is known as the Krylov basis $\ket{\mathcal{K}_n}$ with $n = 0, 1, \ldots$. The action of Hamiltonian on this basis is given by
\begin{align}
    H \ket{\mathcal{K}_{n}} = a_{n} \ket{\mathcal{K}_{n}} + b_{n} \ket{\mathcal{K}_{n-1}} + b_{n+1} \ket{\mathcal{K}_{n+1}}\,,
\label{krylov t-b}    
\end{align}
On such basis, the Hamiltonian takes a symmetric tridiagonal form, where the primary diagonal elements are given by $a_{n}$'s, and the subdiagonal and the superdiagonal elements consist of $b_{n}$'s. By performing the above procedure, we readily express the time evolution of the state on the Krylov basis as
\begin{align}
\ket{\Psi(t)} = \sum_{n} \psi_{n}(t) \ket{\mathcal{K}_{n}}\,,
\end{align}
where $\psi_{n}(t)$'s are known as the Krylov basis functions, and they are complex in general. They are obtained 
by solving the following recursive differential equation
\begin{align}
    i \partial_t \psi_{n} (t) = a_{n} \psi_{n} (t) + b_{n+1} \psi_{n+1} (t) +  b_{n} \psi_{n-1} (t)\,. \label{diffe}
\end{align}
with $b_{0} = 0$. The conservation of probability implies $\sum_{n} |\psi_{n} (t)|^2 =1$. The first element $\psi_{0}(t) = \braket{\Psi(t)|\Psi(0)} \equiv S(t)$ is often called the ``return amplitude'' or the autocorrelation function in the operator context \cite{Parker:2018yvk}. Further, this motivates us to define the complexity as
\begin{align}
    C(t) = \sum_{n} n |\psi_{n} (t)|^2\,, \label{krykomp}
\end{align}
which is minimized by the above choice of Krylov basis \cite{Balasubramanian:2022tpr}. In this sense, complexity acts as a natural ``cost functional.'' For practical purposes, it has to be computed in case-by-case examples. However, if the evolution Hamiltonian possesses some symmetry, then one can directly obtain the corresponding Lanczos coefficients and compute the complexity (see Appendix A). In later sections, we see how to understand such a case having SU(2) symmetry.

\section{$q$-deformed SU(2) algebra and Lanczos coefficients}
\label{secIII}
This section briefly reviews $q$-deformed SU(2) algebra. The notion of $q$ deformations has been studied since the late 1990s. Extensive studies have included studies of spin chains, Dirac oscillators, conformal quantum mechanics, and many others \cite{Biedenharn_1989, Macfarlane_1989, Batchelor_1990, Lavagno:2006jx, Youm:2000yu}. We will focus on $q$ deformations of SU(2)-like algebras. For a better understanding of $q$ deformations and their various applications, we direct the readers' attention to the references \cite{chaichian1996introduction, BONATSOS1999537, schmidt2006q, wess2000q, DING201618, Fujikawa:1996fi, fujikawa1997schwinger, pradeep2020dynamics, DACOSTA20192729, swamy2003deformed, Bonatsos:1996qb}. 

Canonically, $q$ numbers are defined as a version of ordinary numbers parametrized by $q$, with a correspondence
\begin{align}
    [x]_q := \frac{q^x - q^{-x}}{q - q^{-1}}\,, ~~~~~~ \lim_{q \rightarrow 1} [x]_q = x\,,
\end{align}
i.e., the ordinary numbers are recovered in the limit $q \rightarrow 1$. The parameter $q$ could be a real number, or it could be a complex phase. For real case, we write $q = e^{\tau}$,  while $q = e^{i\tau}$ for complex phase. In either case, $\tau \in \textbf{R}$. However, in this paper, we only take it as a positive real number. Some examples of the $q$ numbers are $[0]_q = 0$, $[1]_q = 1$, $[2]_q = q + q^{-1}$ and $[3]_q = 1+ q^2 + q^{-2}$ etc. 

It is important to note that the $q$ numbers can be alternately expressed in terms of Chebyshev polynomials of the second kind \cite{Gavrilik:2009pu, Sinha:2022sdo}
\begin{align}
    [n]_q = U_{n-1} (x)\,, \label{cheb1}
\end{align}
where $x = (q+q^{-1})/2$. The above identity can be easily verified by parametrization of $q$, and using the identity $U_{n-1}(\cos \theta) = \sin (n \theta)/\sin \theta$. It is also evident from the definition that $q$-numbers are symmetric under the transformation $q \rightarrow 1/q$. Therefore, for this work, we restrict ourselves to the region $0 < q \leq 1$. Any value of $q$ outside this region can be mapped into it by $q \rightarrow 1/q$.

The $q$-deformed SU(2), known as $\mathrm{SU}_q(2)$ \cite{Kulish1983}, is generated by three generators $J_{0}$ and $J_{\pm}$, with the following algebra \cite{Biedenharn_1989, Macfarlane_1989}
\begin{align}
    [J_{0}, J_{\pm}] = \pm  J_{\pm}\,, ~~~~~  [J_{+}, J_{-}] =[2 J_{0}]_q\,, \label{suq2}
\end{align}
where $[2 J_{0}]_q$ has to be understood as a $q$-deformed operator. For two cases, where $q$ can be real or complex phase, we expand the second commutation relation as \cite{BONATSOS1999537}
\begin{align}
    [J_{+}, J_{-}] &= \frac{1}{\sinh \tau} \sum_{n=0}^{\infty} \frac{(2 \tau J_{0})^{2n+1}}{(2n+1)!}\,~~~~~~~~~~ \mathrm{for}~~ q=e^{\tau}\,, \nonumber \\
    [J_{+}, J_{-}] &= \frac{1}{\sin \tau} \sum_{n=0}^{\infty} (-1)^{n} \frac{(2 \tau J_{0})^{2n+1}}{(2n+1)!}\,\,~~~~ \mathrm{for}~~ q=e^{i\tau}\,.
\end{align}
Hence, the algebra $\mathrm{SU}_q(2)$ is a non-linear generalization of standard SU(2). The limit $q \rightarrow 1$ implies $\tau \rightarrow 0$, where we recover the usual SU(2) algebra.

\begin{figure}[t]
%\subfigure[]{
\includegraphics[height=5.7cm,width=\linewidth]{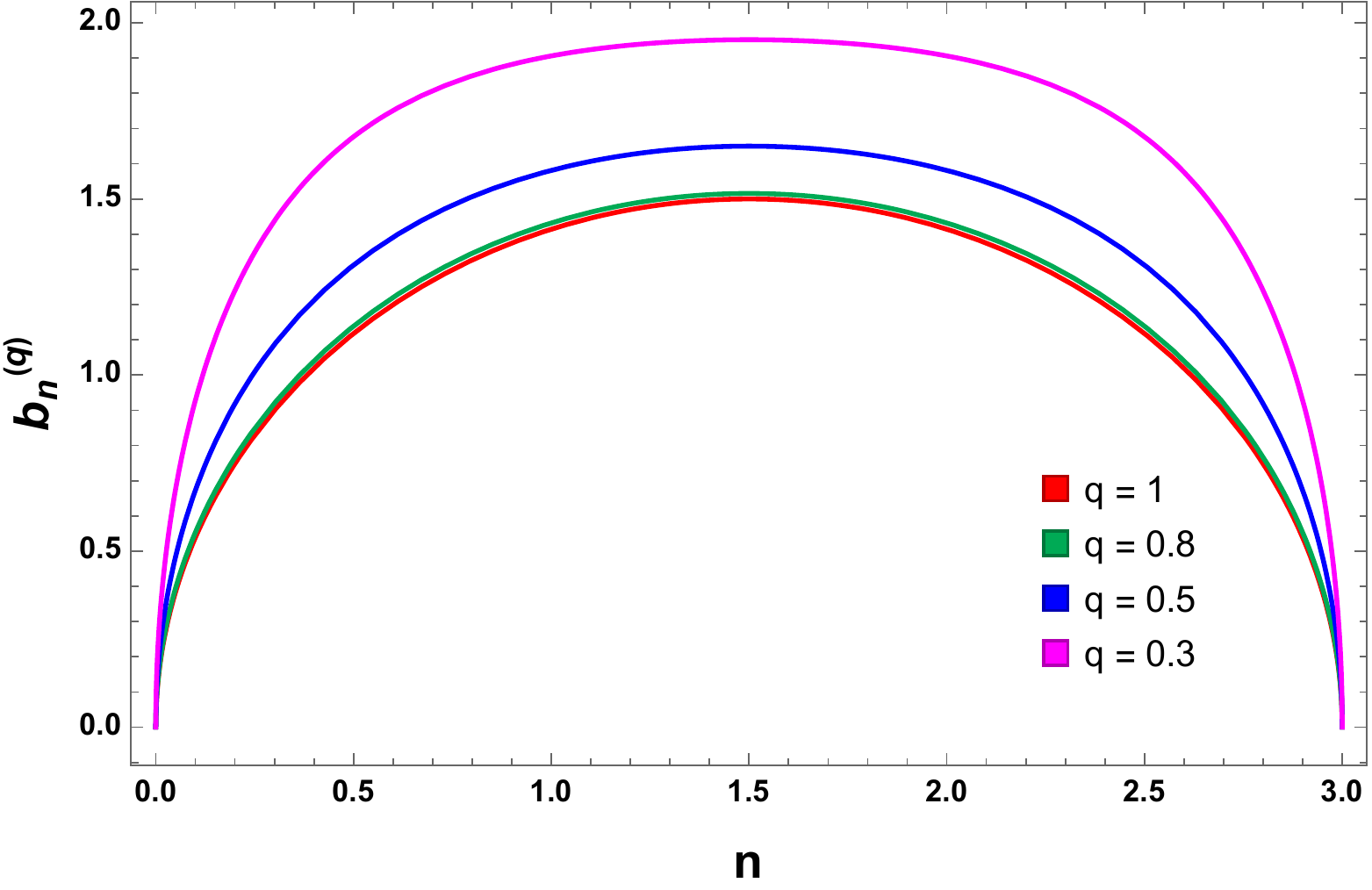}
%}
%\hfill
%\subfigure[]{\includegraphics[height=6cm,width=\linewidth]{bncorsu2.pdf}\label{bncorsu2}}
\caption{Growth of $b_{n}$'s [from Eq.\eqref{cheb1}] for $j = 1$ for various values of $q$. Although the $n$ takes a discrete value, we use a continuum $n$ for the plotting. This visualization will be useful for understanding the behavior of the Lanczos coefficients in later sections.}
\label{bnsu2}
\end{figure}

The natural basis of this algebra is given by $\ket{j, n}$ where $-j \leq n \leq j$. The states can be formed by repeatedly acting the annihilation ladder operator $J_{-}$ on the highest-weight state $\ket{j,j}$ or the creation ladder operator $J_{+}$ on the lowest-weight state $\ket{j,-j}$. Here, we choose the latter, with a slight abuse of notation $n \rightarrow j+n$, to make it consistent with \cite{Caputa:2021sib, Balasubramanian:2022tpr}. We have the states
\begin{align}
\ket{j, - j + n} = \sqrt{\frac{[\Gamma(2j-n+1)]_q}{[n]_q! \,[\Gamma(2j+1)]_q}} \, J_{+}^n \ket{j,-j}\,,
\end{align}
where $n = 0, \cdots, 2j$. The action of the generators is defined as \cite{fujikawa1997schwinger}
\begin{align*}
    J_{0} \ket{j, - j+n } &= (-j+n) \ket{j, - j+n }\,, \nonumber \\
    J_{+} \ket{j, - j+n } &= \sqrt{[n+1]_q\, [2j-n]_q}\, \ket{j, - j+n+1 }\,, \nonumber \\
    J_{-} \ket{j, - j+n } &= \sqrt{[n]_q\, [2j-n+1]_q} \,\ket{j, - j+n-1}\,.
\end{align*}
Here the ordinary numbers are replaced by the $q$ numbers. We consider the Hamiltonian of the form
\begin{align}
    H = \alpha (J_{+} + J_{-}) +  \eta_0 \,J_{0} + \delta 1\,,
\end{align}
where $\alpha$, $\beta$ and $\gamma$ are some model-dependent numbers. The Krylov basis is formed by the basis vectors of $\mathrm{SU}_q(2)$, i.e., $\ket{\mathcal{K}_{n}} = \ket{j, - j+n }$. It is straightforward to compute the Lanczos coefficients as
\begin{align}
    a_{n}^{(q)} =  \eta_0(-j+n) + \delta\,, ~~~ b_{n}^{(q)} = \alpha \sqrt{[n]_q\, [2j-n+1]_q}\,.
\end{align}
Using \eqref{cheb1}, we write them in terms of Chebyshev polynomials as
\begin{align}
    b_{n}^{(q)} &= \alpha \sqrt{U_{n-1}(x)\, U_{2j-n}(x)} = \alpha \bigg[\sum_{k = 0}^{n-1} U_{2j-2n+2k+1}(x) \bigg]^{1/2}\,. \label{chebimp}
\end{align}
where in the last equality, we have used the product formulas of Chebyshev polynomials. It is tempting to call them $q$-Lanczos coefficients, generated from the $q$-deformed algebra. However, for simplicity, we continue to call them Lanczos coefficients and denote them by $b_n$. The Lanczos coefficients for different values of $q$ are shown in Fig.\,\ref{bnsu2}.

\section{Numerical Results}
\label{secIV}
\subsection{Paramagnetic Hamiltonian}

To begin with, we consider the simple paramagnetic Hamiltonian $H_p = \sum_{n=1}^N \sigma^x_{n}$ \cite{PhysRevB.98.155134} with periodic boundary condition. Since we are interested in the time evolution of the $\mathbf{Z}_{2}$ symmetry-broken N\'eel state, we take $N$ to be even. The Hamiltonian can be separated into two different parts as $H_p = H^{+} + H^{-}$ where
\begin{align}
    H^{\pm} = \sum_{n} (\sigma_{2n}^{\pm} + \sigma_{2n-1}^{\mp})\,.
\end{align}

\begin{figure}[t]
\subfigure[]{\includegraphics[height=5.7cm,width=\linewidth]{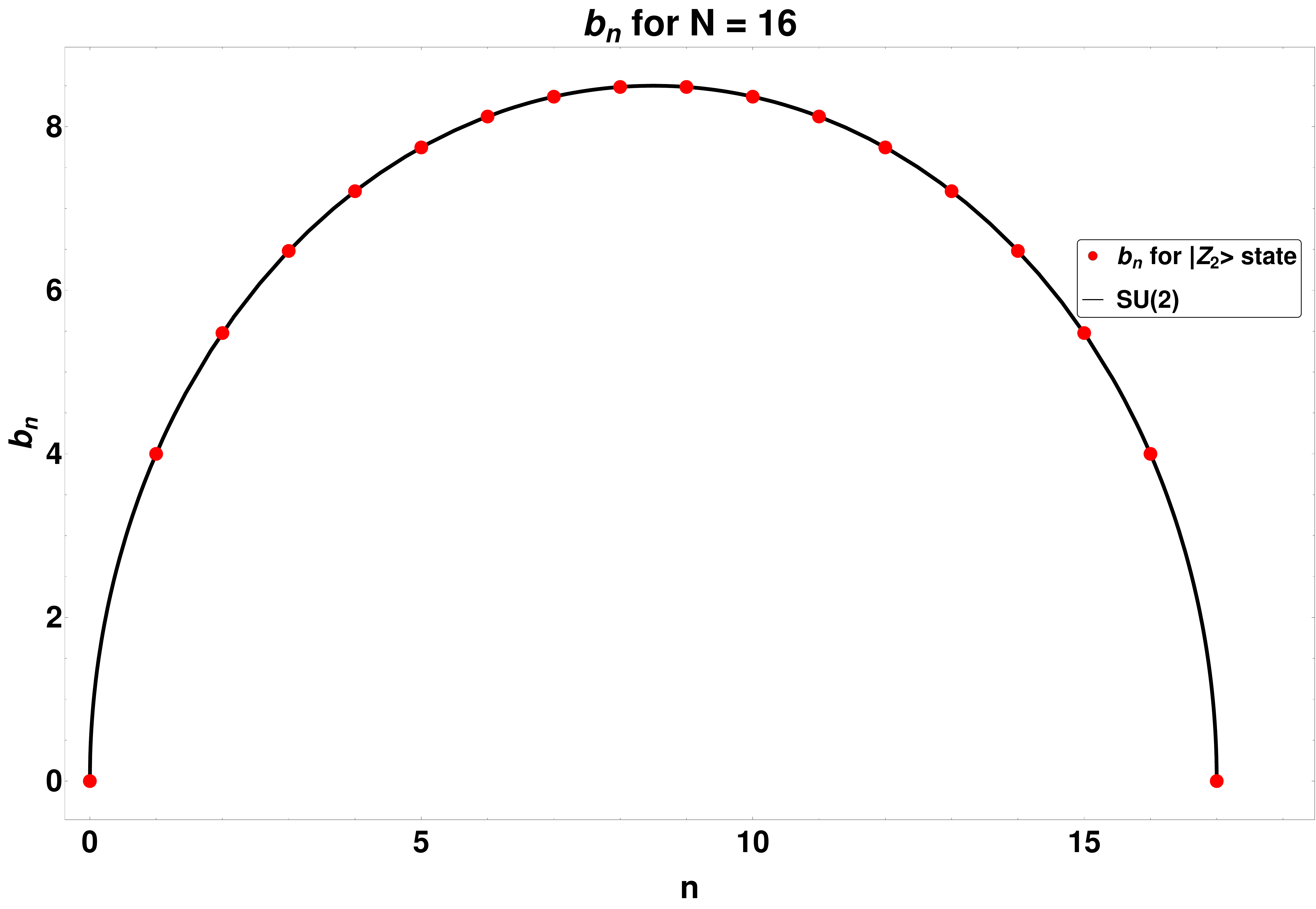}\label{para0}}
\hfill
\subfigure[]{\includegraphics[height=5.7cm,width=\linewidth]{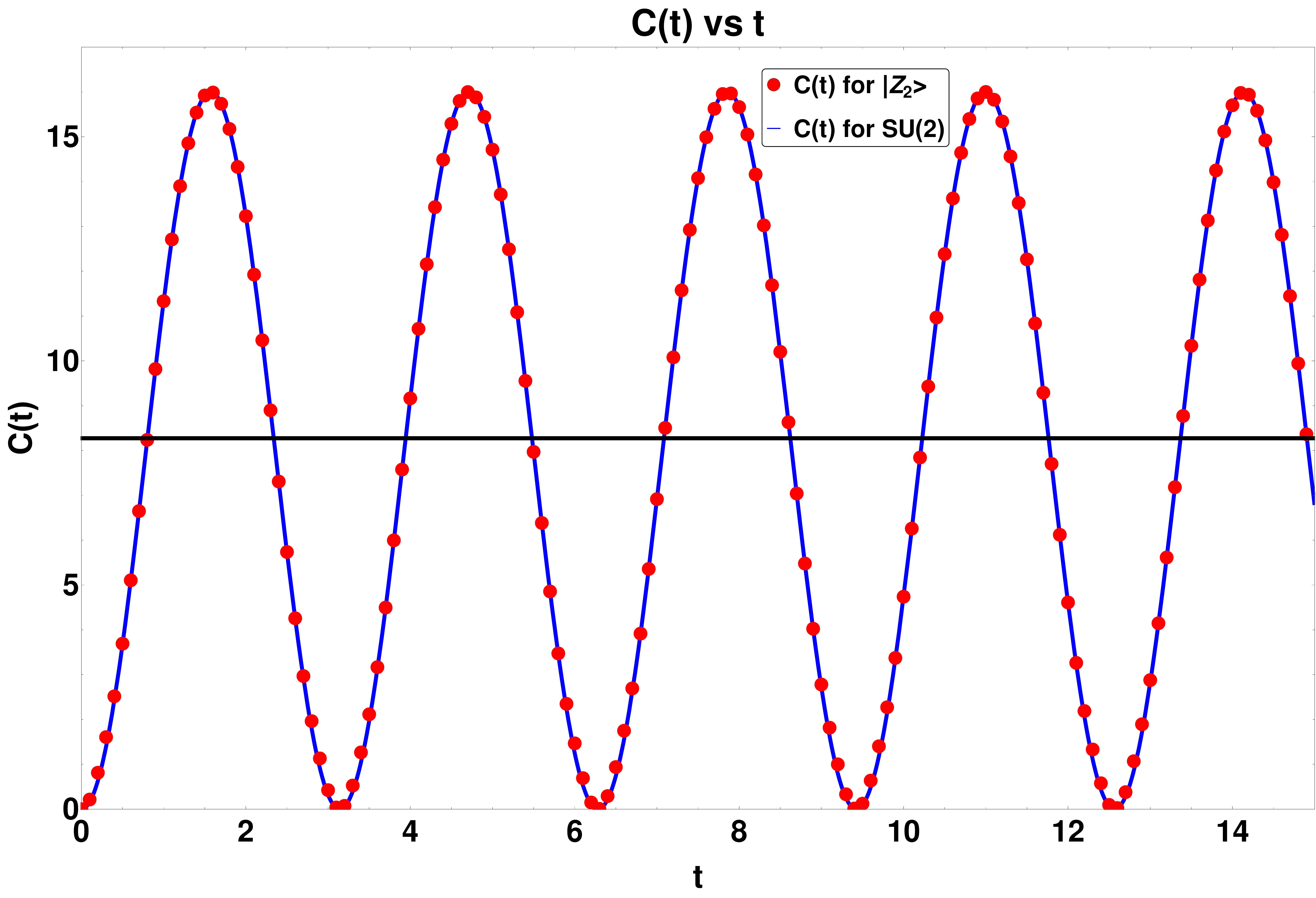}\label{paracomp}}
\caption{(a) Growth of $b_{n}$'s and (b) evolution of complexity $C(t)$ for the paramagnetic Hamiltonian $H_p = \sum_{n=1}^N \sigma^x_{n}$, initialized in the $\ket{\textbf{Z}_2}$ state for a system of lattice size $N=16$. The expression for the Lanczos coefficients and the complexity is given by Eq.\eqref{bnparan} and Eq.\eqref{comparan}, respectively. The Lanczos coefficients exactly terminate at $n = N + 1 =17$, which implies the dimension of the Krylov subspace is $K = 17$.} 
\end{figure}
Here $\sigma^{\pm}_{n} = (\sigma^x_{n} \pm i \sigma^y_{n})/2$. The two parts follow the exact SU(2) algebra, namely, $[H^z, H^{\pm}] = \pm H^{\pm}$ and $[H^{+}, H^{-}] = 2 H^z$ with $H^z =  \sum_{n} (\sigma_{2n}^z - \sigma_{2n-1}^z)/2$, furnishing $\ket{\mathbf{Z}_{2}}$ and $\ket{\mathbf{Z}_{2}'}$ as the lowest- and the highest-weight states, respectively, in the representation of spin $j=N/2$ \cite{PhysRevLett.122.220603}. Choosing the initial state as the N\'eel state $\ket{\mathbf{Z}_{2}}$, the time evolved state $\ket{\Psi(t)} = e^{- i H_p t} \ket{\mathbf{Z}_{2}} = \sum_{n=0}^{N} \psi_{n} (t) \ket{\mathcal{K}_{n}}$ spreads over the Krylov basis, constructed from the basis function $\psi_{n} (t)$ and the basis vectors $\ket{\mathcal{K}_{n}}$. As said before, the lowest- and the highest-weight states are $\ket{\mathcal{K}_{0}} = \ket{\mathbf{Z}_{2}}$ and $\ket{\mathcal{K}_{n}} = \ket{\mathbf{Z}_{2}'}$, respectively. The fidelity for the $\ket{\mathbf{Z}_{2}}$ state shows exact revival. In fact, for any other initial state such as $\ket{\mathbf{Z}_{2}'}$ , $\ket{\mathbf{Z}_{3}}$, $\ket{\mathbf{Z}_{4}}$, etc. we shall see perfect revival in fidelity due to the integrable nature of the Hamiltonian.
%\sout{The fidelity for the $\ket{\mathbf{Z}_{2}}$ state shows exact revival due to the strong overlap with the scar eigenstates}.
The exact representation allows us to infer the Lanczos coefficients (see Appendix A) directly as
\begin{align}
    a_{n} = 0\,, ~~~~ b_{n} = \sqrt{n(N-n + 1)}\,. \label{bnparan}
\end{align}
The Lanczos coefficients $b_{n}$ show the maximum at $n= (N+1)/2$ and terminates at $n = N+1$ (see Fig.\,\ref{para0}), which is the dimension of the Krylov space. The wave functions are given by $\psi_{n} (t) = (^NC_{n})^{1/2} (i \cot t)^{-n} \cos^{N}t$, and they are related by the recursion \eqref{diffe}. Complexity is simply computed using \eqref{krykomp} as
\begin{align}
    C(t) = \sum_{n=0}^{N} n |\psi_{n} (t)|^2 = N \sin^2 t\,, \label{comparan}
\end{align}
which is periodic with a time period $T=\pi$, same as the revival of the survival amplitude $S(t) = \psi_{0} =  \braket{\Psi(t)|\textbf{Z}_{2}} = \cos^{N} t$. Moreover, it reaches maxima of a value $N$ at half-period $T=\pi/2$. This is expected as $\ket{\mathbf{Z}_{2}}$ does not show decoherence, and complexity should not decay with time. The numerical plot is shown in Fig.\,\ref{paracomp}. The long-time average of complexity is $\Bar{C} = N/2$, which
is extensive in system size. All the analytic expressions are perfectly consistent with the numerical results in Figs.\,\ref{para0} and \ref{paracomp}, obtained by the direct implementation of the Lanczos algorithm Eq.\eqref{lancalgo1}-\eqref{lancalgo2}.

%\begin{align}
%    \Bar{C} = \lim_{T \rightarrow \infty} \frac{1}{T} \int_{0}^T C(t)\, dt = \frac{N}{2}\,,
%\end{align}

\subsection{PXP Hamiltonian}
Now we turn to the more complicated PXP Hamiltonian. The Hamiltonian is \cite{Turner_2018}
\begin{align}
    H_{\mathrm{PXP}} = \sum_{m=1}^N P_{m-1} \sigma_{m}^x  P_{m+1}\,, \label{pxp}
\end{align}
where $P = \ket{0}\bra{0}$ is the projector and $\sigma_x = \ket{0}\bra{1} + \ket{1}\bra{0}$ is the Pauli X-matrix. We consider a system of $N$ sites with periodic boundary condition, for which $P_{0} = P_{N}$ and $P_{N+1} = P_1$. The dimension of the Hilbert space can be shown to be $D_{N} = F_{N+1} + F_{N-1} \sim \phi^N$ for large $N$, where $F_{N}$ is the $N$-th Fibonacci number and $\phi = 1.61803\cdots$ is the golden ratio. The Hamiltonian possesses translational, and inversion symmetries \cite{PhysRevLett.122.173401}. The presence of the projectors in the PXP Hamiltonian ensures that no two adjacent sites are in the excited $\ket{1}$ state) (see \cite{PhysRevB.69.075106} for similar hard-boson model), and therefore this Hamiltonian is non-integrable and thermalizing. However, its thermalizing nature is sensitive to the choice of the initial state. It is observed that the $\ket{\mathbf{Z}_{2}}$ state, written as $\ket{010101\cdots}$, shows weak thermalization \cite{Turner_2018}. States without any $\mathbf{Z}$ symmetries are known to thermalize much faster. Even states with larger $\mathbf{Z}$ symmetries, like $\ket{\mathbf{Z}_{3}}$, also thermalize slower than a generic state in the system. However, their thermalizing nature is stronger than the $\ket{\mathbf{Z}_{2}}$ state. These states are related to the many-body scar states (i.e., the weakly entangled eigenstates) of the PXP Hilbert space. Specifically, such $\mathbf{Z}$ states are known to be comprised of superposition of the weakly entangled eigenstates \cite{PhysRevLett.122.220603}.

Similar to the paramagnetic case, we split the Hamiltonian as $H_{\mathrm{PXP}} = H_{+} + H_{-}$ as \cite{PhysRevB.98.155134}
\begin{align}
    H_{\pm} = \sum_{m \in \mathrm{odd}} P_{m-1} \sigma_{m}^{\mp} P_{m+1} + \sum_{m \in \mathrm{even}} P_{m-1} \sigma_{m}^{\pm} P_{m+1}\,.\label{Hpm}
\end{align}
However, in this case, the generators $H_{\pm}$ 
satisfy $[H_{+}, H_{-}] = \sum_{m} (-1)^{m} P_{m-1} \sigma^z_{m} P_{m+1}$, and $[H_z, H_{\pm}] \approx \pm H_{\pm}$, i.e., it does not obey the exact SU(2); it only obeys approximately, marked by the notation ``$\approx$'' \cite{PhysRevB.98.155134}. It is indeed possible to associate the $H_{\pm}$ to a broken SU(2) algebra via appropriate scaling of the $H_{\pm}$ generators. Due to forming such a ``broken'' algebra, the fragmentation of the PXP Hamiltonian is only approximate $H_{\mathrm{PXP}} \approx H_{\mathrm{nonth}} \bigoplus H_{\mathrm{th}}$. Hence, we cannot apply the above analytic tools as we did in the previous section. 

\begin{figure}[h]
\includegraphics[height=5.7cm,width=\linewidth]{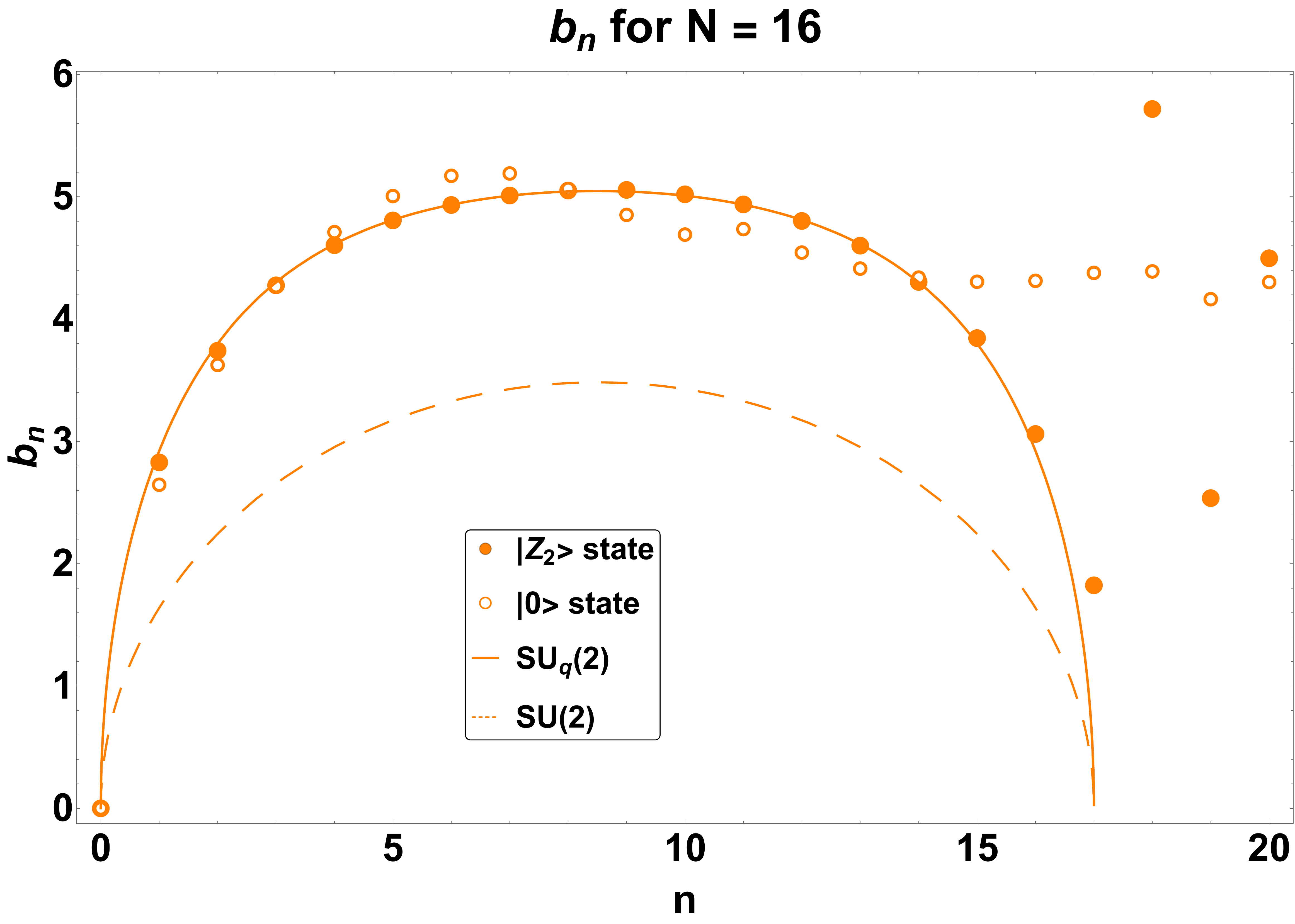} 
\caption{Growth of $b_{n}$'s (disks) for the $\ket{\mathbf{Z}_{2}}$ state versus the $q$-deformed SU(2) (thick line) result for the PXP Hamiltonian \eqref{pxp}. The standard SU(2) result is given for comparison (dashed line). The $b_{n}$'s for a generic state ($\ket{0}$ state; without any $\mathbf{Z}$ symmetry) is also plotted (circles). Here we choose the system size $N = 16$ and $K \sim 20$ Krylov basis vectors. Around $K \sim N+1$, the state is driven out of the approximate Krylov subspace. This is in contrast with the paramagnetic Hamiltonian in Fig.\,\ref{para0}, where the Krylov subspace was exact and shown by the dashed line in this figure.}\label{pxpbnm}
\end{figure}

\begin{figure}[t]
\includegraphics[height=5.7cm,width=\linewidth]{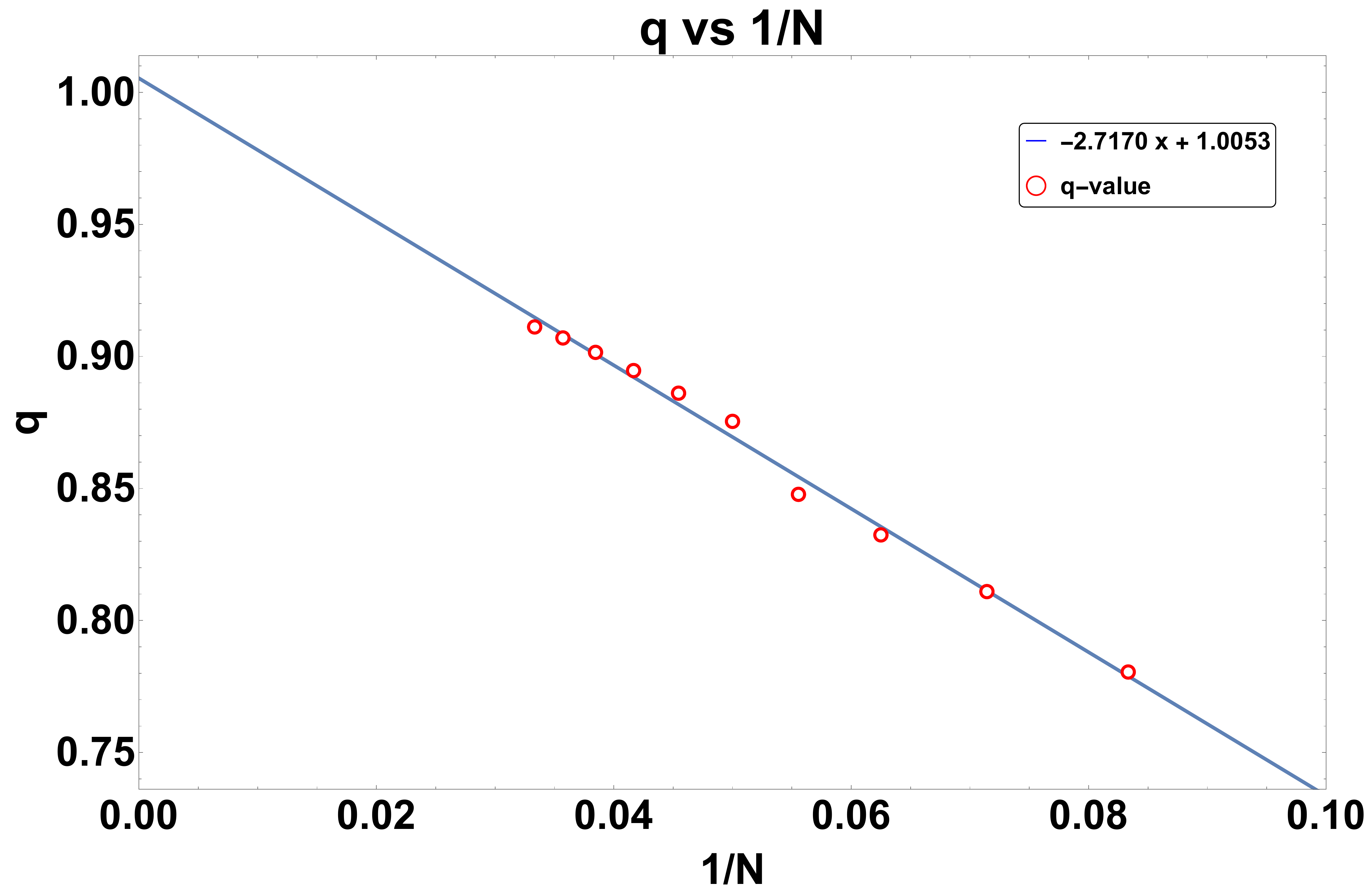}
\caption{$q$ values versus $1/N$ for the PXP model. Possibly due to finite-size effects, the $q$ values are different for different system sizes $N$. The asymptotic value turns out to be $q_\infty = 0.9947 \pm 0.0044$, which is close to the SU(2) value ($q = 1$). The system sizes considered in this are $N = 12, 14, \cdots, 30$. The linear regression fit has an $R^2$ value of $0.992$ and a standard error of $0.0040$.}\label{fig:lin}
\end{figure}
We start from two different initial states, $\ket{\mathbf{Z}_{2}}$ and a generic state $\ket{0}$ without any $\mathbf{Z}$ symmetry. We evolve them by the PXP Hamiltonian. Our results for $b_n$ are presented in Fig.\,\ref{pxpbnm} for $N = 16$. Interestingly enough, we find that the $a_n$'s turn out to be exactly zero numerically which provides us an additional motivation for breaking the Hamiltonian into of $H_{\pm}$ as an approximation of the SU(2) algebra for PXP [see Eq. \eqref{symH} and \eqref{ll}]. We approximate the Lanczos coefficients $b_{n}$'s for $\ket{\mathbf{Z}_{2}}$ initial state by the $q$-deformed SU(2) algebra by \eqref{chebimp}. We find that a good approximation to the observed $b_{n}(\mathrm{PXP})$ is given by $\alpha$ in the range $\{0.400,0.442\}$, and the $q$-value is dependent on the system size, due to finite size effects. One may note that the value of $\alpha$ varies with system sizes. However, the variation does not follow any obvious pattern. This leads us to suspect that the reason behind it may be finite size effects and/or inherent numerical inaccuracies in the Lanczos algorithm.

\begin{table}[t]
%\begin{adjustbox}{padding=1em,bgcolor=white,frame=1pt,center}
%\begin{minipage}{\dimexpr\textwidth-1pt-1em}
\begin{tabular}{ |c|c|c| } 
\hline
 $N$ & $q$-value & $\alpha$\\ 
 \hline 
 $N = 12$ & $q_{12} =0.78047$  & $\alpha_{12} = 0.40059$\\
 $N = 14$ & $q_{14} =0.81093$  & $\alpha_{14} = 0.40759$\\
 $N = 16$ & $q_{16} =0.83240$  & $\alpha_{16} = 0.40971$\\
 $N = 18$ & $q_{18} =0.84775$  & $\alpha_{18} = 0.40762$\\
 $N = 20$ & $q_{20} =0.87539$  & $\alpha_{20} = 0.44092$\\
 $N = 22$ & $q_{22} =0.88607$  & $\alpha_{22} = 0.44219$\\
 $N = 24$ & $q_{24} =0.89463$  & $\alpha_{24} = 0.44212$\\
 $N = 26$ & $q_{26} =0.90152$  & $\alpha_{26} = 0.44069$\\
 $N = 28$ & $q_{28} =0.90698$  & $\alpha_{28} = 0.43765$\\
 $N = 30$ & $q_{30} =0.91115$  & $\alpha_{30} = 0.43256$ \\
 \hline
\end{tabular}
\caption{Table outlining the values of $q$ and $\alpha$ obtained numerically via least-square fitting for system sizes $N = 12$ to $N = 30$.}\label{table}
%\end{minipage}
%\end{adjustbox}
\end{table}
The $q$-value (and $\alpha$) is determined individually for each system size via a least-square fit of the Lanczos coefficients. The function $\sum_{n = 0}^{N+1}|b_{n}^{\text{PXP}} - b_{n}^{(q)}|^{2}$ is minimized for $q$ and $\alpha$. The asymptotic value of $q$ at $N \rightarrow \infty$ is obtained by a linear fit on the $q$ versus $1/N$ plot (Fig.\,\ref{fig:lin}). The extrapolation of the fitting gives the value of $q_{\infty}$ to be $1.0053 \pm 0.0044$. However, since we have chosen the convention that $q$ is restricted between $0$ and $1$, and since the quantity $b^{(q)}_{n}$ remains same under $q \rightarrow 1/q$, we choose $q_{\infty} = 1/1.0053 = 0.9947$.
%\textcolor{blue}{This gives us the value $q_{\infty} = 0.9947 \pm 0.0044$. The fitting data extrapolates to $1.0053$, which is the same as $q_\infty$ due to the $q \rightarrow 1/q$ symmetry of $b^{(q)}_{n}$.}\footnote{\textcolor{blue}{We clarify the point that as $q$ approaches closer to unity from $1^{-}$, $1/q$ also approaches closer to unity from $1^{+}$. The extrapolation to $q=1.0053$ is purely a numerical artifact.}}

To see the extent of the difference that occurs between $q = 0.9947$ and $q = 1$, it is instructive to parametrize $q = e^{\tau}$ and perform a series expansion near $\tau = 0$ for the same. The first term is  $\mathcal{O}(1)$ term which is the usual SU(2) expression. The next $\mathcal{O}(\tau)$ term turns out to be proportional to $N^{5/2}$. Therefore, one can infer that despite the thermodynamic result, the deviation of the Lanczos coefficient from the SU(2) result is infinitely large. The same can be seen numerically, comparing the difference between the $\mathrm{SU}_{q}(2)$ (or even the PXP) Lanczos coefficients and the SU(2) Lanczos coefficients for different values of $N$. While the value of $q$ for increasing $N$ does increase towards unity, the difference between the $q$-deformed and pure SU(2) Lanczos also increases. Thus, the Lanczos coefficients tell us that even though the thermodynamic limit is close to SU(2) in terms of the parameter $q$, in terms of measurable quantities it is indeed very distant from pure SU(2).

The ``broken'' SU(2) algebra satisfies the following commutation relations
\begin{align}
    [J_{+},J_{-}] &= 2 J_{0}\,, \label{comm1}\\
    [J_{0},J_{\pm}] &= \pm J_{\pm} + \text{extra terms}\,, \label{comm2}
\end{align}
where the PXP Hamiltonian is written as $H =\alpha(J_{+} + J_{-})$. The Lanczos coefficients and the wave functions corresponding to the Krylov basis expansion for an unbroken SU(2) algebra are derived in Appendix A. We note the expression for the Lanczos coefficients below
\begin{align}
    b_{n} = \alpha\sqrt{n (2 j - n + 1)}\,.
\end{align}
A $q$-deformation of such an SU(2) algebra would lead us to the expression \eqref{chebimp}. As mentioned, we numerically obtain the condition that $\alpha \in \{0.400,0.442\}$. The additional terms in \eqref{comm2} reflect the deviation from the SU(2) algebra. We find that $q$-deformation of this algebra is a good approximation to Eq.\eqref{comm1}-\eqref{comm2}. However, it is not exact, and there are still additional terms present, as we will see shortly. Therefore, the algebra satisfied by the PXP ladder operators is a ``broken'' $q$-deformed SU(2).

The algebra in Eqs.\eqref{comm1} and \eqref{comm2} can be explicitly seen to be as follows [\cite{PhysRevB.101.165139} and Eq.\eqref{Hpm}]:
\begin{align}
    H_{\pm} = \sum_{n} \tilde{\sigma}^\mp_{2n + 1} + \tilde{\sigma}^\pm_{2n}\,,
\end{align}
where $\tilde{\sigma}^{(*)}_{m} \equiv P_{m - 1} \sigma^{(*)}_{m} P_{m + 1}$. 
We have the following commutator
\begin{align}
    [H_{+}, H_{-}] = \sum_{n} \tilde{\sigma}^z_{2 n} - \tilde{\sigma}^z_{2 n + 1} \equiv  2 H_{0}\,.
\end{align}
It is straightforward to see that the following commutator holds
\begin{align}
    [H_{0}, H_{\pm}] = &\pm \left(\sum_{n} \tilde{\sigma}^{\mp}_{2n + 1} + \tilde{\sigma}^{\pm}_{2n} \right) \pm X_{\pm}\,,
\end{align}
where the ``extra terms'' $X_{\pm}$ are
\begin{align*}
    - X_{\pm} &= \sum_{m \in \text{odd}} P_{m-2}P_{m - 1}\sigma^{\mp}_{m} P_{m + 1} +  P_{m-1}\sigma^{\mp}_{m} P_{m + 1}P_{m + 2} \notag \\
    &+ \sum_{m \in \text{even}} P_{m-2}P_{m - 1}\sigma^{\pm}_{m} P_{m + 1} +  P_{m-1}\sigma^{\pm}_{m} P_{m + 1}P_{m + 2}\,.
\end{align*}
As mentioned in \cite{PhysRevB.101.165139}, this algebra is readily identified as a broken SU(2), where the algebra breaking terms are $X_{\pm}$. However, such an algebra interpreted as SU(2) implies that we must have $\alpha = 1$, which is not the case as seen from Fig.\,\ref{pxpbnm}. Rather, we find the $\alpha$ values mentioned in Table.\,\ref{table} represent a better approximation to the PXP Hamiltonian's Lanczos coefficients. 

To write the algebra given above as a broken SU(2), we begin by writing the PXP Hamiltonian as $H_{\mathrm{PXP}} = \alpha (J_{+} + J_{-})$ (for some constant $\alpha$) where we have 
\begin{align}
    J_{\pm} = \frac{1}{\alpha}\sum_{n} \tilde{\sigma}^\mp_{2n + 1} + \tilde{\sigma}^\pm_{2n}\,.
\end{align}
For the rest of this discussion, we assume that $\alpha$ is a real number. 
The commutation relations between these ladder operators are
\begin{align}
    [J_{+}, J_{-}] = \frac{1}{\alpha^2}[H_{+}, H_{-}] = \frac{2}{\alpha^2}H_{0} \equiv 2 J_{0}\,.
\end{align}
This gives us the relation that 
\begin{align}
    J_{0} = \frac{1}{\alpha^2}H_{0}\,.
\end{align}
We write the commutator between $J_{0}$ and $J_{\pm}$ as
\begin{align}
    [J_{0}, J_{\pm}] %&= \frac{1}{\alpha^3 \epsilon}[H_{0}, H_{\pm}] =  \pm \frac{1}{\alpha^3 \epsilon}H_{\pm} \pm \frac{1}{\alpha^3 \epsilon} X_{\pm} \notag \\
    &= \pm \frac{1}{\alpha^2}J_{\pm} \pm \frac{1}{\alpha^3} X_{\pm}\,. \label{Jpm}
\end{align}
From Eq.\eqref{comm1}-\eqref{comm2}, we note that this is not exactly a broken SU(2) algebra. To cast it into that form, we absorb a term $\frac{1-\alpha^2}{\alpha^2}J_{\pm}$ into $X_{\pm}$. Therefore, the commutator in Eq.\eqref{Jpm} is better written as
\begin{align}
    [J_{0}, J_{\pm}] = \pm J_{\pm} \pm \tilde{X}_{\pm}\,,
\end{align}
where we have $\tilde{X}_{\pm} = \frac{1}{\alpha^{3}} X_{\pm} + \frac{(1- \alpha^2)}{\alpha^{2}}J_{\pm}$. %This suggests that the PXP Hamiltonian can be well represented by $q$-deformation of an $SU(2)$. 

\subsection{Broken $q$-deformed SU(2) algebra for the PXP Hamiltonian}
Interpreting the algebra of PXP Hamiltonian as a version of $q$-deformed algebra, we have
\begin{align}
    [J_{+}, J_{-}] &= [2 J_{0}]_q\,, \label{qcomm1}\\
    [J_{0}, J_{\pm}] &= \pm J_{\pm}\,, \label{qcomm2}
\end{align}
where $[2  J_{0}]_q = \frac{\sinh{2 \tau J_{0}}}{\sinh{\tau}}$, using $q = e^{\tau}$. This relation can, in principle, be inverted, provided the inversion is treated as a series. Therefore, we would have
\begin{align*}
    J_{0} &= \frac{1}{2\tau}\arcsinh([2 J_{0}]_q \sinh\tau) \,, \notag \\ 
    &= \frac{1}{2\tau }\sum_{n = 0}^{\infty}\frac{(-1)^{n}(2n)!}{2^{2n}(n!)^{2}(2n+1)}(\sinh\tau)^{2n+1}([J_{+}, J_{-}])^{2n + 1}\,.
\end{align*}
We evaluate $[J_{0}, J_{\pm}]$ and demonstrate that it is close to $\pm J_{\pm}$. This can be done by considering the series expansion given above term-by-term. We consider only the first order correction to the $J_{0}$
\begin{align}
    J_{0} = \frac{\sinh\tau}{2\tau}[J_{+}, J_{-}] - \frac{(\sinh\tau)^3}{12 \tau }[J_{+}, J_{-}]^{3} + \cdots\,,
\end{align}
Note that in the $\tau \rightarrow 0$ limit, this reduces to the usual commutation relation. 
Splitting the Hamiltonian in (as before) $H_{\mathrm{PXP}} = \alpha (J_{+} + J_{-})$, we know that these follow the commutation relations \eqref{comm1}-\eqref{comm2}
\begin{align}
    \frac{1}{2}[[J_{+}, J_{-}],J_{\pm}] = \pm (J_{\pm} + \tilde{X}_{\pm})\,.
\end{align}
We introduce the notation $Q = [J_{+}, J_{-}]$ and use the relation $[Q, J_{\pm} ] = \pm 2 ( J_{\pm} + \tilde{X}_{\pm})$. Evaluating the commutator $[J_{0}, J_{\pm}]$ gives us
\begin{widetext}
\begin{align}
    [J_{0}, J_{\pm}] %&= \frac{\sinh\tau}{2\tau \epsilon}[Q, J_{\pm}]- \frac{(\sinh\tau)^3}{12 \tau \epsilon}[Q^{3},J_{\pm}] + \cdots \notag \\
    %&= \pm \frac{\sinh\tau}{\tau \epsilon} (\epsilon \delta J_{\pm} + \epsilon \tilde{X}_{\pm}) - \frac{(\sinh\tau)^3}{12 \tau \epsilon}\left(Q^2[Q,J_{\pm}] + Q[Q,J_{\pm}]Q + [Q,J_{\pm}]Q^2\right) + \cdots \notag \\
    &= \pm \frac{\sinh\tau}{\tau} ( J_{\pm} + \tilde{X}_{\pm}) \mp \frac{(\sinh\tau)^3}{6 \tau }\left(Q^2 ( J_{\pm} +  \tilde{X}_{\pm}) + Q(J_{\pm} +  \tilde{X}_{\pm})Q + ( J_{\pm} + \tilde{X}_{\pm})Q^2\right) + \cdots\,. \notag
\end{align}
From here, we can read off the extra terms in the commutator as
\begin{align}
    [J_{0}, J_{\pm}]_{\text{extra}} &= \pm \frac{\sinh\tau}{\tau} \tilde{X}^\pm \mp \frac{(\sinh\tau)^3}{6 \tau }\left(Q^2 ( J_{\pm} +  \tilde{X}_{\pm}) + Q(  J_{\pm} +  \tilde{X}_{\pm})Q + (  J_{\pm} + \tilde{X}_{\pm})Q^2\right) \mp (1 - \frac{\sinh\tau}{\tau})J_{\pm} \cdots\,. \label{Extra}
\end{align}
Therefore, we find that the PXP Hamiltonian, when written in terms of ladder operators $J_{\pm}$, corresponds to a broken version of the $q$-deformed algebra given in \eqref{qcomm1} and \eqref{qcomm2}. The algebra-breaking terms are given in \eqref{Extra}.
\end{widetext}

\subsection{Perturbative correction}

\begin{figure}[t]
\includegraphics[height=5.8cm,width=1\linewidth]{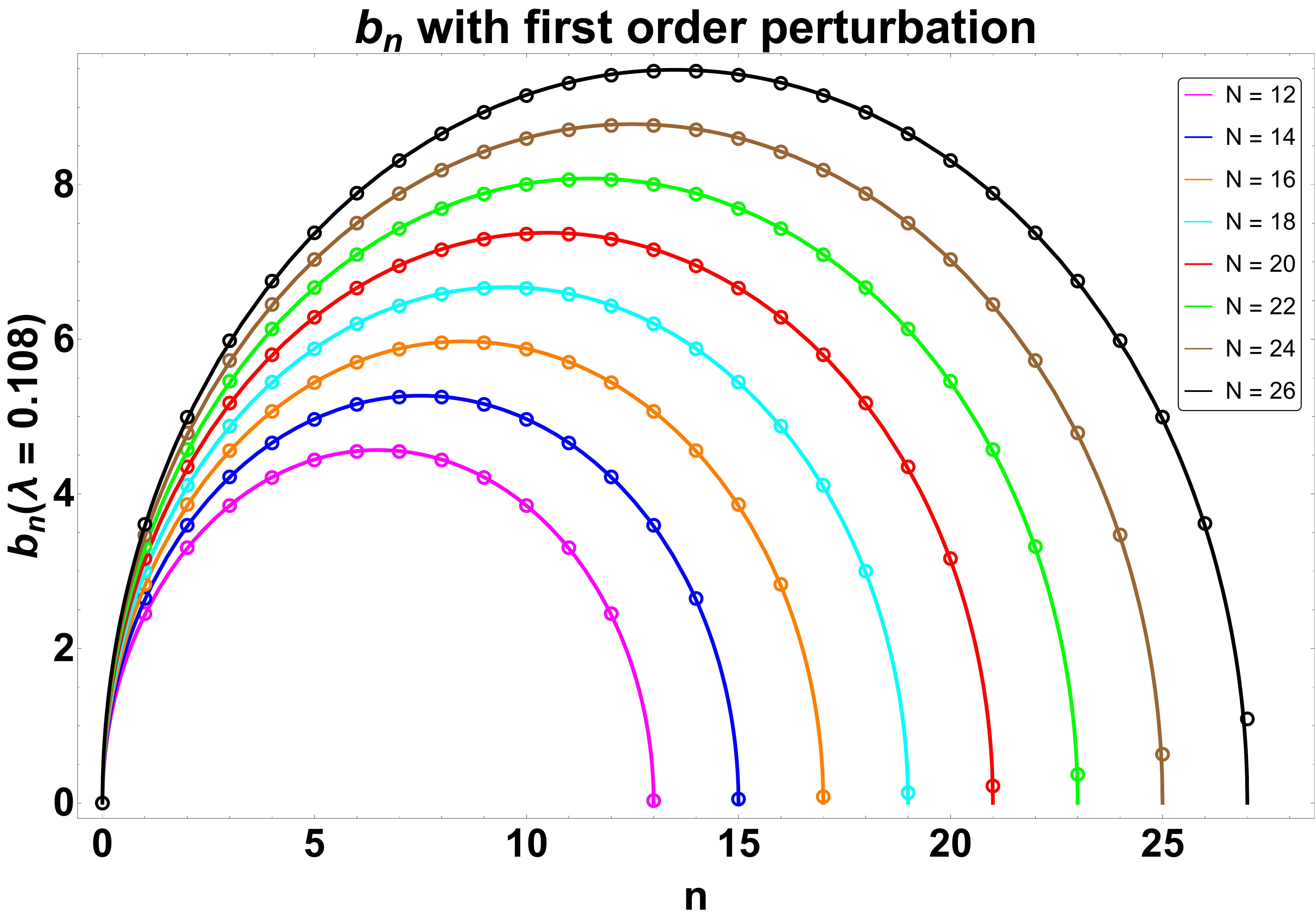}
\caption{Plot of $b_{n}$'s for the $\ket{\mathbf{Z}_{2}}$ state, after adding the perturbation \eqref{pxpc} to the PXP model for different lattice sizes. The dots indicate the numerical results while the line indicates the expression \eqref{bnlam} ($\alpha \approx 0.7025$). Both are in excellent agreement for all system sizes considered.}\label{fig:bnlambda}
\end{figure}

In the previous section, we showed that $q$-deformed algebra enables us to find an analytic form of the Lanczos coefficients within an excellent approximation. The analytical expression holds only within the approximate Krylov subspace. This approximation occurs because the symmetry algebra of the PXP Hamiltonian is not exact $\mathrm{SU}_q(2)$. However, the SU(2) structure can be recovered by adding a suitable term in the PXP Hamiltonian
\begin{align}
   H_{\mathrm{pert}} = \lambda \sum_{m=1}^N  (\mathrm{PXPP} + \mathrm{PPXP})\label{pxpc}\,. 
\end{align}
with $\lambda = 0.108$ \cite{PhysRevB.101.165139} (we assume the same perturbation strength for system sizes $N = 12, \ldots, 26$ and find excellent agreement). Consider the full Hamiltonian as $H_{\mathrm{PXP}}^{(1)} = H_{\mathrm{PXP}} + H_{\mathrm{pert}}$. Here, we use a compact notation PXPP to denote $P_{m-1}  \sigma_{m}^x  P_{m+1}  P_{m+2}$ and similarly for the second one. We plot the Lanczos coefficients (see Fig.\,\ref{fig:bnlambda}) and the first few wave functions $\psi_i(t)$ with $i = 1, \ldots, 4$ [see Fig.\,\ref{fig:psi01234Pert}] corresponding to the Krylov basis expansion for the Hamiltonian $H_{\mathrm{PXP}}^{(1)}$, initialized by the same N\'eel state. We see that the Lanczos coefficients closely follow a ``broken'' SU(2), and the revival of complexity also increases compared to the unperturbed PXP Hamiltonian. It is worth mentioning that $q$ deformation of the ``broken'' SU(2) algebra does not yield a better approximation. In other words, attempting to $q$ deform this algebra and numerically determine the $q$ value ends up giving $q = 1$ for all system sizes considered.

First, we begin by considering the Lanczos coefficients. We consider system sizes from $N = 12$ to $26$ and evolve the $\ket{\mathbf{Z}_{2}}$ state. The Lanczos coefficients show excellent agreement with the ``broken'' SU(2) result (see Fig.\,\ref{fig:bnlambda})
\begin{align}
    b_{n} = \alpha\sqrt{n (N - n + 1)}\,, \label{bnlam}
\end{align}
with $\alpha \approx 0.7025$ for all system sizes considered.

The value of $\alpha$ again suggests that writing $H^{(1)}_{\mathrm{PXP}} = \alpha (J^{(1)}_{+} + J^{(1)}_{-})$, the ladder operators must satisfy the commutation relations in Eqs.\eqref{comm1} and \eqref{comm2}, with numerically fixed $\alpha$. 

Following the discussion on non-perturbed PXP, we write $\alpha J^{(1)}_{\pm} = H^{(1)}_{\pm}$ where
\begin{align}
    H^{(1)}_{\pm} &= \sum_{n} \tilde{\sigma}^{\mp}_{2n + 1} + \tilde{\sigma}^{\pm}_{2n} + \lambda \sum_{n} \tilde{\sigma}^{\mp}_{2n + 1} P_{2n + 3} + \tilde{\sigma}^{\pm}_{2n} P_{2n + 2} \notag \\ &+ \lambda \sum_{n} P_{2n-1}\tilde{\sigma}^{\mp}_{2n + 1} + P_{2n - 2}\tilde{\sigma}^{\pm}_{2n}\,.
\end{align}
The commutator of these ladder operators can be evaluated to give
\begin{align}
   \frac{1}{2}[H^{(1)}_{+}, H^{(1)}_{-}] &= \sum_{n} \tilde{\sigma}^z_{2 n} - \tilde{\sigma}^z_{2 n + 1}\notag \\ 
   &- 2\lambda  \sum_{n} P_{2n - 2}\tilde{\sigma}^z_{2 n} - P_{2n - 1}\tilde{\sigma}^z_{2 n + 1} \notag \\
   &+ 2\lambda  \sum_{n} \tilde{\sigma}^z_{2 n}P_{2n + 2} - \tilde{\sigma}^z_{2 n + 1}P_{2n + 3} \notag \\
   &+ \lambda  \sum_{n} \tilde{\sigma}^{+}_{2 n}\tilde{\sigma}^{-}_{2n + 2}P_{2n + 3} - \tilde{\sigma}^{+}_{2 n + 1}\tilde{\sigma}^{-}_{2n + 3}P_{2n + 4} \notag \\
   &+ \lambda  \sum_{n} \tilde{\sigma}^{-}_{2 n}\tilde{\sigma}^{+}_{2n + 2}P_{2n + 3} - \tilde{\sigma}^{-}_{2 n + 1}\tilde{\sigma}^{+}_{2n + 3}P_{2n + 4}\notag \\ 
   &+ \lambda^{2}Y^{(1)} \,,\label{HpmCommZZ}
\end{align}
where we have only written down the $\mathcal{O}(\lambda^0)$ and $\mathcal{O}(\lambda)$ terms. There are also $\mathcal{O}(\lambda^2)$ terms, which are written collectively as the operator $Y^{(1)}$. Therefore, the right-hand side of Eq.$\eqref{HpmCommZZ}$ can be identified as $\alpha^2 J^{(1)}_{0}$. Using this, the commutator $[J^{(1)}_{0}, J^{(1)}_{\pm}]$ can be evaluated. One can see that it possible to write this commutator\footnote{While these commutators are very tedious to evaluate by hand, it is possible to evaluate them in a few seconds using \cite{Bull}.} as
\begin{align}
    [J^{(1)}_{0}, J^{(1)}_{\pm}] = \pm J^{(1)}_{\pm} \pm \tilde{X}^{(1)}_{\pm} \,, \label{JpmComm2}
\end{align}
where $\tilde{X}^{(1)}_{\pm}$ contains terms up to order $\lambda^3$. Here we again have absorbed a $\frac{1-\alpha^2}{\alpha^2}J^{(1)}_{\pm}$ term into $\tilde{X}^{(1)}_{\pm} $. 

The perturbation strength $\lambda$ is canonically fixed by studying the fidelity and complexity and ensuring that it demonstrates nearly perfect revival at a periodic time interval. We provide an intuitive explanation of the behavior of the wave functions $\psi_{n}(t)$'s and the complexity $C(t)$ in the next section.

In Appendix A, we study the wave functions in the Krylov basis expansion of the $\ket{\mathbf{Z}_{2}}$ state. For Hamiltonians of the type $\alpha (H_{+} + H_{-})$ the wave function $\psi_{n}$'s depend parametrically only on $\alpha$. 
The ``zeroth'' wave function, i.e., $\psi_{0}$ is nothing but fidelity. We plot the same for the $\ket{\mathbf{Z}_{2}}$ state and a generic state and fit it to \eqref{psin}. The results are given in Fig.\,\ref{fig:psi0Pert}. As is clear, there is strong agreement between the perturbed PXP fidelity and \eqref{psin} (for $n = 0$). We also evaluate a few other $\psi_{n} (t)$ ($n = 1,2,3,4$) numerically, and find good agreement with \eqref{psin} in all cases. 

In summary, the PXP Hamiltonian with first-order perturbation $H^{(1)}_\mathrm{PXP}$ corresponds to a broken SU(2) algebra. We find very strong agreement with the Lanczos coefficients and Krylov basis wave functions of unbroken SU(2). The algebra-breaking terms, therefore, have a subleading contribution.
%\textcolor{red}{PN: reached up to here}.

\section{Complexity of the $\ket{\mathbf{Z}_{2}}$ state: Results and a physical picture}
\label{secV}

Application of the Hamiltonian on any initial state can be thought of as a single-particle tight-binding problem on a lattice, when expressed in Krylov basis due to Eq.\eqref{krylov t-b}, where the $n^{\text{th}}$ Krylov basis state can be interpreted as the $n^{\text{th}}$ lattice site. The Lanczos coefficients $b_n$ and $b_{n+1}$ denotes the hopping amplitude from $n^{\text{th}}$ site to $(n-1)^{\text{th}}$ and $(n+1)^{\text{th}}$ sites, respectively, under application of $H$. Further, the square of Krylov basis wave functions $|\psi_{n}(t)|^{2}$ corresponds to the probability of finding the particle at the $n^{\text{th}}$ site. For a given Hamiltonian, if under time evolution of the initial state (achieved by repeated application of the Hamiltonian), we obtain a perfect revival. This implies that we must have a finite number of Krylov basis states (compared to the size of the Hilbert space). In other words, the effective tight-binding model is defined only over a finite number of lattice sites. The particle starts initially from the $0^{\text{th}}$ site and ``hops'' between those finite numbers of lattice sites states under time evolution. This can happen if the Hamiltonian has perfect SU(2) symmetry and can be split into parts which can be used to generate the finite number of Krylov basis states from the initial state, as explained in Sec. \ref{secIV}. For the case of $\ket{\mathbf{Z}_{2}}$ state evolved by the integrable paramagnetic Hamiltonian, the initial $\ket{\mathbf{Z}_{2}}$ state corresponds to the $0^{\text{th}}$ site and $\ket{\mathbf{Z'}_{2}}$ state corresponds to the last site of the finite tight-binding lattice.

For a system of size, $N=16$ the Hilbert-space dimension is $\sim 6.5\times 10^{4}$, yet we only need $17$ Krylov basis states to describe the evolution of $\ket{\mathbf{Z}_{2}}$ in the paramagnetic model. In Fig.\,\ref{para0} we observe $b_n$ is exactly $0$ for $n=0$ and $n=17$, implying that under time evolution, the state can never go beyond the $17^{\text{th}}$ Krylov basis state as the hopping is entirely suppressed. Hence, the time evolution is bounded within the non-thermal sector of the Hilbert space given by the Krylov basis states. The structure of the $b_n$'s also tells that the hopping rates are higher in the middle of the lattice and are gradually suppressed at either end. From a kinetic point of view, we may say that the particle, on average, spends most of the time near the $0^{\text{th}}$ and $17^{\text{th}}$ sites. However, it periodically moves back and forth between the two ends. 

Complexity defined as $C(t) = \sum_{n} n |\psi_{n} (t)|^2 $ can now be easily interpreted as the ``average position of the particle'' under time evolution in the lattice. In terms of Krylov basis states, we can rephrase it by saying that under time evolution, the state is most likely to be ``close'' to $\ket{\mathbf{Z}_{2}}$ or $\ket{\mathbf{Z'}_{2}}$ state, and it periodically oscillates in between, which is nicely captured by the oscillatory behavior of the complexity shown in Fig.\,\ref{paracomp}. The extrema of $C(t)$ occurs at values close to $0$ and $16$, denoting the positions of the states $\ket{\mathbf{Z}_{2}}$ and $\ket{\mathbf{Z'}_{2}}$ in Krylov basis, where it is most likely to be found if observed over some interval of time. 

\begin{figure}[t]
\subfigure[]{\includegraphics[height=5.7cm,width=1\linewidth]{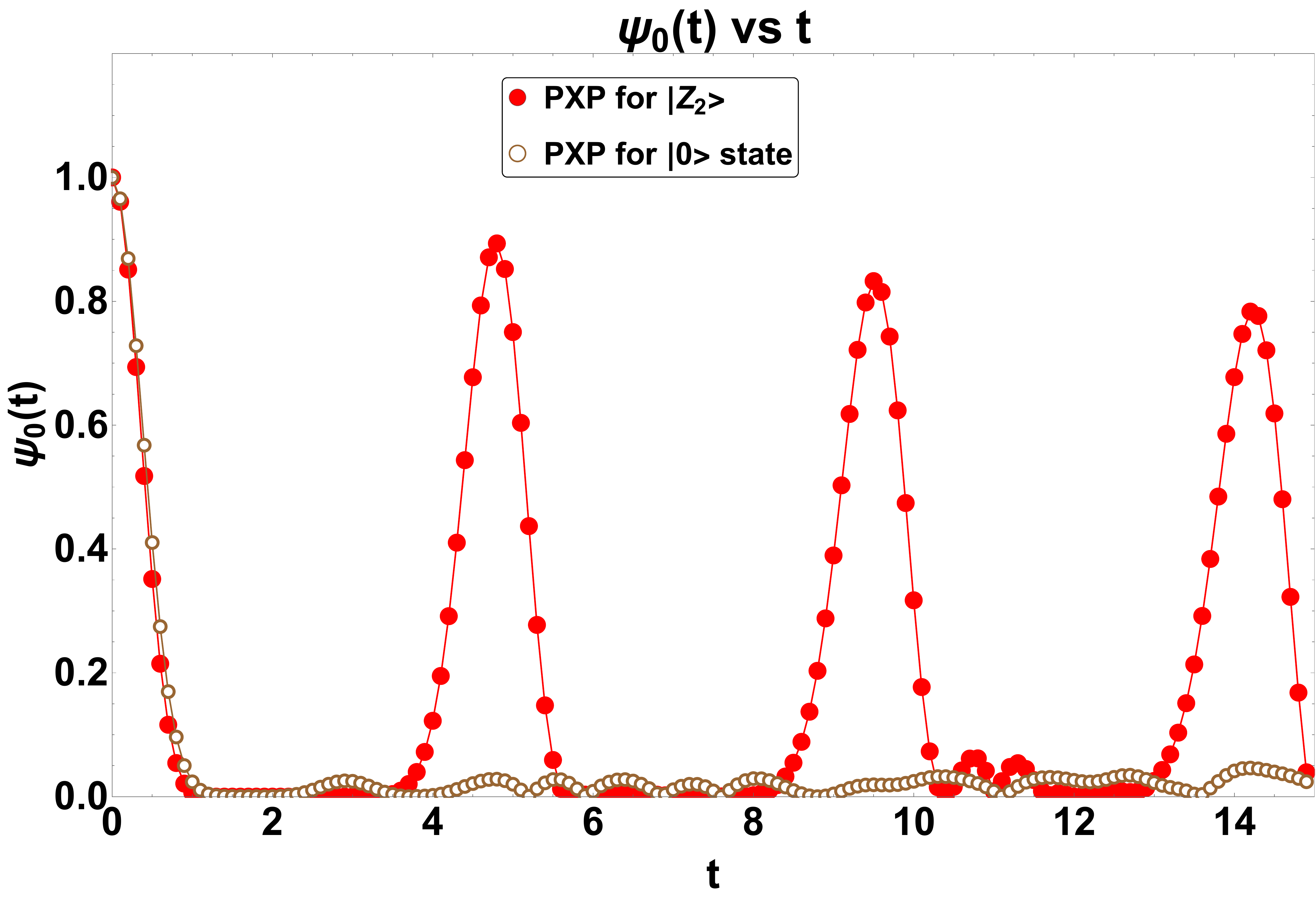}\label{fig:psinPure0}}
\hfill
\subfigure[]{\includegraphics[height=5.7cm,width=1\linewidth]{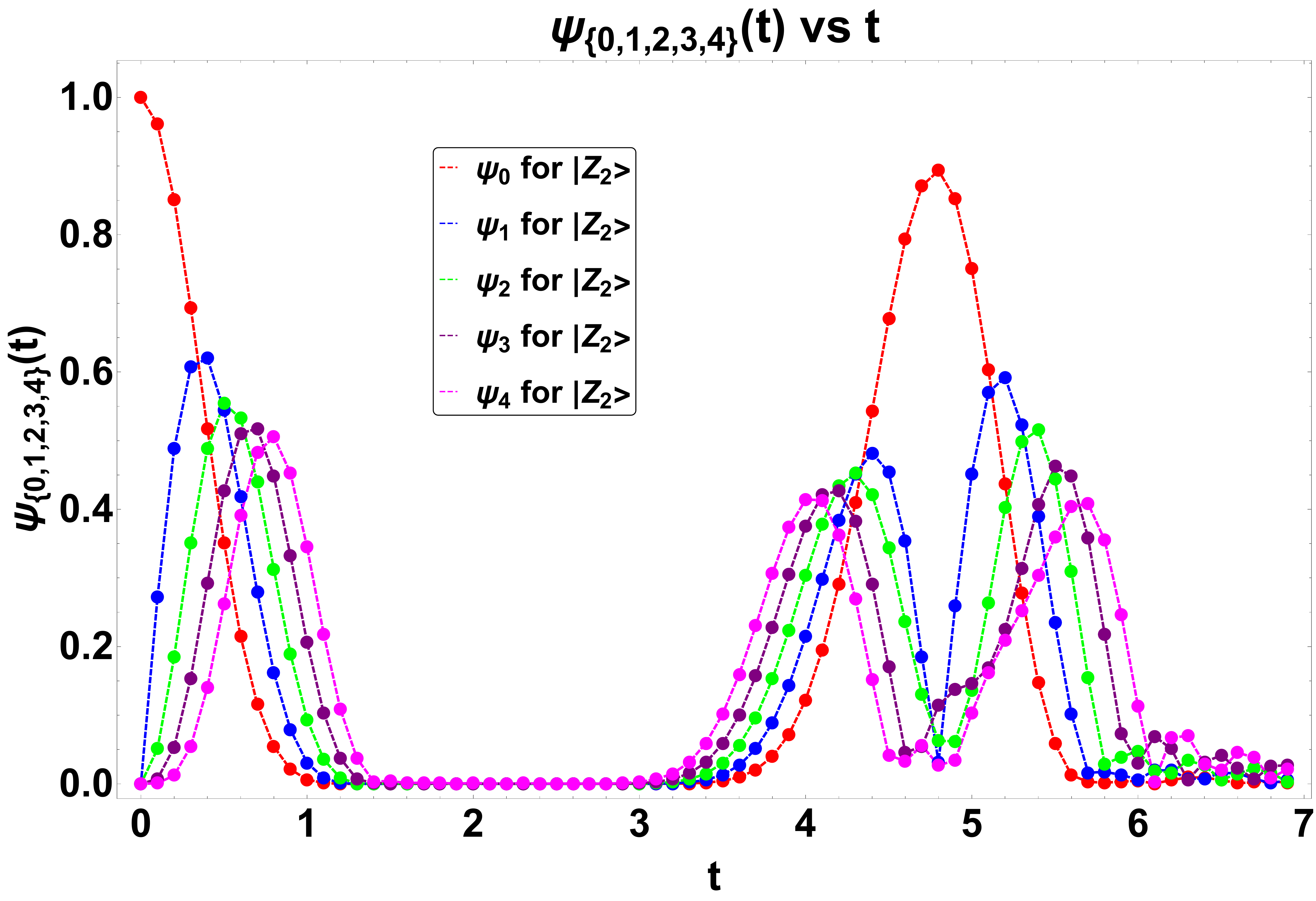}\label{fig:psi01234Pure0}}
\caption{(a) Behavior of $\psi_{0} (t)$ for $\ket{\mathbf{Z}_{2}}$ state (red disk) in PXP model. The brown circles represent $\psi_{0} (t)$ for an arbitrary state that does not possess $\mathbf{Z}$ symmetry ($\ket{0}$ state). (b) $\psi_0 (t)$, $\psi_1 (t)$, $\psi_2 (t)$, $\psi_3 (t)$, and $\psi_4 (t)$ for $\ket{\mathbf{Z}_{2}}$ state in PXP model showing the evolution of initial state in the Krylov basis with a slight decay at revival due to the broken SU(2) symmetry of the model. In all cases, we choose $N = 16$.}
\end{figure}

Regarding the evolution of $\ket{\mathbf{Z}_{2}}$ state under the non-integrable PXP model, we observe in Fig.\,\ref{pxpbnm} that $b_n$'s do not become zero at any finite $n$ although they tend to come close to zero (at $n = 17$ for $N=16$) before shooting off to larger values. This initial tendency is well understood by the weakly broken SU(2) structure of the Hamiltonian and the presence of the scar-states in the Hilbert space. In this case, in the tight-binding picture, the particle is not bound inside the first $17$ sites, although there is a tendency to stay localized near the $0^{\text{th}}$ and $17^{\text{th}}$ site due to the suppressed hopping. Once it reaches the $17^{\text{th}}$ site, it can hop outside the region with a higher amplitude given by the next $b_n$. In the Krylov basis picture, we understand this as the initial $\ket{\mathbf{Z}_{2}}$ state, after sufficient time, ``leaks" into the thermal part of the Hilbert space and thereby deviating from the exact revival over time as shown in the plot for fidelity or $|\psi_0(t)| = |\braket{ \Psi(t)|\mathbf{Z}_{2}}|$ in Fig.\,\ref{fig:psinPure0}. We also show the fidelity of a generic product state, which quickly goes to zero, showing no sign of periodic revival at all. Specifically, we consider the state given by $\ket{0010100100100010}$ (in terms of $\sigma^z$ configuration) for system size $N = 16$.

\begin{figure}[t]
\includegraphics[height=5.7cm,width=1\linewidth]{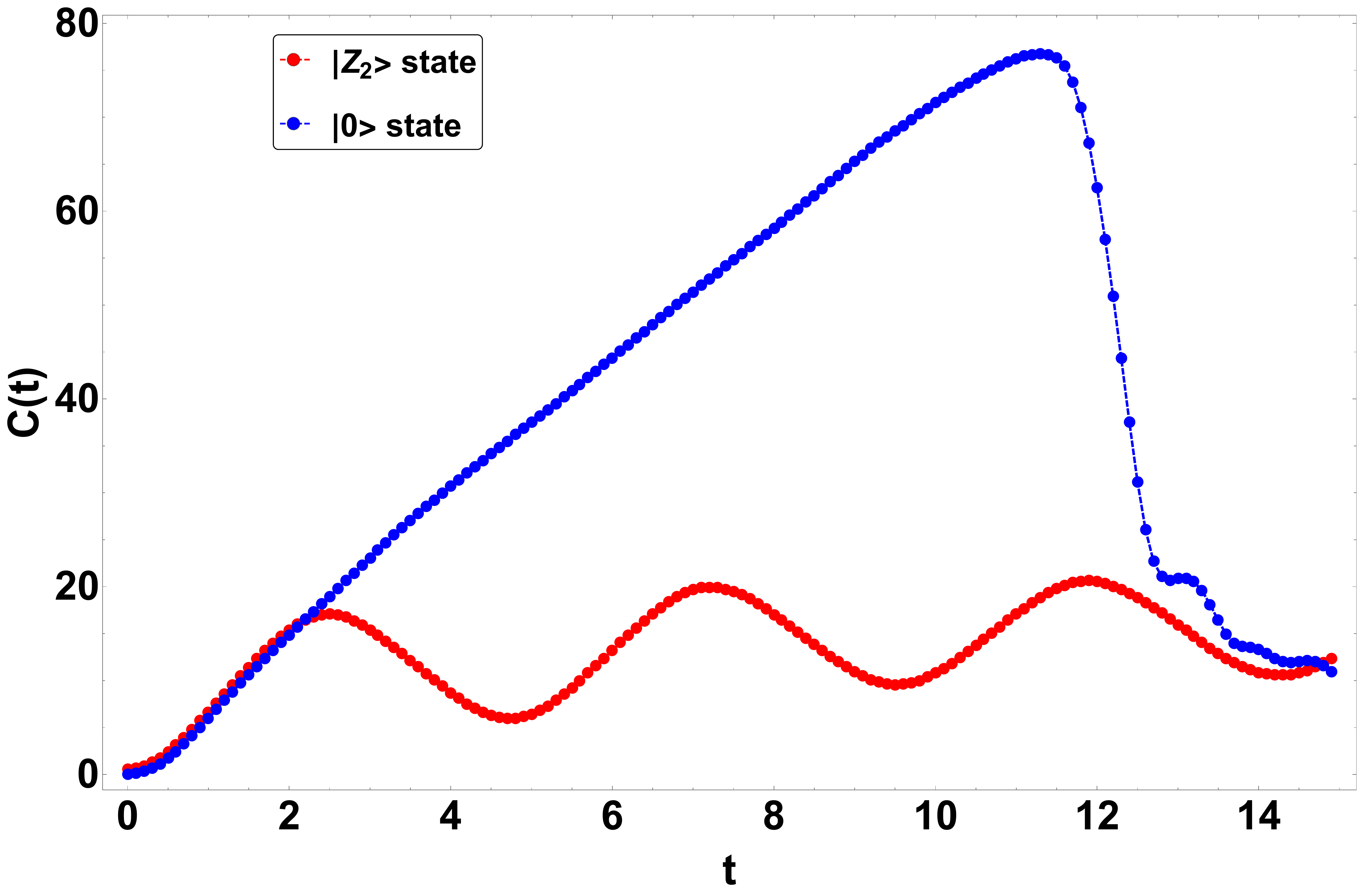}
\caption{Evolution of complexity $C(t)$ for the $\ket{\mathbf{Z}_{2}}$ state (in red) and for a generic state without \textbf{Z} symmetry ($\ket{0}$ state, in blue) in PXP model for system size $N=16$. The complexity for the generic state grows without any constraint, only to be bounded by the finite number of Krylov basis states considered in our computation. This is reflected by the sudden dip in the complexity (shown in blue). Subsequent time evolution only makes it hop towards lower Krylov basis states.}\label{fig:CompZ2rand}
\end{figure}

In Fig.\,\ref{fig:psi01234Pure0} we have shown the behavior of the absolute values of the Krylov wave functions $\psi_0$, $\psi_1$, $\psi_2$, $\psi_3$, and $\psi_4$, whose squares are the probabilities of finding the particle in $0, 1, 2, 3, 4 ^\text{th}$ Krylov bases ``lattice sites'' as functions of time. We observe that as time progresses, the particle gradually moves from the $0^\text{th}$  site to the subsequent sites\footnote{See \cite{Roberts:2014isa} for a similar situation, but with slightly different motivation.}. Before the revival period also, we see that the particle slowly returns to the $0^\text{th}$ site from the latter sites and then, after a reflection at the $0^\text{th}$ site again moves towards the next sites. The reflection can be understood by the presence of the double peaks of $|\psi_n(t)|$'s for $n \neq 0$ and the single peaks of $|\psi_0(t)|$ around the revival period. However, the decreasing amplitude of all the wave functions at revival is due to the broken SU(2) structure and the ``leaking'' of the state into the thermal part of Hilbert space which has already been explained.

The complexity plot in Fig.\,\ref{fig:CompZ2rand} for the $\ket{\mathbf{Z}_{2}}$ state in the case of the PXP model also has interesting signatures of weak ergodicity breaking. Like the case of the integrable and SU(2) symmetric paramagnetic model, the $C(t)$ has an oscillatory behavior. However, it is not bounded between any finite portion of the Krylov basis. Hence along with an oscillatory part, it also grows slowly into higher order Krylov basis states, implying that under time evolution, it never fully comes back to the $\ket{\mathbf{Z}_{2}}$ state. On the other hand, a generic state starts spreading into the Krylov space of basis states right away and only gets bounded because we have only taken a finite number of Krylov basis states for computational convenience. Then, the subsequent application of Hamiltonian on this state only makes it comeback to the lower Krylov basis vectors.

\begin{figure}[t]
\subfigure[]{\includegraphics[height=5.7cm,width=1\linewidth]{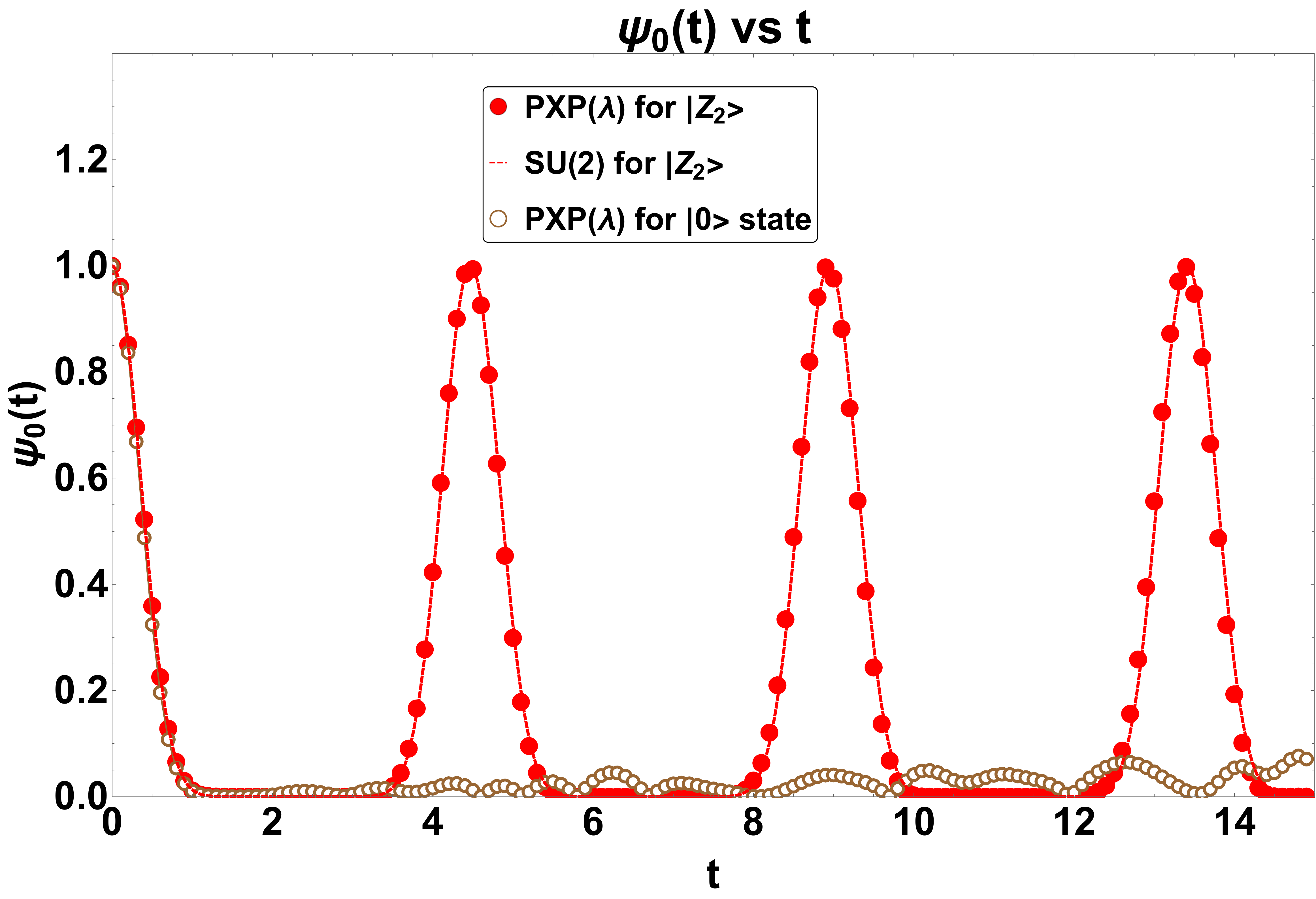}\label{fig:psi0Pert}}
\hfill
\subfigure[]{\includegraphics[height=5.7cm,width=1\linewidth]{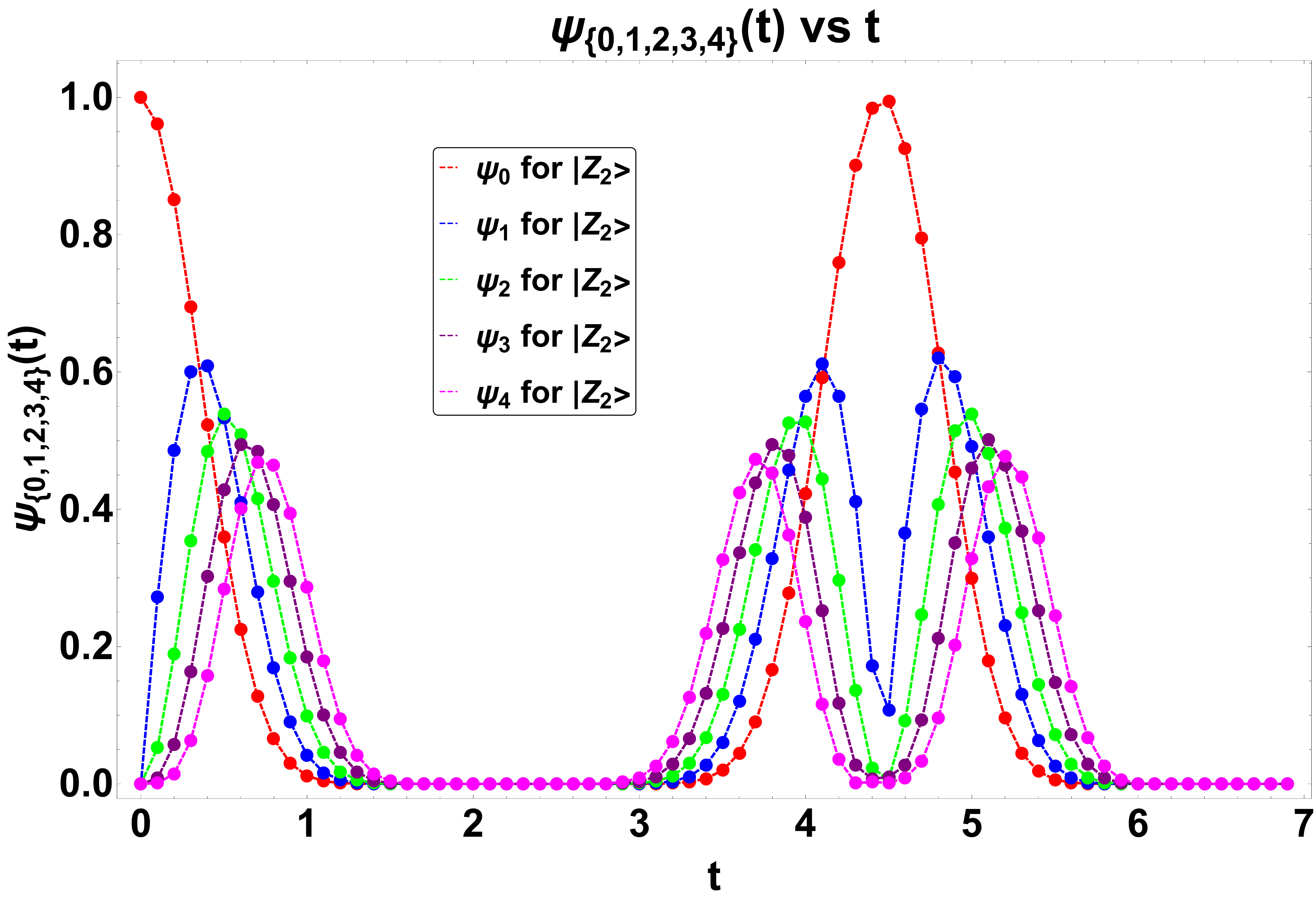}\label{fig:psi01234Pert}}
\caption{(a) Behavior of $\psi_{0} (t)$ for $\lambda = 0.108$ for the $\ket{\mathbf{Z}_{2}}$ state (red disk) compared with the analytical result \eqref{psin} (dashed red line, for $n = 0$). The brown circles represent $\psi_{0} (t)$ for an arbitrary state that does not possess $\mathbf{Z}$ symmetry ($\ket{0}$ state).
(b) $\psi_0 (t)$, $\psi_1 (t)$, $\psi_2 (t)$, $\psi_3 (t)$, and $\psi_4 (t)$ for $\ket{\mathbf{Z}_{2}}$ state in perturbed PXP model showing the evolution of initial state in the Krylov basis. The amplitudes are the same at revival periods because of the recovered SU(2) structure due to the addition of perturbation. In all cases, we choose $N = 16$.}
\end{figure}

The approximate recovery of SU(2) algebra after the addition of a suitable perturbation, for the case of $\ket{\mathbf{Z}_{2}}$ initial state, can be observed directly from the behavior of the Lanczos coefficients, the Krylov wave functions, and the complexity. Fig.\,\ref{fig:bnlambda} shows the bounded nature of $b_n$ for different system sizes. This behavior is almost exactly like the case of the SU(2) paramagnetic model, implying that the Hilbert space division into non-thermal and thermal parts is nearly exact. The perturbation strength is fixed by ensuring that the time-periodic revival of the state is almost perfect without any noticeable reduction of amplitude. This is seen in plot for fidelity or $|\psi_0(t)|$ of the $\ket{\mathbf{Z}_{2}}$ state in Fig.\,\ref{fig:psi0Pert}. We have also shown the fidelity for the same generic state as before given by $\ket{0010100100100010}$ in $\sigma_z$ basis, which decays to zero after a small time interval.

\begin{figure}[t]
\includegraphics[height=5.7cm,width=1\linewidth]{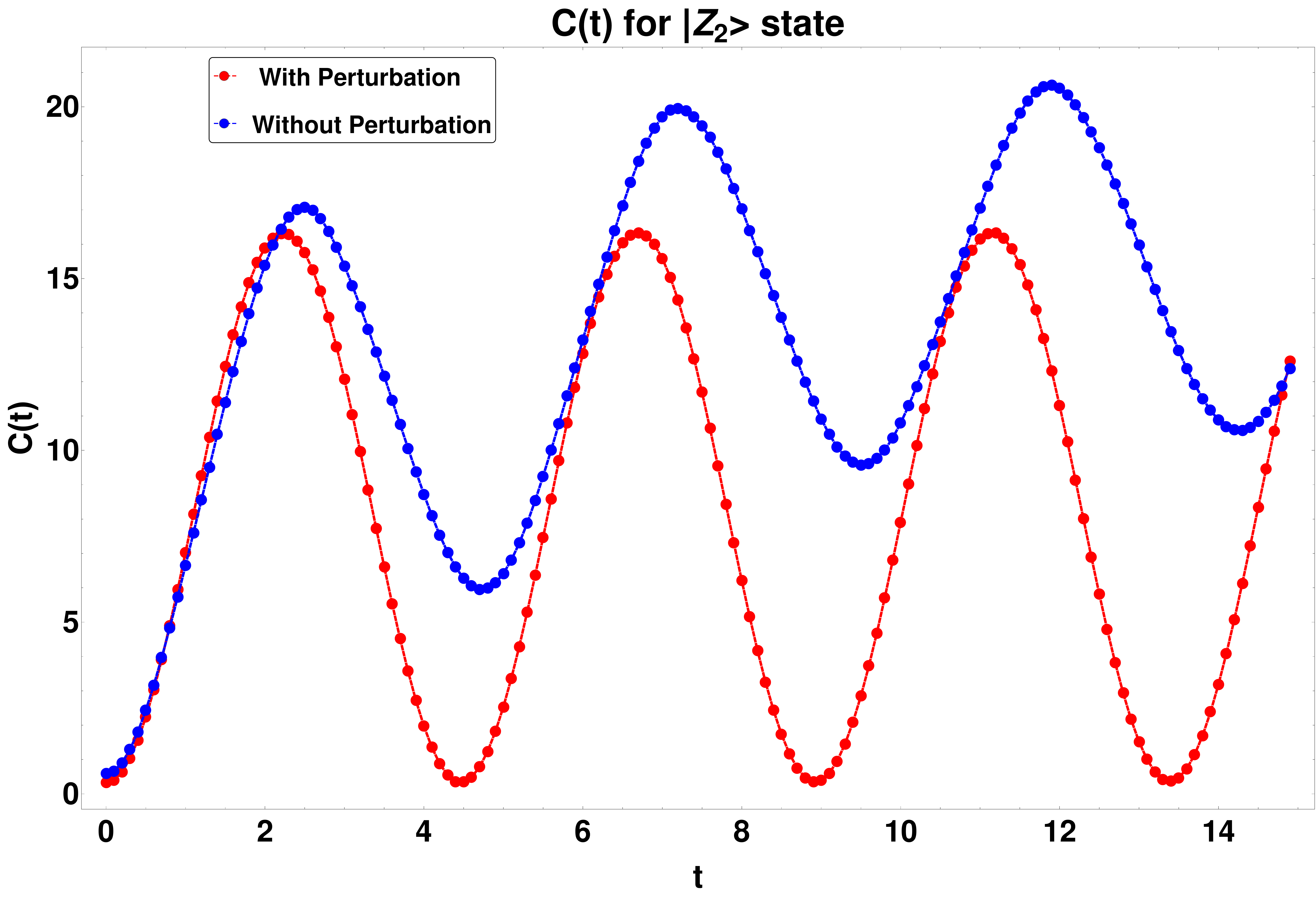}
\caption{Evolution of complexity $C(t)$ for the $\ket{\mathbf{Z}_{2}}$ state in case of PXP model (in blue) and perturbed PXP model (in red) for system size $N=16$.}
\label{fig:CompZ2}
\end{figure}

The plots for the wave function in Fig.\,\ref{fig:psi01234Pert} have a similar feature to the case without perturbation. However, in this case, there is a very negligible amount of the decay of the wave functions at revival, implying the nearly ``loss-less'' dynamics of the initial state in the finite Krylov basis. The complexity also regains its fully oscillatory behavior, very much like the exact SU(2) case. However, there is a slight increase in complexity which can be noticed only after a number of periods. To contrast the two cases, we show the complexity before and after adding the perturbation term in a single plot in Fig.\,\ref{fig:CompZ2}.

Thus, the behavior of $b_n$, the Krylov wave functions, and the periodic and bounded nature of complexity can be invoked to understand the periodic revival of certain initial states for a non-integrable Hamiltonian, which can happen due to the presence of weakly entangled scar states.

\section{Conclusion}
\label{secVI}
%\emph{Conclusion:} 
In this work, we have extended the study of Krylov state (spread) complexity for the time evolution governed by the  PXP Hamiltonian. While the exact SU(2) algebra allows us to infer the closed-form expression of Lanczos coefficients and corresponding complexity, it is not the case for the PXP Hamiltonian. The algebra is not exact SU(2), yet still close to SU(2). We have quantified this closeness by mapping the algebra to a class of well-known $q$-deformed algebra $\mathrm{SU}_q(2)$, where the deformation generically encodes the algebra-breaking terms. Still, as we have shown that PXP cannot be written as exact $\mathrm{SU}_q(2)$ for any $q$, but the approximation is much better compared to the usual SU(2) algebra. However, the crucial point is that expressing as $\mathrm{SU}_q(2)$ allows us to compute the Lanczos coefficients for the initial N\'eel state in terms of Chebyshev polynomials. This was missing in previous studies, and we aim to fill this gap by proving analytic expression, which fits the numerical results in good approximation. The complexity for the N\'eel fails to become completely periodic and oscillates with a much slower value compared to the generic state. The complete periodicity can be well recovered by adding a perturbation to the PXP Hamiltonian, which primarily restricts the generators within a closed algebra in an approximate sense. This, however, no longer holds for a generic state.

Previous studies \cite{Caputa:2022eye} have been focused on the complexity, where one uses the spin-$1/2$ representation of SU(2) algebra. This suggests that only two Krylov bases are possible, and the complexity can be written in terms of fidelity. Hence, complexity and fidelity carry the same information. On the other hand, we have worked with an effective spin-$N/2$ representation, thus allowing us to consider a larger number of Krylov basis vectors. This implies that in our case, complexity, in general, carries more information than fidelity. 

Furthermore, the physical interpretation of Lanczos coefficients, Krylov basis wave functions, and the complexity is general and not specific to the system discussed in the paper. Such a picture will help to understand the behavior of complexity of any initial state evolving under generic Hamiltonian. Specifically, this work opens up possibilities to extend for other \textbf{Z} symmetric state (e.g., like $\ket{\mathbf{Z_3}}$, $\ket{\mathbf{Z_4}}$) with and without perturbation. Another interesting future direction will be to study the behavior of complexity beyond the PXP model such as higher spin \cite{Schecter:2019oej, You:2022hfz}, periodically driven systems \cite{PhysRevB.101.245107, PhysRevB.102.075123, PhysRevResearch.3.L012010, Hudomal:2022vtv}, and hypercube models \cite{PhysRevB.105.245137}.

For the PXP model itself, it will be interesting to consider other possibilities such as $(p,q)$ deformation or general $\Phi$ deformation of SU(2) \cite{Bonatsos:1996qb} satisfied by the Hamiltonian. For the $q$ deformation studied in this work, it would be worth attempting to derive exact analytic expressions for the Krylov basis wave functions, as well as complexity and $K$-entropy, if possible. Additionally, understanding the reason behind the system-size dependence of the $q$ value might shed further light on the true nature of the algebra describing the PXP Hamiltonian.

Finally, it would be interesting to study the effect of transverse magnetic field on the PXP Hamiltonian \cite{PhysRevB.105.125123}
\begin{align}
    H = \sum_m (P_{m-1} \sigma_m^x  P_{m+1} - \chi\, \sigma_m^z)\,.
\end{align}
with $\chi > 0$. This Hamiltonian not only possesses scar states but also contains the critical states near a critical value $\chi = \chi_c \approx 0.655$ \cite{PhysRevB.105.125123}, where an Ising-type phase transition occurs. For such Hamiltonian, $\ket{\mathbf{Z_2}}$ state thermalizes near the critical regime while it fails to thermalize off-criticality. It is tempting to think that, in such a case, the complexity might show unbounded growth even for the $\ket{\mathbf{Z_2}}$ state. We hope to return to some of these questions in future studies.

\section{Acknowledgements}
%\emph{Acknowledgements:}
We would like to thank Aranya Bhattacharya, Hugo A. Camargo, Pawel Caputa, Chethan Krishnan, Silvia Pappalardi, Tanay Pathak, Diptiman Sen, and Masaki Tezuka for useful discussions and comments on the draft. We especially thank Tanay Pathak for collaborations on early stages of the work and Aninda Sinha for making us aware of the $q$-deformed algebra through the Refs. \cite{chaichian1996introduction, BONATSOS1999537, Sinha:2022sdo}. Some numerical computations were done using QuSpin \cite{weinberg2017quspin, weinberg2019quspin}. BB would like to thank the organizers and participants of CHAHOL22 and the Max-Planck-Institut f\"{u}r Physik komplexer Systeme, Dresden, for helpful discussions. P.N. wishes to thank NITheCS and the University of Cape Town for the hospitality during the final stages of the work, where parts of the results were presented. B.B. and S.S. are supported by the Ministry of Human Resource Development (MHRD), Government of India, through the Prime Ministers' Research Fellowship. The work of P.N. is supported by the JSPS Grant-in-Aid for Transformative Research Areas (A) ``Extreme Universe'' No. 21H05190.
\\
\newline
BB and SS contributed equally to this work.

\bibliography{references} 

\begin{thebibliography}{112}%
\makeatletter
\providecommand \@ifxundefined [1]{%
 \@ifx{#1\undefined}
}%
\providecommand \@ifnum [1]{%
 \ifnum #1\expandafter \@firstoftwo
 \else \expandafter \@secondoftwo
 \fi
}%
\providecommand \@ifx [1]{%
 \ifx #1\expandafter \@firstoftwo
 \else \expandafter \@secondoftwo
 \fi
}%
\providecommand \natexlab [1]{#1}%
\providecommand \enquote  [1]{``#1''}%
\providecommand \bibnamefont  [1]{#1}%
\providecommand \bibfnamefont [1]{#1}%
\providecommand \citenamefont [1]{#1}%
\providecommand \href@noop [0]{\@secondoftwo}%
\providecommand \href [0]{\begingroup \@sanitize@url \@href}%
\providecommand \@href[1]{\@@startlink{#1}\@@href}%
\providecommand \@@href[1]{\endgroup#1\@@endlink}%
\providecommand \@sanitize@url [0]{\catcode `\\12\catcode `\$12\catcode
  `\&12\catcode `\#12\catcode `\^12\catcode `\_12\catcode `\%12\relax}%
\providecommand \@@startlink[1]{}%
\providecommand \@@endlink[0]{}%
\providecommand \url  [0]{\begingroup\@sanitize@url \@url }%
\providecommand \@url [1]{\endgroup\@href {#1}{\urlprefix }}%
\providecommand \urlprefix  [0]{URL }%
\providecommand \Eprint [0]{\href }%
\providecommand \doibase [0]{http://dx.doi.org/}%
\providecommand \selectlanguage [0]{\@gobble}%
\providecommand \bibinfo  [0]{\@secondoftwo}%
\providecommand \bibfield  [0]{\@secondoftwo}%
\providecommand \translation [1]{[#1]}%
\providecommand \BibitemOpen [0]{}%
\providecommand \bibitemStop [0]{}%
\providecommand \bibitemNoStop [0]{.\EOS\space}%
\providecommand \EOS [0]{\spacefactor3000\relax}%
\providecommand \BibitemShut  [1]{\csname bibitem#1\endcsname}%
\let\auto@bib@innerbib\@empty
%</preamble>
\bibitem [{\citenamefont {Calabrese}\ and\ \citenamefont
  {Cardy}(2005)}]{Calabrese:2005in}%
  \BibitemOpen
  \bibfield  {author} {\bibinfo {author} {\bibfnamefont {Pasquale}\
  \bibnamefont {Calabrese}}\ and\ \bibinfo {author} {\bibfnamefont {John~L.}\
  \bibnamefont {Cardy}},\ }\bibfield  {title} {\enquote {\bibinfo {title}
  {{Evolution of entanglement entropy in one-dimensional systems}},}\ }\href
  {\doibase 10.1088/1742-5468/2005/04/P04010} {\bibfield  {journal} {\bibinfo
  {journal} {J. Stat. Mech.}\ }\textbf {\bibinfo {volume} {0504}},\ \bibinfo
  {pages} {P04010} (\bibinfo {year} {2005})},\ \Eprint
  {http://arxiv.org/abs/cond-mat/0503393} {arXiv:cond-mat/0503393} \BibitemShut
  {NoStop}%
\bibitem [{\citenamefont {Santos}\ \emph {et~al.}(2012)\citenamefont {Santos},
  \citenamefont {Polkovnikov},\ and\ \citenamefont {Rigol}}]{Santos_2012}%
  \BibitemOpen
  \bibfield  {author} {\bibinfo {author} {\bibfnamefont {Lea~F.}\ \bibnamefont
  {Santos}}, \bibinfo {author} {\bibfnamefont {Anatoli}\ \bibnamefont
  {Polkovnikov}}, \ and\ \bibinfo {author} {\bibfnamefont {Marcos}\
  \bibnamefont {Rigol}},\ }\bibfield  {title} {\enquote {\bibinfo {title} {Weak
  and strong typicality in quantum systems},}\ }\href {\doibase
  10.1103/physreve.86.010102} {\bibfield  {journal} {\bibinfo  {journal}
  {Physical Review E}\ }\textbf {\bibinfo {volume} {86}} (\bibinfo {year}
  {2012}),\ 10.1103/physreve.86.010102}\BibitemShut {NoStop}%
\bibitem [{\citenamefont {Deutsch}\ \emph {et~al.}(2013)\citenamefont
  {Deutsch}, \citenamefont {Li},\ and\ \citenamefont {Sharma}}]{Deutsch_2013}%
  \BibitemOpen
  \bibfield  {author} {\bibinfo {author} {\bibfnamefont {J.~M.}\ \bibnamefont
  {Deutsch}}, \bibinfo {author} {\bibfnamefont {Haibin}\ \bibnamefont {Li}}, \
  and\ \bibinfo {author} {\bibfnamefont {Auditya}\ \bibnamefont {Sharma}},\
  }\bibfield  {title} {\enquote {\bibinfo {title} {Microscopic origin of
  thermodynamic entropy in isolated systems},}\ }\href {\doibase
  10.1103/physreve.87.042135} {\bibfield  {journal} {\bibinfo  {journal}
  {Physical Review E}\ }\textbf {\bibinfo {volume} {87}} (\bibinfo {year}
  {2013}),\ 10.1103/physreve.87.042135}\BibitemShut {NoStop}%
\bibitem [{\citenamefont {Collura}\ \emph {et~al.}(2014)\citenamefont
  {Collura}, \citenamefont {Kormos},\ and\ \citenamefont
  {Calabrese}}]{Collura_2014}%
  \BibitemOpen
  \bibfield  {author} {\bibinfo {author} {\bibfnamefont {Mario}\ \bibnamefont
  {Collura}}, \bibinfo {author} {\bibfnamefont {Márton}\ \bibnamefont
  {Kormos}}, \ and\ \bibinfo {author} {\bibfnamefont {Pasquale}\ \bibnamefont
  {Calabrese}},\ }\bibfield  {title} {\enquote {\bibinfo {title} {Stationary
  entanglement entropies following an interaction quench in 1d bose gas},}\
  }\href {\doibase 10.1088/1742-5468/2014/01/p01009} {\bibfield  {journal}
  {\bibinfo  {journal} {Journal of Statistical Mechanics: Theory and
  Experiment}\ }\textbf {\bibinfo {volume} {2014}},\ \bibinfo {pages} {P01009}
  (\bibinfo {year} {2014})}\BibitemShut {NoStop}%
\bibitem [{\citenamefont {Kaufman}\ \emph {et~al.}(2016)\citenamefont
  {Kaufman}, \citenamefont {Tai}, \citenamefont {Lukin}, \citenamefont
  {Rispoli}, \citenamefont {Schittko}, \citenamefont {Preiss},\ and\
  \citenamefont {Greiner}}]{Kaufman_2016}%
  \BibitemOpen
  \bibfield  {author} {\bibinfo {author} {\bibfnamefont {A.~M.}\ \bibnamefont
  {Kaufman}}, \bibinfo {author} {\bibfnamefont {M.~E.}\ \bibnamefont {Tai}},
  \bibinfo {author} {\bibfnamefont {A.}~\bibnamefont {Lukin}}, \bibinfo
  {author} {\bibfnamefont {M.}~\bibnamefont {Rispoli}}, \bibinfo {author}
  {\bibfnamefont {R.}~\bibnamefont {Schittko}}, \bibinfo {author}
  {\bibfnamefont {P.~M.}\ \bibnamefont {Preiss}}, \ and\ \bibinfo {author}
  {\bibfnamefont {M.}~\bibnamefont {Greiner}},\ }\bibfield  {title} {\enquote
  {\bibinfo {title} {Quantum thermalization through entanglement in an isolated
  many-body system},}\ }\href {\doibase 10.1126/science.aaf6725} {\bibfield
  {journal} {\bibinfo  {journal} {Science}\ }\textbf {\bibinfo {volume}
  {353}},\ \bibinfo {pages} {794–800} (\bibinfo {year} {2016})}\BibitemShut
  {NoStop}%
\bibitem [{\citenamefont {Essler}\ and\ \citenamefont
  {Fagotti}(2016)}]{Essler:2016ufo}%
  \BibitemOpen
  \bibfield  {author} {\bibinfo {author} {\bibfnamefont {Fabian H.~L.}\
  \bibnamefont {Essler}}\ and\ \bibinfo {author} {\bibfnamefont {Maurizio}\
  \bibnamefont {Fagotti}},\ }\bibfield  {title} {\enquote {\bibinfo {title}
  {{Quench dynamics and relaxation in isolated integrable quantum spin
  chains}},}\ }\href {\doibase 10.1088/1742-5468/2016/06/064002} {\bibfield
  {journal} {\bibinfo  {journal} {J. Stat. Mech.}\ }\textbf {\bibinfo {volume}
  {1606}},\ \bibinfo {pages} {064002} (\bibinfo {year} {2016})},\ \Eprint
  {http://arxiv.org/abs/1603.06452} {arXiv:1603.06452 [cond-mat.quant-gas]}
  \BibitemShut {NoStop}%
\bibitem [{\citenamefont {Deutsch}(1991)}]{PhysRevA.43.2046}%
  \BibitemOpen
  \bibfield  {author} {\bibinfo {author} {\bibfnamefont {J.~M.}\ \bibnamefont
  {Deutsch}},\ }\bibfield  {title} {\enquote {\bibinfo {title} {Quantum
  statistical mechanics in a closed system},}\ }\href {\doibase
  10.1103/PhysRevA.43.2046} {\bibfield  {journal} {\bibinfo  {journal} {Phys.
  Rev. A}\ }\textbf {\bibinfo {volume} {43}},\ \bibinfo {pages} {2046--2049}
  (\bibinfo {year} {1991})}\BibitemShut {NoStop}%
\bibitem [{\citenamefont {Srednicki}(1994)}]{PhysRevE.50.888}%
  \BibitemOpen
  \bibfield  {author} {\bibinfo {author} {\bibfnamefont {Mark}\ \bibnamefont
  {Srednicki}},\ }\bibfield  {title} {\enquote {\bibinfo {title} {Chaos and
  quantum thermalization},}\ }\href {\doibase 10.1103/PhysRevE.50.888}
  {\bibfield  {journal} {\bibinfo  {journal} {Phys. Rev. E}\ }\textbf {\bibinfo
  {volume} {50}},\ \bibinfo {pages} {888--901} (\bibinfo {year}
  {1994})}\BibitemShut {NoStop}%
\bibitem [{\citenamefont {Rigol}\ \emph {et~al.}(2007)\citenamefont {Rigol},
  \citenamefont {Dunjko}, \citenamefont {Yurovsky},\ and\ \citenamefont
  {Olshanii}}]{PhysRevLett.98.050405}%
  \BibitemOpen
  \bibfield  {author} {\bibinfo {author} {\bibfnamefont {Marcos}\ \bibnamefont
  {Rigol}}, \bibinfo {author} {\bibfnamefont {Vanja}\ \bibnamefont {Dunjko}},
  \bibinfo {author} {\bibfnamefont {Vladimir}\ \bibnamefont {Yurovsky}}, \ and\
  \bibinfo {author} {\bibfnamefont {Maxim}\ \bibnamefont {Olshanii}},\
  }\bibfield  {title} {\enquote {\bibinfo {title} {Relaxation in a completely
  integrable many-body quantum system: An ab initio study of the dynamics of
  the highly excited states of 1d lattice hard-core bosons},}\ }\href {\doibase
  10.1103/PhysRevLett.98.050405} {\bibfield  {journal} {\bibinfo  {journal}
  {Phys. Rev. Lett.}\ }\textbf {\bibinfo {volume} {98}},\ \bibinfo {pages}
  {050405} (\bibinfo {year} {2007})},\ \Eprint
  {http://arxiv.org/abs/cond-mat/0604476} {arXiv:cond-mat/0604476
  [cond-mat.other]} \BibitemShut {NoStop}%
\bibitem [{\citenamefont {Rigol}\ \emph {et~al.}(2008)\citenamefont {Rigol},
  \citenamefont {Dunjko},\ and\ \citenamefont {Olshanii}}]{Rigol_2008}%
  \BibitemOpen
  \bibfield  {author} {\bibinfo {author} {\bibfnamefont {Marcos}\ \bibnamefont
  {Rigol}}, \bibinfo {author} {\bibfnamefont {Vanja}\ \bibnamefont {Dunjko}}, \
  and\ \bibinfo {author} {\bibfnamefont {Maxim}\ \bibnamefont {Olshanii}},\
  }\bibfield  {title} {\enquote {\bibinfo {title} {Thermalization and its
  mechanism for generic isolated quantum systems},}\ }\href {\doibase
  10.1038/nature06838} {\bibfield  {journal} {\bibinfo  {journal} {Nature}\
  }\textbf {\bibinfo {volume} {452}},\ \bibinfo {pages} {854--858} (\bibinfo
  {year} {2008})},\ \Eprint {http://arxiv.org/abs/0708.1324} {arXiv:0708.1324
  [cond-mat.stat-mech]} \BibitemShut {NoStop}%
\bibitem [{\citenamefont {Caux}\ and\ \citenamefont
  {Mossel}(2011)}]{Caux:2010by}%
  \BibitemOpen
  \bibfield  {author} {\bibinfo {author} {\bibfnamefont {Jean-Sebastien}\
  \bibnamefont {Caux}}\ and\ \bibinfo {author} {\bibfnamefont {Jorn}\
  \bibnamefont {Mossel}},\ }\bibfield  {title} {\enquote {\bibinfo {title}
  {{Remarks on the notion of quantum integrability}},}\ }\href {\doibase
  10.1088/1742-5468/2011/02/P02023} {\bibfield  {journal} {\bibinfo  {journal}
  {J. Stat. Mech.}\ }\textbf {\bibinfo {volume} {1102}},\ \bibinfo {pages}
  {P02023} (\bibinfo {year} {2011})},\ \Eprint {http://arxiv.org/abs/1012.3587}
  {arXiv:1012.3587 [cond-mat.str-el]} \BibitemShut {NoStop}%
\bibitem [{\citenamefont {Basko}\ \emph {et~al.}(2006)\citenamefont {Basko},
  \citenamefont {Aleiner},\ and\ \citenamefont {Altshuler}}]{Basko_2006}%
  \BibitemOpen
  \bibfield  {author} {\bibinfo {author} {\bibfnamefont {D.M.}\ \bibnamefont
  {Basko}}, \bibinfo {author} {\bibfnamefont {I.L.}\ \bibnamefont {Aleiner}}, \
  and\ \bibinfo {author} {\bibfnamefont {B.L.}\ \bibnamefont {Altshuler}},\
  }\bibfield  {title} {\enquote {\bibinfo {title} {Metal{\textendash}insulator
  transition in a weakly interacting many-electron system with localized
  single-particle states},}\ }\href {\doibase 10.1016/j.aop.2005.11.014}
  {\bibfield  {journal} {\bibinfo  {journal} {Annals of Physics}\ }\textbf
  {\bibinfo {volume} {321}},\ \bibinfo {pages} {1126--1205} (\bibinfo {year}
  {2006})}\BibitemShut {NoStop}%
\bibitem [{\citenamefont {Serbyn}\ \emph {et~al.}(2013)\citenamefont {Serbyn},
  \citenamefont {Papi\ifmmode~\acute{c}\else \'{c}\fi{}},\ and\ \citenamefont
  {Abanin}}]{PhysRevLett.111.127201}%
  \BibitemOpen
  \bibfield  {author} {\bibinfo {author} {\bibfnamefont {Maksym}\ \bibnamefont
  {Serbyn}}, \bibinfo {author} {\bibfnamefont {Z.}~\bibnamefont
  {Papi\ifmmode~\acute{c}\else \'{c}\fi{}}}, \ and\ \bibinfo {author}
  {\bibfnamefont {Dmitry~A.}\ \bibnamefont {Abanin}},\ }\bibfield  {title}
  {\enquote {\bibinfo {title} {Local conservation laws and the structure of the
  many-body localized states},}\ }\href {\doibase
  10.1103/PhysRevLett.111.127201} {\bibfield  {journal} {\bibinfo  {journal}
  {Phys. Rev. Lett.}\ }\textbf {\bibinfo {volume} {111}},\ \bibinfo {pages}
  {127201} (\bibinfo {year} {2013})},\ \Eprint {http://arxiv.org/abs/1305.5554}
  {arXiv:1305.5554 [cond-mat.dis-nn]} \BibitemShut {NoStop}%
\bibitem [{\citenamefont {Oganesyan}\ and\ \citenamefont
  {Huse}(2007)}]{PhysRevB.75.155111}%
  \BibitemOpen
  \bibfield  {author} {\bibinfo {author} {\bibfnamefont {Vadim}\ \bibnamefont
  {Oganesyan}}\ and\ \bibinfo {author} {\bibfnamefont {David~A.}\ \bibnamefont
  {Huse}},\ }\bibfield  {title} {\enquote {\bibinfo {title} {Localization of
  interacting fermions at high temperature},}\ }\href {\doibase
  10.1103/PhysRevB.75.155111} {\bibfield  {journal} {\bibinfo  {journal} {Phys.
  Rev. B}\ }\textbf {\bibinfo {volume} {75}},\ \bibinfo {pages} {155111}
  (\bibinfo {year} {2007})},\ \Eprint {http://arxiv.org/abs/cond-mat/0610854}
  {arXiv:cond-mat/0610854 [cond-mat.str-el]} \BibitemShut {NoStop}%
\bibitem [{\citenamefont {Pal}\ and\ \citenamefont
  {Huse}(2010)}]{PhysRevB.82.174411}%
  \BibitemOpen
  \bibfield  {author} {\bibinfo {author} {\bibfnamefont {Arijeet}\ \bibnamefont
  {Pal}}\ and\ \bibinfo {author} {\bibfnamefont {David~A.}\ \bibnamefont
  {Huse}},\ }\bibfield  {title} {\enquote {\bibinfo {title} {Many-body
  localization phase transition},}\ }\href {\doibase
  10.1103/PhysRevB.82.174411} {\bibfield  {journal} {\bibinfo  {journal} {Phys.
  Rev. B}\ }\textbf {\bibinfo {volume} {82}},\ \bibinfo {pages} {174411}
  (\bibinfo {year} {2010})},\ \Eprint {http://arxiv.org/abs/1010.1992}
  {arXiv:1010.1992 [cond-mat.dis-nn]} \BibitemShut {NoStop}%
\bibitem [{\citenamefont {Huse}\ \emph {et~al.}(2014)\citenamefont {Huse},
  \citenamefont {Nandkishore},\ and\ \citenamefont
  {Oganesyan}}]{PhysRevB.90.174202}%
  \BibitemOpen
  \bibfield  {author} {\bibinfo {author} {\bibfnamefont {David~A.}\
  \bibnamefont {Huse}}, \bibinfo {author} {\bibfnamefont {Rahul}\ \bibnamefont
  {Nandkishore}}, \ and\ \bibinfo {author} {\bibfnamefont {Vadim}\ \bibnamefont
  {Oganesyan}},\ }\bibfield  {title} {\enquote {\bibinfo {title} {Phenomenology
  of fully many-body-localized systems},}\ }\href {\doibase
  10.1103/PhysRevB.90.174202} {\bibfield  {journal} {\bibinfo  {journal} {Phys.
  Rev. B}\ }\textbf {\bibinfo {volume} {90}},\ \bibinfo {pages} {174202}
  (\bibinfo {year} {2014})},\ \Eprint {http://arxiv.org/abs/1408.4297}
  {arXiv:1408.4297 [cond-mat.stat-mech]} \BibitemShut {NoStop}%
\bibitem [{\citenamefont {Nandkishore}\ and\ \citenamefont
  {Huse}(2015)}]{nand_huse}%
  \BibitemOpen
  \bibfield  {author} {\bibinfo {author} {\bibfnamefont {Rahul}\ \bibnamefont
  {Nandkishore}}\ and\ \bibinfo {author} {\bibfnamefont {David~A.}\
  \bibnamefont {Huse}},\ }\bibfield  {title} {\enquote {\bibinfo {title}
  {Many-body localization and thermalization in quantum statistical
  mechanics},}\ }\href {\doibase 10.1146/annurev-conmatphys-031214-014726}
  {\bibfield  {journal} {\bibinfo  {journal} {Annual Review of Condensed Matter
  Physics}\ }\textbf {\bibinfo {volume} {6}},\ \bibinfo {pages} {15--38}
  (\bibinfo {year} {2015})},\ \Eprint {http://arxiv.org/abs/1404.0686}
  {arXiv:1404.0686 [cond-mat.stat-mech]} \BibitemShut {NoStop}%
\bibitem [{\citenamefont {Abanin}\ \emph {et~al.}(2019)\citenamefont {Abanin},
  \citenamefont {Altman}, \citenamefont {Bloch},\ and\ \citenamefont
  {Serbyn}}]{RevModPhys.91.021001}%
  \BibitemOpen
  \bibfield  {author} {\bibinfo {author} {\bibfnamefont {Dmitry~A.}\
  \bibnamefont {Abanin}}, \bibinfo {author} {\bibfnamefont {Ehud}\ \bibnamefont
  {Altman}}, \bibinfo {author} {\bibfnamefont {Immanuel}\ \bibnamefont
  {Bloch}}, \ and\ \bibinfo {author} {\bibfnamefont {Maksym}\ \bibnamefont
  {Serbyn}},\ }\bibfield  {title} {\enquote {\bibinfo {title} {Colloquium:
  Many-body localization, thermalization, and entanglement},}\ }\href {\doibase
  10.1103/RevModPhys.91.021001} {\bibfield  {journal} {\bibinfo  {journal}
  {Rev. Mod. Phys.}\ }\textbf {\bibinfo {volume} {91}},\ \bibinfo {pages}
  {021001} (\bibinfo {year} {2019})},\ \Eprint
  {http://arxiv.org/abs/1804.11065} {arXiv:1804.11065 [cond-mat.dis-nn]}
  \BibitemShut {NoStop}%
\bibitem [{\citenamefont {Turner}\ \emph
  {et~al.}(2018{\natexlab{a}})\citenamefont {Turner}, \citenamefont
  {Michailidis}, \citenamefont {Abanin}, \citenamefont {Serbyn},\ and\
  \citenamefont {Papi{\'{c}}}}]{Turner_2018}%
  \BibitemOpen
  \bibfield  {author} {\bibinfo {author} {\bibfnamefont {C.~J.}\ \bibnamefont
  {Turner}}, \bibinfo {author} {\bibfnamefont {A.~A.}\ \bibnamefont
  {Michailidis}}, \bibinfo {author} {\bibfnamefont {D.~A.}\ \bibnamefont
  {Abanin}}, \bibinfo {author} {\bibfnamefont {M.}~\bibnamefont {Serbyn}}, \
  and\ \bibinfo {author} {\bibfnamefont {Z.}~\bibnamefont {Papi{\'{c}}}},\
  }\bibfield  {title} {\enquote {\bibinfo {title} {Weak ergodicity breaking
  from quantum many-body scars},}\ }\href {\doibase 10.1038/s41567-018-0137-5}
  {\bibfield  {journal} {\bibinfo  {journal} {Nature Physics}\ }\textbf
  {\bibinfo {volume} {14}},\ \bibinfo {pages} {745--749} (\bibinfo {year}
  {2018}{\natexlab{a}})}\BibitemShut {NoStop}%
\bibitem [{\citenamefont {Heller}(1984)}]{PhysRevLett.53.1515}%
  \BibitemOpen
  \bibfield  {author} {\bibinfo {author} {\bibfnamefont {Eric~J.}\ \bibnamefont
  {Heller}},\ }\bibfield  {title} {\enquote {\bibinfo {title} {Bound-state
  eigenfunctions of classically chaotic hamiltonian systems: Scars of periodic
  orbits},}\ }\href {\doibase 10.1103/PhysRevLett.53.1515} {\bibfield
  {journal} {\bibinfo  {journal} {Phys. Rev. Lett.}\ }\textbf {\bibinfo
  {volume} {53}},\ \bibinfo {pages} {1515--1518} (\bibinfo {year}
  {1984})}\BibitemShut {NoStop}%
\bibitem [{\citenamefont {Bernien}\ \emph {et~al.}(2017)\citenamefont
  {Bernien}, \citenamefont {Schwartz}, \citenamefont {Keesling}, \citenamefont
  {Levine}, \citenamefont {Omran}, \citenamefont {Pichler}, \citenamefont
  {Choi}, \citenamefont {Zibrov}, \citenamefont {Endres}, \citenamefont
  {Greiner}, \citenamefont {Vuleti{\'{c}}},\ and\ \citenamefont
  {Lukin}}]{Bernien_2017}%
  \BibitemOpen
  \bibfield  {author} {\bibinfo {author} {\bibfnamefont {Hannes}\ \bibnamefont
  {Bernien}}, \bibinfo {author} {\bibfnamefont {Sylvain}\ \bibnamefont
  {Schwartz}}, \bibinfo {author} {\bibfnamefont {Alexander}\ \bibnamefont
  {Keesling}}, \bibinfo {author} {\bibfnamefont {Harry}\ \bibnamefont
  {Levine}}, \bibinfo {author} {\bibfnamefont {Ahmed}\ \bibnamefont {Omran}},
  \bibinfo {author} {\bibfnamefont {Hannes}\ \bibnamefont {Pichler}}, \bibinfo
  {author} {\bibfnamefont {Soonwon}\ \bibnamefont {Choi}}, \bibinfo {author}
  {\bibfnamefont {Alexander~S.}\ \bibnamefont {Zibrov}}, \bibinfo {author}
  {\bibfnamefont {Manuel}\ \bibnamefont {Endres}}, \bibinfo {author}
  {\bibfnamefont {Markus}\ \bibnamefont {Greiner}}, \bibinfo {author}
  {\bibfnamefont {Vladan}\ \bibnamefont {Vuleti{\'{c}}}}, \ and\ \bibinfo
  {author} {\bibfnamefont {Mikhail~D.}\ \bibnamefont {Lukin}},\ }\bibfield
  {title} {\enquote {\bibinfo {title} {Probing many-body dynamics on a 51-atom
  quantum simulator},}\ }\href {\doibase 10.1038/nature24622} {\bibfield
  {journal} {\bibinfo  {journal} {Nature}\ }\textbf {\bibinfo {volume} {551}},\
  \bibinfo {pages} {579--584} (\bibinfo {year} {2017})}\BibitemShut {NoStop}%
\bibitem [{\citenamefont {Scherg}\ \emph {et~al.}(2021)\citenamefont {Scherg},
  \citenamefont {Kohlert}, \citenamefont {Sala}, \citenamefont {Pollmann},
  \citenamefont {Hebbe~Madhusudhana}, \citenamefont {Bloch},\ and\
  \citenamefont {Aidelsburger}}]{Scherg:2020mcp}%
  \BibitemOpen
  \bibfield  {author} {\bibinfo {author} {\bibfnamefont {Sebastian}\
  \bibnamefont {Scherg}}, \bibinfo {author} {\bibfnamefont {Thomas}\
  \bibnamefont {Kohlert}}, \bibinfo {author} {\bibfnamefont {Pablo}\
  \bibnamefont {Sala}}, \bibinfo {author} {\bibfnamefont {Frank}\ \bibnamefont
  {Pollmann}}, \bibinfo {author} {\bibfnamefont {Bharath}\ \bibnamefont
  {Hebbe~Madhusudhana}}, \bibinfo {author} {\bibfnamefont {Immanuel}\
  \bibnamefont {Bloch}}, \ and\ \bibinfo {author} {\bibfnamefont {Monika}\
  \bibnamefont {Aidelsburger}},\ }\bibfield  {title} {\enquote {\bibinfo
  {title} {{Observing non-ergodicity due to kinetic constraints in tilted
  Fermi-Hubbard chains}},}\ }\href {\doibase 10.1038/s41467-021-24726-0}
  {\bibfield  {journal} {\bibinfo  {journal} {Nature Commun.}\ }\textbf
  {\bibinfo {volume} {12}},\ \bibinfo {pages} {4490} (\bibinfo {year}
  {2021})},\ \Eprint {http://arxiv.org/abs/2010.12965} {arXiv:2010.12965
  [cond-mat.quant-gas]} \BibitemShut {NoStop}%
\bibitem [{\citenamefont {Zhao}\ \emph {et~al.}(2020)\citenamefont {Zhao},
  \citenamefont {Vovrosh}, \citenamefont {Mintert},\ and\ \citenamefont
  {Knolle}}]{PhysRevLett.124.160604}%
  \BibitemOpen
  \bibfield  {author} {\bibinfo {author} {\bibfnamefont {Hongzheng}\
  \bibnamefont {Zhao}}, \bibinfo {author} {\bibfnamefont {Joseph}\ \bibnamefont
  {Vovrosh}}, \bibinfo {author} {\bibfnamefont {Florian}\ \bibnamefont
  {Mintert}}, \ and\ \bibinfo {author} {\bibfnamefont {Johannes}\ \bibnamefont
  {Knolle}},\ }\bibfield  {title} {\enquote {\bibinfo {title} {Quantum
  many-body scars in optical lattices},}\ }\href {\doibase
  10.1103/PhysRevLett.124.160604} {\bibfield  {journal} {\bibinfo  {journal}
  {Phys. Rev. Lett.}\ }\textbf {\bibinfo {volume} {124}},\ \bibinfo {pages}
  {160604} (\bibinfo {year} {2020})},\ \Eprint
  {http://arxiv.org/abs/2002.01746} {arXiv:2002.01746 [cond-mat.quant-gas]}
  \BibitemShut {NoStop}%
\bibitem [{\citenamefont {Choi}\ \emph {et~al.}(2019)\citenamefont {Choi},
  \citenamefont {Turner}, \citenamefont {Pichler}, \citenamefont {Ho},
  \citenamefont {Michailidis}, \citenamefont {Papi\ifmmode~\acute{c}\else
  \'{c}\fi{}}, \citenamefont {Serbyn}, \citenamefont {Lukin},\ and\
  \citenamefont {Abanin}}]{PhysRevLett.122.220603}%
  \BibitemOpen
  \bibfield  {author} {\bibinfo {author} {\bibfnamefont {Soonwon}\ \bibnamefont
  {Choi}}, \bibinfo {author} {\bibfnamefont {Christopher~J.}\ \bibnamefont
  {Turner}}, \bibinfo {author} {\bibfnamefont {Hannes}\ \bibnamefont
  {Pichler}}, \bibinfo {author} {\bibfnamefont {Wen~Wei}\ \bibnamefont {Ho}},
  \bibinfo {author} {\bibfnamefont {Alexios~A.}\ \bibnamefont {Michailidis}},
  \bibinfo {author} {\bibfnamefont {Zlatko}\ \bibnamefont
  {Papi\ifmmode~\acute{c}\else \'{c}\fi{}}}, \bibinfo {author} {\bibfnamefont
  {Maksym}\ \bibnamefont {Serbyn}}, \bibinfo {author} {\bibfnamefont
  {Mikhail~D.}\ \bibnamefont {Lukin}}, \ and\ \bibinfo {author} {\bibfnamefont
  {Dmitry~A.}\ \bibnamefont {Abanin}},\ }\bibfield  {title} {\enquote {\bibinfo
  {title} {Emergent su(2) dynamics and perfect quantum many-body scars},}\
  }\href {\doibase 10.1103/PhysRevLett.122.220603} {\bibfield  {journal}
  {\bibinfo  {journal} {Phys. Rev. Lett.}\ }\textbf {\bibinfo {volume} {122}},\
  \bibinfo {pages} {220603} (\bibinfo {year} {2019})},\ \Eprint
  {http://arxiv.org/abs/1812.05561} {arXiv:1812.05561 [quant-ph]} \BibitemShut
  {NoStop}%
\bibitem [{\citenamefont {Yao}\ \emph {et~al.}(2022)\citenamefont {Yao},
  \citenamefont {Pan}, \citenamefont {Liu},\ and\ \citenamefont
  {Zhai}}]{PhysRevB.105.125123}%
  \BibitemOpen
  \bibfield  {author} {\bibinfo {author} {\bibfnamefont {Zhiyuan}\ \bibnamefont
  {Yao}}, \bibinfo {author} {\bibfnamefont {Lei}\ \bibnamefont {Pan}}, \bibinfo
  {author} {\bibfnamefont {Shang}\ \bibnamefont {Liu}}, \ and\ \bibinfo
  {author} {\bibfnamefont {Hui}\ \bibnamefont {Zhai}},\ }\bibfield  {title}
  {\enquote {\bibinfo {title} {Quantum many-body scars and quantum
  criticality},}\ }\href {\doibase 10.1103/PhysRevB.105.125123} {\bibfield
  {journal} {\bibinfo  {journal} {Phys. Rev. B}\ }\textbf {\bibinfo {volume}
  {105}},\ \bibinfo {pages} {125123} (\bibinfo {year} {2022})},\ \Eprint
  {http://arxiv.org/abs/2108.05113} {arXiv:2108.05113 [cond-mat.quant-gas]}
  \BibitemShut {NoStop}%
\bibitem [{\citenamefont {Turner}\ \emph
  {et~al.}(2018{\natexlab{b}})\citenamefont {Turner}, \citenamefont
  {Michailidis}, \citenamefont {Abanin}, \citenamefont {Serbyn},\ and\
  \citenamefont {Papi\ifmmode~\acute{c}\else \'{c}\fi{}}}]{PhysRevB.98.155134}%
  \BibitemOpen
  \bibfield  {author} {\bibinfo {author} {\bibfnamefont {C.~J.}\ \bibnamefont
  {Turner}}, \bibinfo {author} {\bibfnamefont {A.~A.}\ \bibnamefont
  {Michailidis}}, \bibinfo {author} {\bibfnamefont {D.~A.}\ \bibnamefont
  {Abanin}}, \bibinfo {author} {\bibfnamefont {M.}~\bibnamefont {Serbyn}}, \
  and\ \bibinfo {author} {\bibfnamefont {Z.}~\bibnamefont
  {Papi\ifmmode~\acute{c}\else \'{c}\fi{}}},\ }\bibfield  {title} {\enquote
  {\bibinfo {title} {Quantum scarred eigenstates in a rydberg atom chain:
  Entanglement, breakdown of thermalization, and stability to perturbations},}\
  }\href {\doibase 10.1103/PhysRevB.98.155134} {\bibfield  {journal} {\bibinfo
  {journal} {Phys. Rev. B}\ }\textbf {\bibinfo {volume} {98}},\ \bibinfo
  {pages} {155134} (\bibinfo {year} {2018}{\natexlab{b}})},\ \Eprint
  {http://arxiv.org/abs/1806.10933} {arXiv:1806.10933 [cond-mat.quant-gas]}
  \BibitemShut {NoStop}%
\bibitem [{\citenamefont {Bull}\ \emph {et~al.}(2020)\citenamefont {Bull},
  \citenamefont {Desaules},\ and\ \citenamefont {Papi\ifmmode~\acute{c}\else
  \'{c}\fi{}}}]{PhysRevB.101.165139}%
  \BibitemOpen
  \bibfield  {author} {\bibinfo {author} {\bibfnamefont {Kieran}\ \bibnamefont
  {Bull}}, \bibinfo {author} {\bibfnamefont {Jean-Yves}\ \bibnamefont
  {Desaules}}, \ and\ \bibinfo {author} {\bibfnamefont {Zlatko}\ \bibnamefont
  {Papi\ifmmode~\acute{c}\else \'{c}\fi{}}},\ }\bibfield  {title} {\enquote
  {\bibinfo {title} {Quantum scars as embeddings of weakly broken lie algebra
  representations},}\ }\href {\doibase 10.1103/PhysRevB.101.165139} {\bibfield
  {journal} {\bibinfo  {journal} {Phys. Rev. B}\ }\textbf {\bibinfo {volume}
  {101}},\ \bibinfo {pages} {165139} (\bibinfo {year} {2020})},\ \Eprint
  {http://arxiv.org/abs/2001.08232} {arXiv:2001.08232 [cond-mat.str-el]}
  \BibitemShut {NoStop}%
\bibitem [{\citenamefont {Khemani}\ \emph {et~al.}(2019)\citenamefont
  {Khemani}, \citenamefont {Laumann},\ and\ \citenamefont
  {Chandran}}]{PhysRevB.99.161101}%
  \BibitemOpen
  \bibfield  {author} {\bibinfo {author} {\bibfnamefont {Vedika}\ \bibnamefont
  {Khemani}}, \bibinfo {author} {\bibfnamefont {Chris~R.}\ \bibnamefont
  {Laumann}}, \ and\ \bibinfo {author} {\bibfnamefont {Anushya}\ \bibnamefont
  {Chandran}},\ }\bibfield  {title} {\enquote {\bibinfo {title} {Signatures of
  integrability in the dynamics of rydberg-blockaded chains},}\ }\href
  {\doibase 10.1103/PhysRevB.99.161101} {\bibfield  {journal} {\bibinfo
  {journal} {Phys. Rev. B}\ }\textbf {\bibinfo {volume} {99}},\ \bibinfo
  {pages} {161101} (\bibinfo {year} {2019})},\ \Eprint
  {http://arxiv.org/abs/1807.02108} {arXiv:1807.02108 [cond-mat.str-el]}
  \BibitemShut {NoStop}%
\bibitem [{\citenamefont {Lin}\ and\ \citenamefont
  {Motrunich}(2019)}]{PhysRevLett.122.173401}%
  \BibitemOpen
  \bibfield  {author} {\bibinfo {author} {\bibfnamefont {Cheng-Ju}\
  \bibnamefont {Lin}}\ and\ \bibinfo {author} {\bibfnamefont {Olexei~I.}\
  \bibnamefont {Motrunich}},\ }\bibfield  {title} {\enquote {\bibinfo {title}
  {Exact quantum many-body scar states in the rydberg-blockaded atom chain},}\
  }\href {\doibase 10.1103/PhysRevLett.122.173401} {\bibfield  {journal}
  {\bibinfo  {journal} {Phys. Rev. Lett.}\ }\textbf {\bibinfo {volume} {122}},\
  \bibinfo {pages} {173401} (\bibinfo {year} {2019})},\ \Eprint
  {http://arxiv.org/abs/1810.00888} {arXiv:1810.00888 [cond-mat.quant-gas]}
  \BibitemShut {NoStop}%
\bibitem [{\citenamefont {Zhao}\ \emph {et~al.}(2021)\citenamefont {Zhao},
  \citenamefont {Smith}, \citenamefont {Mintert},\ and\ \citenamefont
  {Knolle}}]{PhysRevLett.127.150601}%
  \BibitemOpen
  \bibfield  {author} {\bibinfo {author} {\bibfnamefont {Hongzheng}\
  \bibnamefont {Zhao}}, \bibinfo {author} {\bibfnamefont {Adam}\ \bibnamefont
  {Smith}}, \bibinfo {author} {\bibfnamefont {Florian}\ \bibnamefont
  {Mintert}}, \ and\ \bibinfo {author} {\bibfnamefont {Johannes}\ \bibnamefont
  {Knolle}},\ }\bibfield  {title} {\enquote {\bibinfo {title} {Orthogonal
  quantum many-body scars},}\ }\href {\doibase 10.1103/PhysRevLett.127.150601}
  {\bibfield  {journal} {\bibinfo  {journal} {Phys. Rev. Lett.}\ }\textbf
  {\bibinfo {volume} {127}},\ \bibinfo {pages} {150601} (\bibinfo {year}
  {2021})},\ \Eprint {http://arxiv.org/abs/2102.07672} {arXiv:2102.07672
  [cond-mat.stat-mech]} \BibitemShut {NoStop}%
\bibitem [{\citenamefont {Desaules}\ \emph
  {et~al.}(2022{\natexlab{a}})\citenamefont {Desaules}, \citenamefont
  {Pietracaprina}, \citenamefont {Papi\'c}, \citenamefont {Goold},\ and\
  \citenamefont {Pappalardi}}]{Desaules:2021acx}%
  \BibitemOpen
  \bibfield  {author} {\bibinfo {author} {\bibfnamefont {Jean-Yves}\
  \bibnamefont {Desaules}}, \bibinfo {author} {\bibfnamefont {Francesca}\
  \bibnamefont {Pietracaprina}}, \bibinfo {author} {\bibfnamefont {Zlatko}\
  \bibnamefont {Papi\'c}}, \bibinfo {author} {\bibfnamefont {John}\
  \bibnamefont {Goold}}, \ and\ \bibinfo {author} {\bibfnamefont {Silvia}\
  \bibnamefont {Pappalardi}},\ }\bibfield  {title} {\enquote {\bibinfo {title}
  {{Extensive Multipartite Entanglement from su(2) Quantum Many-Body Scars}},}\
  }\href {\doibase 10.1103/PhysRevLett.129.020601} {\bibfield  {journal}
  {\bibinfo  {journal} {Phys. Rev. Lett.}\ }\textbf {\bibinfo {volume} {129}},\
  \bibinfo {pages} {020601} (\bibinfo {year} {2022}{\natexlab{a}})},\ \Eprint
  {http://arxiv.org/abs/2109.09724} {arXiv:2109.09724 [quant-ph]} \BibitemShut
  {NoStop}%
\bibitem [{\citenamefont {Langlett}\ \emph {et~al.}(2022)\citenamefont
  {Langlett}, \citenamefont {Yang}, \citenamefont {Wildeboer}, \citenamefont
  {Gorshkov}, \citenamefont {Iadecola},\ and\ \citenamefont
  {Xu}}]{PhysRevB.105.L060301}%
  \BibitemOpen
  \bibfield  {author} {\bibinfo {author} {\bibfnamefont {Christopher~M.}\
  \bibnamefont {Langlett}}, \bibinfo {author} {\bibfnamefont {Zhi-Cheng}\
  \bibnamefont {Yang}}, \bibinfo {author} {\bibfnamefont {Julia}\ \bibnamefont
  {Wildeboer}}, \bibinfo {author} {\bibfnamefont {Alexey~V.}\ \bibnamefont
  {Gorshkov}}, \bibinfo {author} {\bibfnamefont {Thomas}\ \bibnamefont
  {Iadecola}}, \ and\ \bibinfo {author} {\bibfnamefont {Shenglong}\
  \bibnamefont {Xu}},\ }\bibfield  {title} {\enquote {\bibinfo {title} {Rainbow
  scars: From area to volume law},}\ }\href {\doibase
  10.1103/PhysRevB.105.L060301} {\bibfield  {journal} {\bibinfo  {journal}
  {Phys. Rev. B}\ }\textbf {\bibinfo {volume} {105}},\ \bibinfo {pages}
  {L060301} (\bibinfo {year} {2022})},\ \Eprint
  {http://arxiv.org/abs/2107.03416} {arXiv:2107.03416 [cond-mat.str-el]}
  \BibitemShut {NoStop}%
\bibitem [{\citenamefont {O'Dea}\ \emph {et~al.}(2020)\citenamefont {O'Dea},
  \citenamefont {Burnell}, \citenamefont {Chandran},\ and\ \citenamefont
  {Khemani}}]{ODea:2020ooe}%
  \BibitemOpen
  \bibfield  {author} {\bibinfo {author} {\bibfnamefont {Nicholas}\
  \bibnamefont {O'Dea}}, \bibinfo {author} {\bibfnamefont {Fiona}\ \bibnamefont
  {Burnell}}, \bibinfo {author} {\bibfnamefont {Anushya}\ \bibnamefont
  {Chandran}}, \ and\ \bibinfo {author} {\bibfnamefont {Vedika}\ \bibnamefont
  {Khemani}},\ }\bibfield  {title} {\enquote {\bibinfo {title} {{From tunnels
  to towers: quantum scars from Lie Algebras and q-deformed Lie Algebras}},}\
  }\href {\doibase 10.1103/PhysRevResearch.2.043305} {\bibfield  {journal}
  {\bibinfo  {journal} {Phys. Rev. Res.}\ }\textbf {\bibinfo {volume} {2}},\
  \bibinfo {pages} {043305} (\bibinfo {year} {2020})},\ \Eprint
  {http://arxiv.org/abs/2007.16207} {arXiv:2007.16207 [cond-mat.stat-mech]}
  \BibitemShut {NoStop}%
\bibitem [{\citenamefont {Chandran}\ \emph {et~al.}(2022)\citenamefont
  {Chandran}, \citenamefont {Iadecola}, \citenamefont {Khemani},\ and\
  \citenamefont {Moessner}}]{Chandran:2022jtd}%
  \BibitemOpen
  \bibfield  {author} {\bibinfo {author} {\bibfnamefont {Anushya}\ \bibnamefont
  {Chandran}}, \bibinfo {author} {\bibfnamefont {Thomas}\ \bibnamefont
  {Iadecola}}, \bibinfo {author} {\bibfnamefont {Vedika}\ \bibnamefont
  {Khemani}}, \ and\ \bibinfo {author} {\bibfnamefont {Roderich}\ \bibnamefont
  {Moessner}},\ }\bibfield  {title} {\enquote {\bibinfo {title} {{Quantum
  Many-Body Scars: A Quasiparticle Perspective}},}\ }\href@noop {} {\
  (\bibinfo {year} {2022})},\ \Eprint {http://arxiv.org/abs/2206.11528}
  {arXiv:2206.11528 [cond-mat.str-el]} \BibitemShut {NoStop}%
\bibitem [{\citenamefont {Tang}\ \emph {et~al.}(2022)\citenamefont {Tang},
  \citenamefont {O'Dea},\ and\ \citenamefont {Chandran}}]{tang2021multi}%
  \BibitemOpen
  \bibfield  {author} {\bibinfo {author} {\bibfnamefont {Long-Hin}\
  \bibnamefont {Tang}}, \bibinfo {author} {\bibfnamefont {Nicholas}\
  \bibnamefont {O'Dea}}, \ and\ \bibinfo {author} {\bibfnamefont {Anushya}\
  \bibnamefont {Chandran}},\ }\bibfield  {title} {\enquote {\bibinfo {title}
  {Multimagnon quantum many-body scars from tensor operators},}\ }\href
  {\doibase 10.1103/PhysRevResearch.4.043006} {\bibfield  {journal} {\bibinfo
  {journal} {Phys. Rev. Research}\ }\textbf {\bibinfo {volume} {4}},\ \bibinfo
  {pages} {043006} (\bibinfo {year} {2022})},\ \Eprint
  {http://arxiv.org/abs/2110.11448} {arXiv:2110.11448 [cond-mat.str-el]}
  \BibitemShut {NoStop}%
\bibitem [{\citenamefont {Shibata}\ \emph {et~al.}(2020)\citenamefont
  {Shibata}, \citenamefont {Yoshioka},\ and\ \citenamefont
  {Katsura}}]{PhysRevLett.124.180604}%
  \BibitemOpen
  \bibfield  {author} {\bibinfo {author} {\bibfnamefont {Naoyuki}\ \bibnamefont
  {Shibata}}, \bibinfo {author} {\bibfnamefont {Nobuyuki}\ \bibnamefont
  {Yoshioka}}, \ and\ \bibinfo {author} {\bibfnamefont {Hosho}\ \bibnamefont
  {Katsura}},\ }\bibfield  {title} {\enquote {\bibinfo {title} {Onsager's scars
  in disordered spin chains},}\ }\href {\doibase
  10.1103/PhysRevLett.124.180604} {\bibfield  {journal} {\bibinfo  {journal}
  {Phys. Rev. Lett.}\ }\textbf {\bibinfo {volume} {124}},\ \bibinfo {pages}
  {180604} (\bibinfo {year} {2020})},\ \Eprint
  {http://arxiv.org/abs/1912.13399} {arXiv:1912.13399 [quant-ph]} \BibitemShut
  {NoStop}%
\bibitem [{\citenamefont {Pakrouski}\ \emph {et~al.}(2020)\citenamefont
  {Pakrouski}, \citenamefont {Pallegar}, \citenamefont {Popov},\ and\
  \citenamefont {Klebanov}}]{Pakrouski:2020hym}%
  \BibitemOpen
  \bibfield  {author} {\bibinfo {author} {\bibfnamefont {Kiryl}\ \bibnamefont
  {Pakrouski}}, \bibinfo {author} {\bibfnamefont {Preethi~N.}\ \bibnamefont
  {Pallegar}}, \bibinfo {author} {\bibfnamefont {Fedor~K.}\ \bibnamefont
  {Popov}}, \ and\ \bibinfo {author} {\bibfnamefont {Igor~R.}\ \bibnamefont
  {Klebanov}},\ }\bibfield  {title} {\enquote {\bibinfo {title} {{Many Body
  Scars as a Group Invariant Sector of Hilbert Space}},}\ }\href {\doibase
  10.1103/PhysRevLett.125.230602} {\bibfield  {journal} {\bibinfo  {journal}
  {Phys. Rev. Lett.}\ }\textbf {\bibinfo {volume} {125}},\ \bibinfo {pages}
  {230602} (\bibinfo {year} {2020})},\ \Eprint
  {http://arxiv.org/abs/2007.00845} {arXiv:2007.00845 [cond-mat.str-el]}
  \BibitemShut {NoStop}%
\bibitem [{\citenamefont {Pakrouski}\ \emph {et~al.}(2021)\citenamefont
  {Pakrouski}, \citenamefont {Pallegar}, \citenamefont {Popov},\ and\
  \citenamefont {Klebanov}}]{Pakrouski:2021jon}%
  \BibitemOpen
  \bibfield  {author} {\bibinfo {author} {\bibfnamefont {Kiryl}\ \bibnamefont
  {Pakrouski}}, \bibinfo {author} {\bibfnamefont {Preethi~N.}\ \bibnamefont
  {Pallegar}}, \bibinfo {author} {\bibfnamefont {Fedor~K.}\ \bibnamefont
  {Popov}}, \ and\ \bibinfo {author} {\bibfnamefont {Igor~R.}\ \bibnamefont
  {Klebanov}},\ }\bibfield  {title} {\enquote {\bibinfo {title} {{Group
  theoretic approach to many-body scar states in fermionic lattice models}},}\
  }\href {\doibase 10.1103/PhysRevResearch.3.043156} {\bibfield  {journal}
  {\bibinfo  {journal} {Phys. Rev. Res.}\ }\textbf {\bibinfo {volume} {3}},\
  \bibinfo {pages} {043156} (\bibinfo {year} {2021})},\ \Eprint
  {http://arxiv.org/abs/2106.10300} {arXiv:2106.10300 [cond-mat.str-el]}
  \BibitemShut {NoStop}%
\bibitem [{\citenamefont {Iadecola}\ and\ \citenamefont
  {Schecter}(2020)}]{PhysRevB.101.024306}%
  \BibitemOpen
  \bibfield  {author} {\bibinfo {author} {\bibfnamefont {Thomas}\ \bibnamefont
  {Iadecola}}\ and\ \bibinfo {author} {\bibfnamefont {Michael}\ \bibnamefont
  {Schecter}},\ }\bibfield  {title} {\enquote {\bibinfo {title} {Quantum
  many-body scar states with emergent kinetic constraints and
  finite-entanglement revivals},}\ }\href {\doibase
  10.1103/PhysRevB.101.024306} {\bibfield  {journal} {\bibinfo  {journal}
  {Phys. Rev. B}\ }\textbf {\bibinfo {volume} {101}},\ \bibinfo {pages}
  {024306} (\bibinfo {year} {2020})},\ \Eprint
  {http://arxiv.org/abs/1910.11350} {arXiv:1910.11350 [cond-mat.str-el]}
  \BibitemShut {NoStop}%
\bibitem [{\citenamefont {Dooley}\ \emph {et~al.}(2022)\citenamefont {Dooley},
  \citenamefont {Pappalardi},\ and\ \citenamefont {Goold}}]{Dooley:2022kge}%
  \BibitemOpen
  \bibfield  {author} {\bibinfo {author} {\bibfnamefont {Shane}\ \bibnamefont
  {Dooley}}, \bibinfo {author} {\bibfnamefont {Silvia}\ \bibnamefont
  {Pappalardi}}, \ and\ \bibinfo {author} {\bibfnamefont {John}\ \bibnamefont
  {Goold}},\ }\bibfield  {title} {\enquote {\bibinfo {title} {{Entanglement
  enhanced metrology with quantum many-body scars}},}\ }\href@noop {} {\
  (\bibinfo {year} {2022})},\ \Eprint {http://arxiv.org/abs/2207.13521}
  {arXiv:2207.13521 [quant-ph]} \BibitemShut {NoStop}%
\bibitem [{\citenamefont {Desaules}\ \emph
  {et~al.}(2022{\natexlab{b}})\citenamefont {Desaules}, \citenamefont
  {Banerjee}, \citenamefont {Hudomal}, \citenamefont {Papi\'c}, \citenamefont
  {Sen},\ and\ \citenamefont {Halimeh}}]{Desaules:2022ibp}%
  \BibitemOpen
  \bibfield  {author} {\bibinfo {author} {\bibfnamefont {Jean-Yves}\
  \bibnamefont {Desaules}}, \bibinfo {author} {\bibfnamefont {Debasish}\
  \bibnamefont {Banerjee}}, \bibinfo {author} {\bibfnamefont {Ana}\
  \bibnamefont {Hudomal}}, \bibinfo {author} {\bibfnamefont {Zlatko}\
  \bibnamefont {Papi\'c}}, \bibinfo {author} {\bibfnamefont {Arnab}\
  \bibnamefont {Sen}}, \ and\ \bibinfo {author} {\bibfnamefont {Jad~C.}\
  \bibnamefont {Halimeh}},\ }\bibfield  {title} {\enquote {\bibinfo {title}
  {{Weak Ergodicity Breaking in the Schwinger Model}},}\ }\href@noop {} {\
  (\bibinfo {year} {2022}{\natexlab{b}})},\ \Eprint
  {http://arxiv.org/abs/2203.08830} {arXiv:2203.08830 [cond-mat.str-el]}
  \BibitemShut {NoStop}%
\bibitem [{\citenamefont {Su}\ \emph {et~al.}(2022)\citenamefont {Su},
  \citenamefont {Sun}, \citenamefont {Hudomal}, \citenamefont {Desaules},
  \citenamefont {Zhou}, \citenamefont {Yang}, \citenamefont {Halimeh},
  \citenamefont {Yuan}, \citenamefont {Papi\'c},\ and\ \citenamefont
  {Pan}}]{Su:2022glk}%
  \BibitemOpen
  \bibfield  {author} {\bibinfo {author} {\bibfnamefont {Guo-Xian}\
  \bibnamefont {Su}}, \bibinfo {author} {\bibfnamefont {Hui}\ \bibnamefont
  {Sun}}, \bibinfo {author} {\bibfnamefont {Ana}\ \bibnamefont {Hudomal}},
  \bibinfo {author} {\bibfnamefont {Jean-Yves}\ \bibnamefont {Desaules}},
  \bibinfo {author} {\bibfnamefont {Zhao-Yu}\ \bibnamefont {Zhou}}, \bibinfo
  {author} {\bibfnamefont {Bing}\ \bibnamefont {Yang}}, \bibinfo {author}
  {\bibfnamefont {Jad~C.}\ \bibnamefont {Halimeh}}, \bibinfo {author}
  {\bibfnamefont {Zhen-Sheng}\ \bibnamefont {Yuan}}, \bibinfo {author}
  {\bibfnamefont {Zlatko}\ \bibnamefont {Papi\'c}}, \ and\ \bibinfo {author}
  {\bibfnamefont {Jian-Wei}\ \bibnamefont {Pan}},\ }\bibfield  {title}
  {\enquote {\bibinfo {title} {{Observation of unconventional many-body
  scarring in a quantum simulator}},}\ }\href@noop {} {\  (\bibinfo {year}
  {2022})},\ \Eprint {http://arxiv.org/abs/2201.00821} {arXiv:2201.00821
  [cond-mat.quant-gas]} \BibitemShut {NoStop}%
\bibitem [{\citenamefont {Desaules}\ \emph
  {et~al.}(2022{\natexlab{c}})\citenamefont {Desaules}, \citenamefont
  {Hudomal}, \citenamefont {Banerjee}, \citenamefont {Sen}, \citenamefont
  {Papi\'c},\ and\ \citenamefont {Halimeh}}]{Desaules:2022kse}%
  \BibitemOpen
  \bibfield  {author} {\bibinfo {author} {\bibfnamefont {Jean-Yves}\
  \bibnamefont {Desaules}}, \bibinfo {author} {\bibfnamefont {Ana}\
  \bibnamefont {Hudomal}}, \bibinfo {author} {\bibfnamefont {Debasish}\
  \bibnamefont {Banerjee}}, \bibinfo {author} {\bibfnamefont {Arnab}\
  \bibnamefont {Sen}}, \bibinfo {author} {\bibfnamefont {Zlatko}\ \bibnamefont
  {Papi\'c}}, \ and\ \bibinfo {author} {\bibfnamefont {Jad~C.}\ \bibnamefont
  {Halimeh}},\ }\bibfield  {title} {\enquote {\bibinfo {title} {{Prominent
  quantum many-body scars in a truncated Schwinger model}},}\ }\href@noop {} {\
   (\bibinfo {year} {2022}{\natexlab{c}})},\ \Eprint
  {http://arxiv.org/abs/2204.01745} {arXiv:2204.01745 [cond-mat.quant-gas]}
  \BibitemShut {NoStop}%
\bibitem [{\citenamefont {Magnifico}\ \emph {et~al.}(2020)\citenamefont
  {Magnifico}, \citenamefont {Dalmonte}, \citenamefont {Facchi}, \citenamefont
  {Pascazio}, \citenamefont {Pepe},\ and\ \citenamefont
  {Ercolessi}}]{Magnifico:2019kyj}%
  \BibitemOpen
  \bibfield  {author} {\bibinfo {author} {\bibfnamefont {Giuseppe}\
  \bibnamefont {Magnifico}}, \bibinfo {author} {\bibfnamefont {Marcello}\
  \bibnamefont {Dalmonte}}, \bibinfo {author} {\bibfnamefont {Paolo}\
  \bibnamefont {Facchi}}, \bibinfo {author} {\bibfnamefont {Saverio}\
  \bibnamefont {Pascazio}}, \bibinfo {author} {\bibfnamefont {Francesco~V.}\
  \bibnamefont {Pepe}}, \ and\ \bibinfo {author} {\bibfnamefont {Elisa}\
  \bibnamefont {Ercolessi}},\ }\bibfield  {title} {\enquote {\bibinfo {title}
  {{Real Time Dynamics and Confinement in the $Z_n$ Schwinger-Weyl lattice
  model for 1+1 QED}},}\ }\href {\doibase 10.22331/q-2020-06-15-281} {\bibfield
   {journal} {\bibinfo  {journal} {Quantum}\ }\textbf {\bibinfo {volume} {4}},\
  \bibinfo {pages} {281} (\bibinfo {year} {2020})},\ \Eprint
  {http://arxiv.org/abs/1909.04821} {arXiv:1909.04821 [quant-ph]} \BibitemShut
  {NoStop}%
\bibitem [{\citenamefont {Serbyn}\ \emph {et~al.}(2021)\citenamefont {Serbyn},
  \citenamefont {Abanin},\ and\ \citenamefont {Papi\'c}}]{Serbyn:2020wys}%
  \BibitemOpen
  \bibfield  {author} {\bibinfo {author} {\bibfnamefont {Maksym}\ \bibnamefont
  {Serbyn}}, \bibinfo {author} {\bibfnamefont {Dmitry~A.}\ \bibnamefont
  {Abanin}}, \ and\ \bibinfo {author} {\bibfnamefont {Zlatko}\ \bibnamefont
  {Papi\'c}},\ }\bibfield  {title} {\enquote {\bibinfo {title} {{Quantum
  many-body scars and weak breaking of ergodicity}},}\ }\href {\doibase
  10.1038/s41567-021-01230-2} {\bibfield  {journal} {\bibinfo  {journal}
  {Nature Phys.}\ }\textbf {\bibinfo {volume} {17}},\ \bibinfo {pages}
  {675--685} (\bibinfo {year} {2021})},\ \Eprint
  {http://arxiv.org/abs/2011.09486} {arXiv:2011.09486 [quant-ph]} \BibitemShut
  {NoStop}%
\bibitem [{\citenamefont {Moudgalya}\ \emph {et~al.}(2018)\citenamefont
  {Moudgalya}, \citenamefont {Regnault},\ and\ \citenamefont
  {Bernevig}}]{PhysRevB.98.235156}%
  \BibitemOpen
  \bibfield  {author} {\bibinfo {author} {\bibfnamefont {Sanjay}\ \bibnamefont
  {Moudgalya}}, \bibinfo {author} {\bibfnamefont {Nicolas}\ \bibnamefont
  {Regnault}}, \ and\ \bibinfo {author} {\bibfnamefont {B.~Andrei}\
  \bibnamefont {Bernevig}},\ }\bibfield  {title} {\enquote {\bibinfo {title}
  {{Entanglement of exact excited states of Affleck-Kennedy-Lieb-Tasaki models:
  Exact results, many-body scars, and violation of the strong eigenstate
  thermalization hypothesis}},}\ }\href {\doibase 10.1103/PhysRevB.98.235156}
  {\bibfield  {journal} {\bibinfo  {journal} {Phys. Rev. B}\ }\textbf {\bibinfo
  {volume} {98}},\ \bibinfo {pages} {235156} (\bibinfo {year} {2018})},\
  \Eprint {http://arxiv.org/abs/1806.09624} {arXiv:1806.09624
  [cond-mat.str-el]} \BibitemShut {NoStop}%
\bibitem [{\citenamefont {Papi\'c}()}]{papic}%
  \BibitemOpen
  \bibfield  {author} {\bibinfo {author} {\bibfnamefont {Zlatko}\ \bibnamefont
  {Papi\'c}},\ }\bibfield  {title} {\enquote {\bibinfo {title} {Weak ergodicity
  breaking through the lens of quantum entanglement},}\ }\href@noop {} {\
  }\Eprint {http://arxiv.org/abs/2108.03460} {arXiv:2108.03460
  [cond-mat.quant-gas]} \BibitemShut {NoStop}%
\bibitem [{\citenamefont {Moudgalya}\ and\ \citenamefont
  {Motrunich}(2022)}]{Moudgalya:2021ixk}%
  \BibitemOpen
  \bibfield  {author} {\bibinfo {author} {\bibfnamefont {Sanjay}\ \bibnamefont
  {Moudgalya}}\ and\ \bibinfo {author} {\bibfnamefont {Olexei~I.}\ \bibnamefont
  {Motrunich}},\ }\bibfield  {title} {\enquote {\bibinfo {title} {{Hilbert
  Space Fragmentation and Commutant Algebras}},}\ }\href {\doibase
  10.1103/PhysRevX.12.011050} {\bibfield  {journal} {\bibinfo  {journal} {Phys.
  Rev. X}\ }\textbf {\bibinfo {volume} {12}},\ \bibinfo {pages} {011050}
  (\bibinfo {year} {2022})},\ \Eprint {http://arxiv.org/abs/2108.10324}
  {arXiv:2108.10324 [cond-mat.stat-mech]} \BibitemShut {NoStop}%
\bibitem [{\citenamefont {Moudgalya}\ \emph {et~al.}(2019)\citenamefont
  {Moudgalya}, \citenamefont {Prem}, \citenamefont {Nandkishore}, \citenamefont
  {Regnault},\ and\ \citenamefont {Bernevig}}]{moudgalya2022thermalization}%
  \BibitemOpen
  \bibfield  {author} {\bibinfo {author} {\bibfnamefont {Sanjay}\ \bibnamefont
  {Moudgalya}}, \bibinfo {author} {\bibfnamefont {Abhinav}\ \bibnamefont
  {Prem}}, \bibinfo {author} {\bibfnamefont {Rahul}\ \bibnamefont
  {Nandkishore}}, \bibinfo {author} {\bibfnamefont {Nicolas}\ \bibnamefont
  {Regnault}}, \ and\ \bibinfo {author} {\bibfnamefont {B~Andrei}\ \bibnamefont
  {Bernevig}},\ }\bibfield  {title} {\enquote {\bibinfo {title}
  {{Thermalization and its absence within Krylov subspaces of a constrained
  Hamiltonian}},}\ }\href@noop {} {\  (\bibinfo {year} {2019})},\ \Eprint
  {http://arxiv.org/abs/1910.14048} {arXiv:1910.14048 [cond-mat.str-el]}
  \BibitemShut {NoStop}%
\bibitem [{\citenamefont {Parker}\ \emph {et~al.}(2019)\citenamefont {Parker},
  \citenamefont {Cao}, \citenamefont {Avdoshkin}, \citenamefont {Scaffidi},\
  and\ \citenamefont {Altman}}]{Parker:2018yvk}%
  \BibitemOpen
  \bibfield  {author} {\bibinfo {author} {\bibfnamefont {Daniel~E.}\
  \bibnamefont {Parker}}, \bibinfo {author} {\bibfnamefont {Xiangyu}\
  \bibnamefont {Cao}}, \bibinfo {author} {\bibfnamefont {Alexander}\
  \bibnamefont {Avdoshkin}}, \bibinfo {author} {\bibfnamefont {Thomas}\
  \bibnamefont {Scaffidi}}, \ and\ \bibinfo {author} {\bibfnamefont {Ehud}\
  \bibnamefont {Altman}},\ }\bibfield  {title} {\enquote {\bibinfo {title} {{A
  Universal Operator Growth Hypothesis}},}\ }\href {\doibase
  10.1103/PhysRevX.9.041017} {\bibfield  {journal} {\bibinfo  {journal} {Phys.
  Rev. X}\ }\textbf {\bibinfo {volume} {9}},\ \bibinfo {pages} {041017}
  (\bibinfo {year} {2019})},\ \Eprint {http://arxiv.org/abs/1812.08657}
  {arXiv:1812.08657 [cond-mat.stat-mech]} \BibitemShut {NoStop}%
\bibitem [{\citenamefont {Avdoshkin}\ and\ \citenamefont
  {Dymarsky}(2020)}]{Avdoshkin:2019trj}%
  \BibitemOpen
  \bibfield  {author} {\bibinfo {author} {\bibfnamefont {Alexander}\
  \bibnamefont {Avdoshkin}}\ and\ \bibinfo {author} {\bibfnamefont {Anatoly}\
  \bibnamefont {Dymarsky}},\ }\bibfield  {title} {\enquote {\bibinfo {title}
  {{Euclidean operator growth and quantum chaos}},}\ }\href {\doibase
  10.1103/PhysRevResearch.2.043234} {\bibfield  {journal} {\bibinfo  {journal}
  {Phys. Rev. Res.}\ }\textbf {\bibinfo {volume} {2}},\ \bibinfo {pages}
  {043234} (\bibinfo {year} {2020})},\ \Eprint
  {http://arxiv.org/abs/1911.09672} {arXiv:1911.09672 [cond-mat.stat-mech]}
  \BibitemShut {NoStop}%
\bibitem [{\citenamefont {Dymarsky}\ and\ \citenamefont
  {Gorsky}(2020)}]{Dymarsky:2019elm}%
  \BibitemOpen
  \bibfield  {author} {\bibinfo {author} {\bibfnamefont {Anatoly}\ \bibnamefont
  {Dymarsky}}\ and\ \bibinfo {author} {\bibfnamefont {Alexander}\ \bibnamefont
  {Gorsky}},\ }\bibfield  {title} {\enquote {\bibinfo {title} {{Quantum chaos
  as delocalization in Krylov space}},}\ }\href {\doibase
  10.1103/PhysRevB.102.085137} {\bibfield  {journal} {\bibinfo  {journal}
  {Phys. Rev. B}\ }\textbf {\bibinfo {volume} {102}},\ \bibinfo {pages}
  {085137} (\bibinfo {year} {2020})},\ \Eprint
  {http://arxiv.org/abs/1912.12227} {arXiv:1912.12227 [cond-mat.stat-mech]}
  \BibitemShut {NoStop}%
\bibitem [{\citenamefont {Bhattacharyya}\ \emph {et~al.}(2021)\citenamefont
  {Bhattacharyya}, \citenamefont {Chemissany}, \citenamefont {Haque},
  \citenamefont {Murugan},\ and\ \citenamefont {Yan}}]{Bhattacharyya:2020art}%
  \BibitemOpen
  \bibfield  {author} {\bibinfo {author} {\bibfnamefont {Arpan}\ \bibnamefont
  {Bhattacharyya}}, \bibinfo {author} {\bibfnamefont {Wissam}\ \bibnamefont
  {Chemissany}}, \bibinfo {author} {\bibfnamefont {S.~Shajidul}\ \bibnamefont
  {Haque}}, \bibinfo {author} {\bibfnamefont {Jeff}\ \bibnamefont {Murugan}}, \
  and\ \bibinfo {author} {\bibfnamefont {Bin}\ \bibnamefont {Yan}},\ }\bibfield
   {title} {\enquote {\bibinfo {title} {{The Multi-faceted Inverted Harmonic
  Oscillator: Chaos and Complexity}},}\ }\href {\doibase
  10.21468/SciPostPhysCore.4.1.002} {\bibfield  {journal} {\bibinfo  {journal}
  {SciPost Phys. Core}\ }\textbf {\bibinfo {volume} {4}},\ \bibinfo {pages}
  {002} (\bibinfo {year} {2021})},\ \Eprint {http://arxiv.org/abs/2007.01232}
  {arXiv:2007.01232 [hep-th]} \BibitemShut {NoStop}%
\bibitem [{\citenamefont {Rabinovici}\ \emph
  {et~al.}(2022{\natexlab{a}})\citenamefont {Rabinovici}, \citenamefont
  {S\'anchez-Garrido}, \citenamefont {Shir},\ and\ \citenamefont
  {Sonner}}]{Rabinovici:2021qqt}%
  \BibitemOpen
  \bibfield  {author} {\bibinfo {author} {\bibfnamefont {E.}~\bibnamefont
  {Rabinovici}}, \bibinfo {author} {\bibfnamefont {A.}~\bibnamefont
  {S\'anchez-Garrido}}, \bibinfo {author} {\bibfnamefont {R.}~\bibnamefont
  {Shir}}, \ and\ \bibinfo {author} {\bibfnamefont {J.}~\bibnamefont
  {Sonner}},\ }\bibfield  {title} {\enquote {\bibinfo {title} {{Krylov
  localization and suppression of complexity}},}\ }\href {\doibase
  10.1007/JHEP03(2022)211} {\bibfield  {journal} {\bibinfo  {journal} {JHEP}\
  }\textbf {\bibinfo {volume} {03}},\ \bibinfo {pages} {211} (\bibinfo {year}
  {2022}{\natexlab{a}})},\ \Eprint {http://arxiv.org/abs/2112.12128}
  {arXiv:2112.12128 [hep-th]} \BibitemShut {NoStop}%
\bibitem [{\citenamefont {Barb\'on}\ \emph {et~al.}(2019)\citenamefont
  {Barb\'on}, \citenamefont {Rabinovici}, \citenamefont {Shir},\ and\
  \citenamefont {Sinha}}]{Barbon:2019wsy}%
  \BibitemOpen
  \bibfield  {author} {\bibinfo {author} {\bibfnamefont {J.~L.~F.}\
  \bibnamefont {Barb\'on}}, \bibinfo {author} {\bibfnamefont {E.}~\bibnamefont
  {Rabinovici}}, \bibinfo {author} {\bibfnamefont {R.}~\bibnamefont {Shir}}, \
  and\ \bibinfo {author} {\bibfnamefont {R.}~\bibnamefont {Sinha}},\ }\bibfield
   {title} {\enquote {\bibinfo {title} {{On The Evolution Of Operator
  Complexity Beyond Scrambling}},}\ }\href {\doibase 10.1007/JHEP10(2019)264}
  {\bibfield  {journal} {\bibinfo  {journal} {JHEP}\ }\textbf {\bibinfo
  {volume} {10}},\ \bibinfo {pages} {264} (\bibinfo {year} {2019})},\ \Eprint
  {http://arxiv.org/abs/1907.05393} {arXiv:1907.05393 [hep-th]} \BibitemShut
  {NoStop}%
\bibitem [{\citenamefont {Bhattacharjee}\ \emph {et~al.}(2022)\citenamefont
  {Bhattacharjee}, \citenamefont {Cao}, \citenamefont {Nandy},\ and\
  \citenamefont {Pathak}}]{Bhattacharjee:2022vlt}%
  \BibitemOpen
  \bibfield  {author} {\bibinfo {author} {\bibfnamefont {Budhaditya}\
  \bibnamefont {Bhattacharjee}}, \bibinfo {author} {\bibfnamefont {Xiangyu}\
  \bibnamefont {Cao}}, \bibinfo {author} {\bibfnamefont {Pratik}\ \bibnamefont
  {Nandy}}, \ and\ \bibinfo {author} {\bibfnamefont {Tanay}\ \bibnamefont
  {Pathak}},\ }\bibfield  {title} {\enquote {\bibinfo {title} {{Krylov
  complexity in saddle-dominated scrambling}},}\ }\href {\doibase
  10.1007/JHEP05(2022)174} {\bibfield  {journal} {\bibinfo  {journal} {JHEP}\
  }\textbf {\bibinfo {volume} {05}},\ \bibinfo {pages} {174} (\bibinfo {year}
  {2022})},\ \Eprint {http://arxiv.org/abs/2203.03534} {arXiv:2203.03534
  [quant-ph]} \BibitemShut {NoStop}%
\bibitem [{\citenamefont {Balasubramanian}\ \emph {et~al.}(2022)\citenamefont
  {Balasubramanian}, \citenamefont {Caputa}, \citenamefont {Magan},\ and\
  \citenamefont {Wu}}]{Balasubramanian:2022tpr}%
  \BibitemOpen
  \bibfield  {author} {\bibinfo {author} {\bibfnamefont {Vijay}\ \bibnamefont
  {Balasubramanian}}, \bibinfo {author} {\bibfnamefont {Pawel}\ \bibnamefont
  {Caputa}}, \bibinfo {author} {\bibfnamefont {Javier~M.}\ \bibnamefont
  {Magan}}, \ and\ \bibinfo {author} {\bibfnamefont {Qingyue}\ \bibnamefont
  {Wu}},\ }\bibfield  {title} {\enquote {\bibinfo {title} {{Quantum chaos and
  the complexity of spread of states}},}\ }\href {\doibase
  10.1103/PhysRevD.106.046007} {\bibfield  {journal} {\bibinfo  {journal}
  {Phys. Rev. D}\ }\textbf {\bibinfo {volume} {106}},\ \bibinfo {pages}
  {046007} (\bibinfo {year} {2022})},\ \Eprint
  {http://arxiv.org/abs/2202.06957} {arXiv:2202.06957 [hep-th]} \BibitemShut
  {NoStop}%
\bibitem [{\citenamefont {Jefferson}\ and\ \citenamefont
  {Myers}(2017)}]{Jefferson:2017sdb}%
  \BibitemOpen
  \bibfield  {author} {\bibinfo {author} {\bibfnamefont {Ro}~\bibnamefont
  {Jefferson}}\ and\ \bibinfo {author} {\bibfnamefont {Robert~C.}\ \bibnamefont
  {Myers}},\ }\bibfield  {title} {\enquote {\bibinfo {title} {{Circuit
  complexity in quantum field theory}},}\ }\href {\doibase
  10.1007/JHEP10(2017)107} {\bibfield  {journal} {\bibinfo  {journal} {JHEP}\
  }\textbf {\bibinfo {volume} {10}},\ \bibinfo {pages} {107} (\bibinfo {year}
  {2017})},\ \Eprint {http://arxiv.org/abs/1707.08570} {arXiv:1707.08570
  [hep-th]} \BibitemShut {NoStop}%
\bibitem [{\citenamefont {Chapman}\ \emph {et~al.}(2018)\citenamefont
  {Chapman}, \citenamefont {Heller}, \citenamefont {Marrochio},\ and\
  \citenamefont {Pastawski}}]{Chapman:2017rqy}%
  \BibitemOpen
  \bibfield  {author} {\bibinfo {author} {\bibfnamefont {Shira}\ \bibnamefont
  {Chapman}}, \bibinfo {author} {\bibfnamefont {Michal~P.}\ \bibnamefont
  {Heller}}, \bibinfo {author} {\bibfnamefont {Hugo}\ \bibnamefont
  {Marrochio}}, \ and\ \bibinfo {author} {\bibfnamefont {Fernando}\
  \bibnamefont {Pastawski}},\ }\bibfield  {title} {\enquote {\bibinfo {title}
  {{Toward a Definition of Complexity for Quantum Field Theory States}},}\
  }\href {\doibase 10.1103/PhysRevLett.120.121602} {\bibfield  {journal}
  {\bibinfo  {journal} {Phys. Rev. Lett.}\ }\textbf {\bibinfo {volume} {120}},\
  \bibinfo {pages} {121602} (\bibinfo {year} {2018})},\ \Eprint
  {http://arxiv.org/abs/1707.08582} {arXiv:1707.08582 [hep-th]} \BibitemShut
  {NoStop}%
\bibitem [{\citenamefont {Caputa}\ and\ \citenamefont
  {Liu}(2022)}]{Caputa:2022eye}%
  \BibitemOpen
  \bibfield  {author} {\bibinfo {author} {\bibfnamefont {Pawel}\ \bibnamefont
  {Caputa}}\ and\ \bibinfo {author} {\bibfnamefont {Sinong}\ \bibnamefont
  {Liu}},\ }\bibfield  {title} {\enquote {\bibinfo {title} {{Quantum complexity
  and topological phases of matter}},}\ }\href@noop {} {\  (\bibinfo {year}
  {2022})},\ \Eprint {http://arxiv.org/abs/2205.05688} {arXiv:2205.05688
  [hep-th]} \BibitemShut {NoStop}%
\bibitem [{\citenamefont {Viswanath}\ and\ \citenamefont
  {M{\"u}ller}(1994)}]{viswanath1994recursion}%
  \BibitemOpen
  \bibfield  {author} {\bibinfo {author} {\bibfnamefont {V.S.}\ \bibnamefont
  {Viswanath}}\ and\ \bibinfo {author} {\bibfnamefont {G.}~\bibnamefont
  {M{\"u}ller}},\ }\href {https://books.google.co.in/books?id=X2Ug4w17rnMC}
  {\emph {\bibinfo {title} {The Recursion Method: Application to Many Body
  Dynamics}}},\ Lecture Notes in Physics Monographs\ (\bibinfo  {publisher}
  {Springer Berlin Heidelberg},\ \bibinfo {year} {1994})\BibitemShut {NoStop}%
\bibitem [{\citenamefont {Jian}\ \emph {et~al.}(2021)\citenamefont {Jian},
  \citenamefont {Swingle},\ and\ \citenamefont {Xian}}]{Jian:2020qpp}%
  \BibitemOpen
  \bibfield  {author} {\bibinfo {author} {\bibfnamefont {Shao-Kai}\
  \bibnamefont {Jian}}, \bibinfo {author} {\bibfnamefont {Brian}\ \bibnamefont
  {Swingle}}, \ and\ \bibinfo {author} {\bibfnamefont {Zhuo-Yu}\ \bibnamefont
  {Xian}},\ }\bibfield  {title} {\enquote {\bibinfo {title} {{Complexity growth
  of operators in the SYK model and in JT gravity}},}\ }\href {\doibase
  10.1007/JHEP03(2021)014} {\bibfield  {journal} {\bibinfo  {journal} {JHEP}\
  }\textbf {\bibinfo {volume} {03}},\ \bibinfo {pages} {014} (\bibinfo {year}
  {2021})},\ \Eprint {http://arxiv.org/abs/2008.12274} {arXiv:2008.12274
  [hep-th]} \BibitemShut {NoStop}%
\bibitem [{\citenamefont {Rabinovici}\ \emph {et~al.}(2021)\citenamefont
  {Rabinovici}, \citenamefont {S\'anchez-Garrido}, \citenamefont {Shir},\ and\
  \citenamefont {Sonner}}]{Rabinovici:2020ryf}%
  \BibitemOpen
  \bibfield  {author} {\bibinfo {author} {\bibfnamefont {E.}~\bibnamefont
  {Rabinovici}}, \bibinfo {author} {\bibfnamefont {A.}~\bibnamefont
  {S\'anchez-Garrido}}, \bibinfo {author} {\bibfnamefont {R.}~\bibnamefont
  {Shir}}, \ and\ \bibinfo {author} {\bibfnamefont {J.}~\bibnamefont
  {Sonner}},\ }\bibfield  {title} {\enquote {\bibinfo {title} {{Operator
  complexity: a journey to the edge of Krylov space}},}\ }\href {\doibase
  10.1007/JHEP06(2021)062} {\bibfield  {journal} {\bibinfo  {journal} {JHEP}\
  }\textbf {\bibinfo {volume} {06}},\ \bibinfo {pages} {062} (\bibinfo {year}
  {2021})},\ \Eprint {http://arxiv.org/abs/2009.01862} {arXiv:2009.01862
  [hep-th]} \BibitemShut {NoStop}%
\bibitem [{\citenamefont {Noh}(2021)}]{PhysRevE.104.034112}%
  \BibitemOpen
  \bibfield  {author} {\bibinfo {author} {\bibfnamefont {Jae~Dong}\
  \bibnamefont {Noh}},\ }\bibfield  {title} {\enquote {\bibinfo {title}
  {Operator growth in the transverse-field ising spin chain with
  integrability-breaking longitudinal field},}\ }\href {\doibase
  10.1103/PhysRevE.104.034112} {\bibfield  {journal} {\bibinfo  {journal}
  {Phys. Rev. E}\ }\textbf {\bibinfo {volume} {104}},\ \bibinfo {pages}
  {034112} (\bibinfo {year} {2021})},\ \Eprint
  {http://arxiv.org/abs/2107.08287} {arXiv:2107.08287 [quant-ph]} \BibitemShut
  {NoStop}%
\bibitem [{\citenamefont {Cao}(2021)}]{Cao:2020zls}%
  \BibitemOpen
  \bibfield  {author} {\bibinfo {author} {\bibfnamefont {Xiangyu}\ \bibnamefont
  {Cao}},\ }\bibfield  {title} {\enquote {\bibinfo {title} {{A statistical
  mechanism for operator growth}},}\ }\href {\doibase 10.1088/1751-8121/abe77c}
  {\bibfield  {journal} {\bibinfo  {journal} {J. Phys. A}\ }\textbf {\bibinfo
  {volume} {54}},\ \bibinfo {pages} {144001} (\bibinfo {year} {2021})},\
  \Eprint {http://arxiv.org/abs/2012.06544} {arXiv:2012.06544
  [cond-mat.stat-mech]} \BibitemShut {NoStop}%
\bibitem [{\citenamefont {Yates}\ and\ \citenamefont
  {Mitra}(2021)}]{Yates:2021asz}%
  \BibitemOpen
  \bibfield  {author} {\bibinfo {author} {\bibfnamefont {Daniel~J.}\
  \bibnamefont {Yates}}\ and\ \bibinfo {author} {\bibfnamefont {Aditi}\
  \bibnamefont {Mitra}},\ }\bibfield  {title} {\enquote {\bibinfo {title}
  {{Strong and almost strong modes of Floquet spin chains in Krylov
  subspaces}},}\ }\href {\doibase 10.1103/PhysRevB.104.195121} {\bibfield
  {journal} {\bibinfo  {journal} {Phys. Rev. B}\ }\textbf {\bibinfo {volume}
  {104}},\ \bibinfo {pages} {195121} (\bibinfo {year} {2021})},\ \Eprint
  {http://arxiv.org/abs/2105.13246} {arXiv:2105.13246 [cond-mat.str-el]}
  \BibitemShut {NoStop}%
\bibitem [{\citenamefont {Kim}\ \emph {et~al.}(2022)\citenamefont {Kim},
  \citenamefont {Murugan}, \citenamefont {Olle},\ and\ \citenamefont
  {Rosa}}]{Kim:2021okd}%
  \BibitemOpen
  \bibfield  {author} {\bibinfo {author} {\bibfnamefont {Joonho}\ \bibnamefont
  {Kim}}, \bibinfo {author} {\bibfnamefont {Jeff}\ \bibnamefont {Murugan}},
  \bibinfo {author} {\bibfnamefont {Jan}\ \bibnamefont {Olle}}, \ and\ \bibinfo
  {author} {\bibfnamefont {Dario}\ \bibnamefont {Rosa}},\ }\bibfield  {title}
  {\enquote {\bibinfo {title} {{Operator delocalization in quantum
  networks}},}\ }\href {\doibase 10.1103/PhysRevA.105.L010201} {\bibfield
  {journal} {\bibinfo  {journal} {Phys. Rev. A}\ }\textbf {\bibinfo {volume}
  {105}},\ \bibinfo {pages} {L010201} (\bibinfo {year} {2022})},\ \Eprint
  {http://arxiv.org/abs/2109.05301} {arXiv:2109.05301 [quant-ph]} \BibitemShut
  {NoStop}%
\bibitem [{\citenamefont {Caputa}\ and\ \citenamefont
  {Datta}(2021)}]{Caputa:2021ori}%
  \BibitemOpen
  \bibfield  {author} {\bibinfo {author} {\bibfnamefont {Pawel}\ \bibnamefont
  {Caputa}}\ and\ \bibinfo {author} {\bibfnamefont {Shouvik}\ \bibnamefont
  {Datta}},\ }\bibfield  {title} {\enquote {\bibinfo {title} {{Operator growth
  in 2d CFT}},}\ }\href {\doibase 10.1007/JHEP12(2021)188} {\bibfield
  {journal} {\bibinfo  {journal} {JHEP}\ }\textbf {\bibinfo {volume} {12}},\
  \bibinfo {pages} {188} (\bibinfo {year} {2021})},\ \Eprint
  {http://arxiv.org/abs/2110.10519} {arXiv:2110.10519 [hep-th]} \BibitemShut
  {NoStop}%
\bibitem [{\citenamefont {Trigueros}\ and\ \citenamefont
  {Lin}(2022)}]{Trigueros:2021rwj}%
  \BibitemOpen
  \bibfield  {author} {\bibinfo {author} {\bibfnamefont {Fabian~Ballar}\
  \bibnamefont {Trigueros}}\ and\ \bibinfo {author} {\bibfnamefont {Cheng-Ju}\
  \bibnamefont {Lin}},\ }\bibfield  {title} {\enquote {\bibinfo {title}
  {{Krylov complexity of many-body localization: Operator localization in
  Krylov basis}},}\ }\href {\doibase 10.21468/SciPostPhys.13.2.037} {\bibfield
  {journal} {\bibinfo  {journal} {SciPost Phys.}\ }\textbf {\bibinfo {volume}
  {13}},\ \bibinfo {pages} {037} (\bibinfo {year} {2022})},\ \Eprint
  {http://arxiv.org/abs/2112.04722} {arXiv:2112.04722 [cond-mat.dis-nn]}
  \BibitemShut {NoStop}%
\bibitem [{\citenamefont {Caputa}\ \emph {et~al.}(2022)\citenamefont {Caputa},
  \citenamefont {Magan},\ and\ \citenamefont {Patramanis}}]{Caputa:2021sib}%
  \BibitemOpen
  \bibfield  {author} {\bibinfo {author} {\bibfnamefont {Pawel}\ \bibnamefont
  {Caputa}}, \bibinfo {author} {\bibfnamefont {Javier~M.}\ \bibnamefont
  {Magan}}, \ and\ \bibinfo {author} {\bibfnamefont {Dimitrios}\ \bibnamefont
  {Patramanis}},\ }\bibfield  {title} {\enquote {\bibinfo {title} {{Geometry of
  Krylov complexity}},}\ }\href {\doibase 10.1103/PhysRevResearch.4.013041}
  {\bibfield  {journal} {\bibinfo  {journal} {Phys. Rev. Res.}\ }\textbf
  {\bibinfo {volume} {4}},\ \bibinfo {pages} {013041} (\bibinfo {year}
  {2022})},\ \Eprint {http://arxiv.org/abs/2109.03824} {arXiv:2109.03824
  [hep-th]} \BibitemShut {NoStop}%
\bibitem [{\citenamefont {Patramanis}(2022)}]{Patramanis:2021lkx}%
  \BibitemOpen
  \bibfield  {author} {\bibinfo {author} {\bibfnamefont {Dimitrios}\
  \bibnamefont {Patramanis}},\ }\bibfield  {title} {\enquote {\bibinfo {title}
  {{Probing the entanglement of operator growth}},}\ }\href {\doibase
  10.1093/ptep/ptac081} {\bibfield  {journal} {\bibinfo  {journal} {PTEP}\
  }\textbf {\bibinfo {volume} {2022}},\ \bibinfo {pages} {063A01} (\bibinfo
  {year} {2022})},\ \Eprint {http://arxiv.org/abs/2111.03424} {arXiv:2111.03424
  [hep-th]} \BibitemShut {NoStop}%
\bibitem [{\citenamefont {H\"ornedal}\ \emph {et~al.}(2022)\citenamefont
  {H\"ornedal}, \citenamefont {Carabba}, \citenamefont {Matsoukas-Roubeas},\
  and\ \citenamefont {del Campo}}]{Hornedal:2022pkc}%
  \BibitemOpen
  \bibfield  {author} {\bibinfo {author} {\bibfnamefont {Niklas}\ \bibnamefont
  {H\"ornedal}}, \bibinfo {author} {\bibfnamefont {Nicoletta}\ \bibnamefont
  {Carabba}}, \bibinfo {author} {\bibfnamefont {Apollonas~S.}\ \bibnamefont
  {Matsoukas-Roubeas}}, \ and\ \bibinfo {author} {\bibfnamefont {Adolfo}\
  \bibnamefont {del Campo}},\ }\bibfield  {title} {\enquote {\bibinfo {title}
  {{Ultimate Physical Limits to the Growth of Operator Complexity}},}\ }\href
  {\doibase 10.1038/s42005-022-00985-1} {\bibfield  {journal} {\bibinfo
  {journal} {Commun. Phys.}\ }\textbf {\bibinfo {volume} {5}},\ \bibinfo
  {pages} {207} (\bibinfo {year} {2022})},\ \Eprint
  {http://arxiv.org/abs/2202.05006} {arXiv:2202.05006 [quant-ph]} \BibitemShut
  {NoStop}%
\bibitem [{\citenamefont {Heveling}\ \emph
  {et~al.}(2022{\natexlab{a}})\citenamefont {Heveling}, \citenamefont {Wang},\
  and\ \citenamefont {Gemmer}}]{Heveling:2022hth}%
  \BibitemOpen
  \bibfield  {author} {\bibinfo {author} {\bibfnamefont {Robin}\ \bibnamefont
  {Heveling}}, \bibinfo {author} {\bibfnamefont {Jiaozi}\ \bibnamefont {Wang}},
  \ and\ \bibinfo {author} {\bibfnamefont {Jochen}\ \bibnamefont {Gemmer}},\
  }\bibfield  {title} {\enquote {\bibinfo {title} {{Numerically probing the
  universal operator growth hypothesis}},}\ }\href {\doibase
  10.1103/PhysRevE.106.014152} {\bibfield  {journal} {\bibinfo  {journal}
  {Phys. Rev. E}\ }\textbf {\bibinfo {volume} {106}},\ \bibinfo {pages}
  {014152} (\bibinfo {year} {2022}{\natexlab{a}})},\ \Eprint
  {http://arxiv.org/abs/2203.00533} {arXiv:2203.00533 [cond-mat.stat-mech]}
  \BibitemShut {NoStop}%
\bibitem [{\citenamefont {Heveling}\ \emph
  {et~al.}(2022{\natexlab{b}})\citenamefont {Heveling}, \citenamefont {Wang},
  \citenamefont {Bartsch},\ and\ \citenamefont {Gemmer}}]{Heveling:2022orr}%
  \BibitemOpen
  \bibfield  {author} {\bibinfo {author} {\bibfnamefont {Robin}\ \bibnamefont
  {Heveling}}, \bibinfo {author} {\bibfnamefont {Jiaozi}\ \bibnamefont {Wang}},
  \bibinfo {author} {\bibfnamefont {Christian}\ \bibnamefont {Bartsch}}, \ and\
  \bibinfo {author} {\bibfnamefont {Jochen}\ \bibnamefont {Gemmer}},\
  }\bibfield  {title} {\enquote {\bibinfo {title} {{Stability of Exponentially
  Damped Oscillations under Perturbations of the Mori-Chain}},}\ }\href
  {\doibase 10.1088/2399-6528/ac863b} {\bibfield  {journal} {\bibinfo
  {journal} {J. Phys. Comm.}\ }\textbf {\bibinfo {volume} {6}},\ \bibinfo
  {pages} {085009} (\bibinfo {year} {2022}{\natexlab{b}})},\ \Eprint
  {http://arxiv.org/abs/2204.06903} {arXiv:2204.06903 [quant-ph]} \BibitemShut
  {NoStop}%
\bibitem [{\citenamefont {Bhattacharya}\ \emph {et~al.}(2022)\citenamefont
  {Bhattacharya}, \citenamefont {Nandy}, \citenamefont {Nath},\ and\
  \citenamefont {Sahu}}]{Bhattacharya:2022gbz}%
  \BibitemOpen
  \bibfield  {author} {\bibinfo {author} {\bibfnamefont {Aranya}\ \bibnamefont
  {Bhattacharya}}, \bibinfo {author} {\bibfnamefont {Pratik}\ \bibnamefont
  {Nandy}}, \bibinfo {author} {\bibfnamefont {Pingal~Pratyush}\ \bibnamefont
  {Nath}}, \ and\ \bibinfo {author} {\bibfnamefont {Himanshu}\ \bibnamefont
  {Sahu}},\ }\bibfield  {title} {\enquote {\bibinfo {title} {{Operator growth
  and Krylov construction in dissipative open quantum systems}},}\ }\href@noop
  {} {\  (\bibinfo {year} {2022})},\ \Eprint {http://arxiv.org/abs/2207.05347}
  {arXiv:2207.05347 [quant-ph]} \BibitemShut {NoStop}%
\bibitem [{\citenamefont {Banerjee}\ \emph {et~al.}(2022)\citenamefont
  {Banerjee}, \citenamefont {Bhattacharyya}, \citenamefont {Drashni},\ and\
  \citenamefont {Pawar}}]{Banerjee:2022ime}%
  \BibitemOpen
  \bibfield  {author} {\bibinfo {author} {\bibfnamefont {Aritra}\ \bibnamefont
  {Banerjee}}, \bibinfo {author} {\bibfnamefont {Arpan}\ \bibnamefont
  {Bhattacharyya}}, \bibinfo {author} {\bibfnamefont {Priya}\ \bibnamefont
  {Drashni}}, \ and\ \bibinfo {author} {\bibfnamefont {Srinidhi}\ \bibnamefont
  {Pawar}},\ }\bibfield  {title} {\enquote {\bibinfo {title} {{CFT to BMS:
  Complexity and OTOC}},}\ }\href@noop {} {\  (\bibinfo {year} {2022})},\
  \Eprint {http://arxiv.org/abs/2205.15338} {arXiv:2205.15338 [hep-th]}
  \BibitemShut {NoStop}%
\bibitem [{\citenamefont {Rabinovici}\ \emph
  {et~al.}(2022{\natexlab{b}})\citenamefont {Rabinovici}, \citenamefont
  {S\'anchez-Garrido}, \citenamefont {Shir},\ and\ \citenamefont
  {Sonner}}]{Rabinovici:2022beu}%
  \BibitemOpen
  \bibfield  {author} {\bibinfo {author} {\bibfnamefont {E.}~\bibnamefont
  {Rabinovici}}, \bibinfo {author} {\bibfnamefont {A.}~\bibnamefont
  {S\'anchez-Garrido}}, \bibinfo {author} {\bibfnamefont {R.}~\bibnamefont
  {Shir}}, \ and\ \bibinfo {author} {\bibfnamefont {J.}~\bibnamefont
  {Sonner}},\ }\bibfield  {title} {\enquote {\bibinfo {title} {{Krylov
  complexity from integrability to chaos}},}\ }\href {\doibase
  10.1007/JHEP07(2022)151} {\bibfield  {journal} {\bibinfo  {journal} {JHEP}\
  }\textbf {\bibinfo {volume} {07}},\ \bibinfo {pages} {151} (\bibinfo {year}
  {2022}{\natexlab{b}})},\ \Eprint {http://arxiv.org/abs/2207.07701}
  {arXiv:2207.07701 [hep-th]} \BibitemShut {NoStop}%
\bibitem [{\citenamefont {Liu}\ \emph {et~al.}(2022)\citenamefont {Liu},
  \citenamefont {Tang},\ and\ \citenamefont {Zhai}}]{Liu:2022god}%
  \BibitemOpen
  \bibfield  {author} {\bibinfo {author} {\bibfnamefont {Chang}\ \bibnamefont
  {Liu}}, \bibinfo {author} {\bibfnamefont {Haifeng}\ \bibnamefont {Tang}}, \
  and\ \bibinfo {author} {\bibfnamefont {Hui}\ \bibnamefont {Zhai}},\
  }\bibfield  {title} {\enquote {\bibinfo {title} {{Krylov Complexity in Open
  Quantum Systems}},}\ }\href@noop {} {\  (\bibinfo {year} {2022})},\ \Eprint
  {http://arxiv.org/abs/2207.13603} {arXiv:2207.13603 [cond-mat.str-el]}
  \BibitemShut {NoStop}%
\bibitem [{\citenamefont {Polkovnikov}\ \emph {et~al.}(2011)\citenamefont
  {Polkovnikov}, \citenamefont {Sengupta}, \citenamefont {Silva},\ and\
  \citenamefont {Vengalattore}}]{RevModPhys.83.863}%
  \BibitemOpen
  \bibfield  {author} {\bibinfo {author} {\bibfnamefont {Anatoli}\ \bibnamefont
  {Polkovnikov}}, \bibinfo {author} {\bibfnamefont {Krishnendu}\ \bibnamefont
  {Sengupta}}, \bibinfo {author} {\bibfnamefont {Alessandro}\ \bibnamefont
  {Silva}}, \ and\ \bibinfo {author} {\bibfnamefont {Mukund}\ \bibnamefont
  {Vengalattore}},\ }\bibfield  {title} {\enquote {\bibinfo {title}
  {Colloquium: Nonequilibrium dynamics of closed interacting quantum
  systems},}\ }\href {\doibase 10.1103/RevModPhys.83.863} {\bibfield  {journal}
  {\bibinfo  {journal} {Rev. Mod. Phys.}\ }\textbf {\bibinfo {volume} {83}},\
  \bibinfo {pages} {863--883} (\bibinfo {year} {2011})},\ \Eprint
  {http://arxiv.org/abs/1007.5331} {arXiv:1007.5331 [cond-mat.stat-mech]}
  \BibitemShut {NoStop}%
\bibitem [{\citenamefont {Biedenharn}(1989)}]{Biedenharn_1989}%
  \BibitemOpen
  \bibfield  {author} {\bibinfo {author} {\bibfnamefont {L~C}\ \bibnamefont
  {Biedenharn}},\ }\bibfield  {title} {\enquote {\bibinfo {title} {The quantum
  group suq(2) and a q-analogue of the boson operators},}\ }\href {\doibase
  10.1088/0305-4470/22/18/004} {\bibfield  {journal} {\bibinfo  {journal}
  {Journal of Physics A: Mathematical and General}\ }\textbf {\bibinfo {volume}
  {22}},\ \bibinfo {pages} {L873--L878} (\bibinfo {year} {1989})}\BibitemShut
  {NoStop}%
\bibitem [{\citenamefont {Macfarlane}(1989)}]{Macfarlane_1989}%
  \BibitemOpen
  \bibfield  {author} {\bibinfo {author} {\bibfnamefont {A~J}\ \bibnamefont
  {Macfarlane}},\ }\bibfield  {title} {\enquote {\bibinfo {title} {On
  q-analogues of the quantum harmonic oscillator and the quantum group
  suq(2)},}\ }\href {\doibase 10.1088/0305-4470/22/21/020} {\bibfield
  {journal} {\bibinfo  {journal} {Journal of Physics A: Mathematical and
  General}\ }\textbf {\bibinfo {volume} {22}},\ \bibinfo {pages} {4581--4588}
  (\bibinfo {year} {1989})}\BibitemShut {NoStop}%
\bibitem [{\citenamefont {Batchelor}\ \emph {et~al.}(1990)\citenamefont
  {Batchelor}, \citenamefont {Mezincescu}, \citenamefont {Nepomechie},\ and\
  \citenamefont {Rittenberg}}]{Batchelor_1990}%
  \BibitemOpen
  \bibfield  {author} {\bibinfo {author} {\bibfnamefont {M~T}\ \bibnamefont
  {Batchelor}}, \bibinfo {author} {\bibfnamefont {L}~\bibnamefont
  {Mezincescu}}, \bibinfo {author} {\bibfnamefont {R~I}\ \bibnamefont
  {Nepomechie}}, \ and\ \bibinfo {author} {\bibfnamefont {V}~\bibnamefont
  {Rittenberg}},\ }\bibfield  {title} {\enquote {\bibinfo {title}
  {q-deformations of the o(3) symmetric spin-1 heisenberg chain},}\ }\href
  {\doibase 10.1088/0305-4470/23/4/003} {\bibfield  {journal} {\bibinfo
  {journal} {Journal of Physics A: Mathematical and General}\ }\textbf
  {\bibinfo {volume} {23}},\ \bibinfo {pages} {L141--L144} (\bibinfo {year}
  {1990})}\BibitemShut {NoStop}%
\bibitem [{\citenamefont {Lavagno}\ \emph {et~al.}(2006)\citenamefont
  {Lavagno}, \citenamefont {Scarfone},\ and\ \citenamefont
  {Swamy}}]{Lavagno:2006jx}%
  \BibitemOpen
  \bibfield  {author} {\bibinfo {author} {\bibfnamefont {A.}~\bibnamefont
  {Lavagno}}, \bibinfo {author} {\bibfnamefont {A.~M.}\ \bibnamefont
  {Scarfone}}, \ and\ \bibinfo {author} {\bibfnamefont {P.~Narayana}\
  \bibnamefont {Swamy}},\ }\bibfield  {title} {\enquote {\bibinfo {title}
  {{Classical and quantum q-deformed physical systems}},}\ }\href {\doibase
  10.1140/epjc/s2006-02557-y} {\bibfield  {journal} {\bibinfo  {journal} {Eur.
  Phys. J. C}\ }\textbf {\bibinfo {volume} {47}},\ \bibinfo {pages} {253--261}
  (\bibinfo {year} {2006})},\ \Eprint {http://arxiv.org/abs/quant-ph/0605026}
  {arXiv:quant-ph/0605026} \BibitemShut {NoStop}%
\bibitem [{\citenamefont {Youm}(2000)}]{Youm:2000yu}%
  \BibitemOpen
  \bibfield  {author} {\bibinfo {author} {\bibfnamefont {Donam}\ \bibnamefont
  {Youm}},\ }\bibfield  {title} {\enquote {\bibinfo {title} {{Q-deformed
  conformal quantum mechanics}},}\ }\href {\doibase 10.1103/PhysRevD.62.095009}
  {\bibfield  {journal} {\bibinfo  {journal} {Phys. Rev. D}\ }\textbf {\bibinfo
  {volume} {62}},\ \bibinfo {pages} {095009} (\bibinfo {year} {2000})},\
  \Eprint {http://arxiv.org/abs/hep-th/0007114} {arXiv:hep-th/0007114}
  \BibitemShut {NoStop}%
\bibitem [{\citenamefont {Chaichian}\ and\ \citenamefont
  {Demichev}(1996)}]{chaichian1996introduction}%
  \BibitemOpen
  \bibfield  {author} {\bibinfo {author} {\bibfnamefont {M.}~\bibnamefont
  {Chaichian}}\ and\ \bibinfo {author} {\bibfnamefont {A.}~\bibnamefont
  {Demichev}},\ }\href {https://books.google.co.in/books?id=\_vbsCgAAQBAJ}
  {\emph {\bibinfo {title} {Introduction To Quantum Groups}}}\ (\bibinfo
  {publisher} {World Scientific Publishing Company},\ \bibinfo {year}
  {1996})\BibitemShut {NoStop}%
\bibitem [{\citenamefont {Bonatsos}\ and\ \citenamefont
  {Daskaloyannis}(1999)}]{BONATSOS1999537}%
  \BibitemOpen
  \bibfield  {author} {\bibinfo {author} {\bibfnamefont {D.}~\bibnamefont
  {Bonatsos}}\ and\ \bibinfo {author} {\bibfnamefont {C.}~\bibnamefont
  {Daskaloyannis}},\ }\bibfield  {title} {\enquote {\bibinfo {title} {Quantum
  groups and their applications in nuclear physics},}\ }\href {\doibase
  https://doi.org/10.1016/S0146-6410(99)00100-3} {\bibfield  {journal}
  {\bibinfo  {journal} {Progress in Particle and Nuclear Physics}\ }\textbf
  {\bibinfo {volume} {43}},\ \bibinfo {pages} {537--618} (\bibinfo {year}
  {1999})}\BibitemShut {NoStop}%
\bibitem [{\citenamefont {Schmidt}\ and\ \citenamefont
  {Wachter}(2006)}]{schmidt2006q}%
  \BibitemOpen
  \bibfield  {author} {\bibinfo {author} {\bibfnamefont {Alexander}\
  \bibnamefont {Schmidt}}\ and\ \bibinfo {author} {\bibfnamefont {Hartmut}\
  \bibnamefont {Wachter}},\ }\bibfield  {title} {\enquote {\bibinfo {title}
  {q-deformed quantum lie algebras},}\ }\href@noop {} {\bibfield  {journal}
  {\bibinfo  {journal} {Journal of Geometry and Physics}\ }\textbf {\bibinfo
  {volume} {56}},\ \bibinfo {pages} {2289--2325} (\bibinfo {year}
  {2006})}\BibitemShut {NoStop}%
\bibitem [{\citenamefont {Wess}(2000)}]{wess2000q}%
  \BibitemOpen
  \bibfield  {author} {\bibinfo {author} {\bibfnamefont {Julius}\ \bibnamefont
  {Wess}},\ }\bibfield  {title} {\enquote {\bibinfo {title} {q-deformed
  heisenberg algebras},}\ }in\ \href@noop {} {\emph {\bibinfo {booktitle}
  {Geometry and Quantum Physics}}}\ (\bibinfo  {publisher} {Springer},\
  \bibinfo {year} {2000})\ pp.\ \bibinfo {pages} {311--382}\BibitemShut
  {NoStop}%
\bibitem [{\citenamefont {Ding}\ \emph {et~al.}(2016)\citenamefont {Ding},
  \citenamefont {Jia}, \citenamefont {Wu}, \citenamefont {Yan},\ and\
  \citenamefont {Zhao}}]{DING201618}%
  \BibitemOpen
  \bibfield  {author} {\bibinfo {author} {\bibfnamefont {Lu}~\bibnamefont
  {Ding}}, \bibinfo {author} {\bibfnamefont {Xiao-Yu}\ \bibnamefont {Jia}},
  \bibinfo {author} {\bibfnamefont {Ke}~\bibnamefont {Wu}}, \bibinfo {author}
  {\bibfnamefont {Zhao-Wen}\ \bibnamefont {Yan}}, \ and\ \bibinfo {author}
  {\bibfnamefont {Wei-Zhong}\ \bibnamefont {Zhao}},\ }\bibfield  {title}
  {\enquote {\bibinfo {title} {On q-deformed infinite-dimensional n-algebra},}\
  }\href {\doibase https://doi.org/10.1016/j.nuclphysb.2016.01.003} {\bibfield
  {journal} {\bibinfo  {journal} {Nuclear Physics B}\ }\textbf {\bibinfo
  {volume} {904}},\ \bibinfo {pages} {18--38} (\bibinfo {year} {2016})},\
  \Eprint {http://arxiv.org/abs/1404.0464} {arXiv:1404.0464 [hep-th]}
  \BibitemShut {NoStop}%
\bibitem [{\citenamefont {Fujikawa}(1996)}]{Fujikawa:1996fi}%
  \BibitemOpen
  \bibfield  {author} {\bibinfo {author} {\bibfnamefont {Kazuo}\ \bibnamefont
  {Fujikawa}},\ }\bibfield  {title} {\enquote {\bibinfo {title} {{Phase
  operator problem and an index theorem for Q deformed oscillator}},}\ }in\
  \href@noop {} {\emph {\bibinfo {booktitle} {{Frontiers in Quantum Field
  Theory in Honor of the 60th Birthday of Prof. K. Kikkawa}}}}\ (\bibinfo
  {year} {1996})\ pp.\ \bibinfo {pages} {354--366},\ \Eprint
  {http://arxiv.org/abs/hep-th/9603130} {arXiv:hep-th/9603130} \BibitemShut
  {NoStop}%
\bibitem [{\citenamefont {Fujikawa}\ \emph {et~al.}(1997)\citenamefont
  {Fujikawa}, \citenamefont {Kubo},\ and\ \citenamefont
  {Oh}}]{fujikawa1997schwinger}%
  \BibitemOpen
  \bibfield  {author} {\bibinfo {author} {\bibfnamefont {Kazuo}\ \bibnamefont
  {Fujikawa}}, \bibinfo {author} {\bibfnamefont {Harunobu}\ \bibnamefont
  {Kubo}}, \ and\ \bibinfo {author} {\bibfnamefont {CH}~\bibnamefont {Oh}},\
  }\bibfield  {title} {\enquote {\bibinfo {title} {A schwinger term in
  q-deformed su (2) algebra},}\ }\href@noop {} {\bibfield  {journal} {\bibinfo
  {journal} {Modern Physics Letters A}\ }\textbf {\bibinfo {volume} {12}},\
  \bibinfo {pages} {403--409} (\bibinfo {year} {1997})}\BibitemShut {NoStop}%
\bibitem [{\citenamefont {Pradeep}\ \emph {et~al.}(2020)\citenamefont
  {Pradeep}, \citenamefont {Anupama},\ and\ \citenamefont
  {Sudheesh}}]{pradeep2020dynamics}%
  \BibitemOpen
  \bibfield  {author} {\bibinfo {author} {\bibfnamefont {Aditi}\ \bibnamefont
  {Pradeep}}, \bibinfo {author} {\bibfnamefont {Sasidharan}\ \bibnamefont
  {Anupama}}, \ and\ \bibinfo {author} {\bibfnamefont {Chethil}\ \bibnamefont
  {Sudheesh}},\ }\bibfield  {title} {\enquote {\bibinfo {title} {Dynamics of
  observables in a q-deformed harmonic oscillator},}\ }\href@noop {} {\bibfield
   {journal} {\bibinfo  {journal} {The European Physical Journal D}\ }\textbf
  {\bibinfo {volume} {74}},\ \bibinfo {pages} {1--8} (\bibinfo {year}
  {2020})}\BibitemShut {NoStop}%
\bibitem [{\citenamefont {{da Costa}}\ and\ \citenamefont
  {Borges}(2019)}]{DACOSTA20192729}%
  \BibitemOpen
  \bibfield  {author} {\bibinfo {author} {\bibfnamefont {Bruno~G.}\
  \bibnamefont {{da Costa}}}\ and\ \bibinfo {author} {\bibfnamefont
  {Ernesto~P.}\ \bibnamefont {Borges}},\ }\bibfield  {title} {\enquote
  {\bibinfo {title} {Nonlinear quantum mechanics in a q-deformed hilbert
  space},}\ }\href {\doibase https://doi.org/10.1016/j.physleta.2019.05.056}
  {\bibfield  {journal} {\bibinfo  {journal} {Physics Letters A}\ }\textbf
  {\bibinfo {volume} {383}},\ \bibinfo {pages} {2729--2738} (\bibinfo {year}
  {2019})}\BibitemShut {NoStop}%
\bibitem [{\citenamefont {Swamy}(2003)}]{swamy2003deformed}%
  \BibitemOpen
  \bibfield  {author} {\bibinfo {author} {\bibfnamefont {P~Narayana}\
  \bibnamefont {Swamy}},\ }\bibfield  {title} {\enquote {\bibinfo {title}
  {Deformed heisenberg algebra: origin of q-calculus},}\ }\href@noop {}
  {\bibfield  {journal} {\bibinfo  {journal} {Physica A: Statistical Mechanics
  and its Applications}\ }\textbf {\bibinfo {volume} {328}},\ \bibinfo {pages}
  {145--153} (\bibinfo {year} {2003})}\BibitemShut {NoStop}%
\bibitem [{\citenamefont {Bonatsos}\ \emph {et~al.}(1996)\citenamefont
  {Bonatsos}, \citenamefont {Kolokotronis}, \citenamefont {Daskaloyannis},
  \citenamefont {Ludu},\ and\ \citenamefont {Quesne}}]{Bonatsos:1996qb}%
  \BibitemOpen
  \bibfield  {author} {\bibinfo {author} {\bibfnamefont {D.}~\bibnamefont
  {Bonatsos}}, \bibinfo {author} {\bibfnamefont {P.}~\bibnamefont
  {Kolokotronis}}, \bibinfo {author} {\bibfnamefont {C.}~\bibnamefont
  {Daskaloyannis}}, \bibinfo {author} {\bibfnamefont {A.}~\bibnamefont {Ludu}},
  \ and\ \bibinfo {author} {\bibfnamefont {C.}~\bibnamefont {Quesne}},\
  }\bibfield  {title} {\enquote {\bibinfo {title} {{Nonlinear deformed SU(2)
  algebras involving two deforming functions}},}\ }\href {\doibase
  10.1007/BF01690332} {\bibfield  {journal} {\bibinfo  {journal} {Czech. J.
  Phys.}\ }\textbf {\bibinfo {volume} {46}},\ \bibinfo {pages} {1189--1196}
  (\bibinfo {year} {1996})},\ \Eprint {http://arxiv.org/abs/q-alg/9701030}
  {arXiv:q-alg/9701030} \BibitemShut {NoStop}%
\bibitem [{\citenamefont {Gavrilik}\ and\ \citenamefont
  {Pavlyuk}(2010)}]{Gavrilik:2009pu}%
  \BibitemOpen
  \bibfield  {author} {\bibinfo {author} {\bibfnamefont {A.~M.}\ \bibnamefont
  {Gavrilik}}\ and\ \bibinfo {author} {\bibfnamefont {A.~M.}\ \bibnamefont
  {Pavlyuk}},\ }\bibfield  {title} {\enquote {\bibinfo {title} {{On Chebyshev
  polynomials and torus knots}},}\ }\href@noop {} {\bibfield  {journal}
  {\bibinfo  {journal} {Ukr. J. Phys.}\ }\textbf {\bibinfo {volume} {55}},\
  \bibinfo {pages} {129--134} (\bibinfo {year} {2010})},\ \Eprint
  {http://arxiv.org/abs/0912.4674} {arXiv:0912.4674 [math-ph]} \BibitemShut
  {NoStop}%
\bibitem [{\citenamefont {Sinha}(2022)}]{Sinha:2022sdo}%
  \BibitemOpen
  \bibfield  {author} {\bibinfo {author} {\bibfnamefont {Aninda}\ \bibnamefont
  {Sinha}},\ }\bibfield  {title} {\enquote {\bibinfo {title} {{Dispersion
  relations and knot theory}},}\ }\href@noop {} {\  (\bibinfo {year} {2022})},\
  \Eprint {http://arxiv.org/abs/2204.13986} {arXiv:2204.13986 [hep-th]}
  \BibitemShut {NoStop}%
\bibitem [{\citenamefont {Kulish}\ and\ \citenamefont
  {Reshetikhin}(1983)}]{Kulish1983}%
  \BibitemOpen
  \bibfield  {author} {\bibinfo {author} {\bibfnamefont {P.~P.}\ \bibnamefont
  {Kulish}}\ and\ \bibinfo {author} {\bibfnamefont {N.~Yu.}\ \bibnamefont
  {Reshetikhin}},\ }\bibfield  {title} {\enquote {\bibinfo {title} {Quantum
  linear problem for the sine-gordon equation and higher representations},}\
  }\href {\doibase 10.1007/BF01084171} {\bibfield  {journal} {\bibinfo
  {journal} {Journal of Soviet Mathematics}\ }\textbf {\bibinfo {volume}
  {23}},\ \bibinfo {pages} {2435--2441} (\bibinfo {year} {1983})}\BibitemShut
  {NoStop}%
\bibitem [{\citenamefont {Fendley}\ \emph {et~al.}(2004)\citenamefont
  {Fendley}, \citenamefont {Sengupta},\ and\ \citenamefont
  {Sachdev}}]{PhysRevB.69.075106}%
  \BibitemOpen
  \bibfield  {author} {\bibinfo {author} {\bibfnamefont {Paul}\ \bibnamefont
  {Fendley}}, \bibinfo {author} {\bibfnamefont {K.}~\bibnamefont {Sengupta}}, \
  and\ \bibinfo {author} {\bibfnamefont {Subir}\ \bibnamefont {Sachdev}},\
  }\bibfield  {title} {\enquote {\bibinfo {title} {Competing density-wave
  orders in a one-dimensional hard-boson model},}\ }\href {\doibase
  10.1103/PhysRevB.69.075106} {\bibfield  {journal} {\bibinfo  {journal} {Phys.
  Rev. B}\ }\textbf {\bibinfo {volume} {69}},\ \bibinfo {pages} {075106}
  (\bibinfo {year} {2004})},\ \Eprint {http://arxiv.org/abs/cond-mat/0309438}
  {arXiv:cond-mat/0309438 [cond-mat.str-el]} \BibitemShut {NoStop}%
\bibitem [{\citenamefont {Bull}()}]{Bull}%
  \BibitemOpen
  \bibfield  {author} {\bibinfo {author} {\bibfnamefont {K.}~\bibnamefont
  {Bull}},\ }\href {\doibase https://github.com/Cable273/comP} {\
  https://github.com/Cable273/comP}\BibitemShut {NoStop}%
\bibitem [{\citenamefont {Roberts}\ \emph {et~al.}(2015)\citenamefont
  {Roberts}, \citenamefont {Stanford},\ and\ \citenamefont
  {Susskind}}]{Roberts:2014isa}%
  \BibitemOpen
  \bibfield  {author} {\bibinfo {author} {\bibfnamefont {Daniel~A.}\
  \bibnamefont {Roberts}}, \bibinfo {author} {\bibfnamefont {Douglas}\
  \bibnamefont {Stanford}}, \ and\ \bibinfo {author} {\bibfnamefont {Leonard}\
  \bibnamefont {Susskind}},\ }\bibfield  {title} {\enquote {\bibinfo {title}
  {{Localized shocks}},}\ }\href {\doibase 10.1007/JHEP03(2015)051} {\bibfield
  {journal} {\bibinfo  {journal} {JHEP}\ }\textbf {\bibinfo {volume} {03}},\
  \bibinfo {pages} {051} (\bibinfo {year} {2015})},\ \Eprint
  {http://arxiv.org/abs/1409.8180} {arXiv:1409.8180 [hep-th]} \BibitemShut
  {NoStop}%
\bibitem [{\citenamefont {Schecter}\ and\ \citenamefont
  {Iadecola}(2019)}]{Schecter:2019oej}%
  \BibitemOpen
  \bibfield  {author} {\bibinfo {author} {\bibfnamefont {Michael}\ \bibnamefont
  {Schecter}}\ and\ \bibinfo {author} {\bibfnamefont {Thomas}\ \bibnamefont
  {Iadecola}},\ }\bibfield  {title} {\enquote {\bibinfo {title} {{Weak
  Ergodicity Breaking and Quantum Many-Body Scars in Spin-1 XY Magnets}},}\
  }\href {\doibase 10.1103/PhysRevLett.123.147201} {\bibfield  {journal}
  {\bibinfo  {journal} {Phys. Rev. Lett.}\ }\textbf {\bibinfo {volume} {123}},\
  \bibinfo {pages} {147201} (\bibinfo {year} {2019})},\ \Eprint
  {http://arxiv.org/abs/1906.10131} {arXiv:1906.10131 [cond-mat.str-el]}
  \BibitemShut {NoStop}%
\bibitem [{\citenamefont {You}\ \emph {et~al.}(2022)\citenamefont {You},
  \citenamefont {Zhao}, \citenamefont {Ren}, \citenamefont {Sun}, \citenamefont
  {Li},\ and\ \citenamefont {Ole\'s}}]{You:2022hfz}%
  \BibitemOpen
  \bibfield  {author} {\bibinfo {author} {\bibfnamefont {Wen-Long}\
  \bibnamefont {You}}, \bibinfo {author} {\bibfnamefont {Zhuan}\ \bibnamefont
  {Zhao}}, \bibinfo {author} {\bibfnamefont {Jie}\ \bibnamefont {Ren}},
  \bibinfo {author} {\bibfnamefont {Gaoyong}\ \bibnamefont {Sun}}, \bibinfo
  {author} {\bibfnamefont {Liangsheng}\ \bibnamefont {Li}}, \ and\ \bibinfo
  {author} {\bibfnamefont {Andrzej~M.}\ \bibnamefont {Ole\'s}},\ }\bibfield
  {title} {\enquote {\bibinfo {title} {{Quantum many-body scars in spin-1
  Kitaev chains}},}\ }\href {\doibase 10.1103/PhysRevResearch.4.013103}
  {\bibfield  {journal} {\bibinfo  {journal} {Phys. Rev. Res.}\ }\textbf
  {\bibinfo {volume} {4}},\ \bibinfo {pages} {013103} (\bibinfo {year}
  {2022})},\ \Eprint {http://arxiv.org/abs/2201.09220} {arXiv:2201.09220
  [cond-mat.str-el]} \BibitemShut {NoStop}%
\bibitem [{\citenamefont {Mukherjee}\ \emph
  {et~al.}(2020{\natexlab{a}})\citenamefont {Mukherjee}, \citenamefont {Nandy},
  \citenamefont {Sen}, \citenamefont {Sen},\ and\ \citenamefont
  {Sengupta}}]{PhysRevB.101.245107}%
  \BibitemOpen
  \bibfield  {author} {\bibinfo {author} {\bibfnamefont {Bhaskar}\ \bibnamefont
  {Mukherjee}}, \bibinfo {author} {\bibfnamefont {Sourav}\ \bibnamefont
  {Nandy}}, \bibinfo {author} {\bibfnamefont {Arnab}\ \bibnamefont {Sen}},
  \bibinfo {author} {\bibfnamefont {Diptiman}\ \bibnamefont {Sen}}, \ and\
  \bibinfo {author} {\bibfnamefont {K.}~\bibnamefont {Sengupta}},\ }\bibfield
  {title} {\enquote {\bibinfo {title} {Collapse and revival of quantum
  many-body scars via floquet engineering},}\ }\href {\doibase
  10.1103/PhysRevB.101.245107} {\bibfield  {journal} {\bibinfo  {journal}
  {Phys. Rev. B}\ }\textbf {\bibinfo {volume} {101}},\ \bibinfo {pages}
  {245107} (\bibinfo {year} {2020}{\natexlab{a}})},\ \Eprint
  {http://arxiv.org/abs/1907.08212} {arXiv:1907.08212 [quant-ph]} \BibitemShut
  {NoStop}%
\bibitem [{\citenamefont {Mukherjee}\ \emph
  {et~al.}(2020{\natexlab{b}})\citenamefont {Mukherjee}, \citenamefont {Sen},
  \citenamefont {Sen},\ and\ \citenamefont {Sengupta}}]{PhysRevB.102.075123}%
  \BibitemOpen
  \bibfield  {author} {\bibinfo {author} {\bibfnamefont {Bhaskar}\ \bibnamefont
  {Mukherjee}}, \bibinfo {author} {\bibfnamefont {Arnab}\ \bibnamefont {Sen}},
  \bibinfo {author} {\bibfnamefont {Diptiman}\ \bibnamefont {Sen}}, \ and\
  \bibinfo {author} {\bibfnamefont {K.}~\bibnamefont {Sengupta}},\ }\bibfield
  {title} {\enquote {\bibinfo {title} {Dynamics of the vacuum state in a
  periodically driven rydberg chain},}\ }\href {\doibase
  10.1103/PhysRevB.102.075123} {\bibfield  {journal} {\bibinfo  {journal}
  {Phys. Rev. B}\ }\textbf {\bibinfo {volume} {102}},\ \bibinfo {pages}
  {075123} (\bibinfo {year} {2020}{\natexlab{b}})},\ \Eprint
  {http://arxiv.org/abs/2005.07715} {arXiv:2005.07715 [cond-mat.str-el]}
  \BibitemShut {NoStop}%
\bibitem [{\citenamefont {Sugiura}\ \emph {et~al.}(2021)\citenamefont
  {Sugiura}, \citenamefont {Kuwahara},\ and\ \citenamefont
  {Saito}}]{PhysRevResearch.3.L012010}%
  \BibitemOpen
  \bibfield  {author} {\bibinfo {author} {\bibfnamefont {Sho}\ \bibnamefont
  {Sugiura}}, \bibinfo {author} {\bibfnamefont {Tomotaka}\ \bibnamefont
  {Kuwahara}}, \ and\ \bibinfo {author} {\bibfnamefont {Keiji}\ \bibnamefont
  {Saito}},\ }\bibfield  {title} {\enquote {\bibinfo {title} {Many-body scar
  state intrinsic to periodically driven system},}\ }\href {\doibase
  10.1103/PhysRevResearch.3.L012010} {\bibfield  {journal} {\bibinfo  {journal}
  {Phys. Rev. Research}\ }\textbf {\bibinfo {volume} {3}},\ \bibinfo {pages}
  {L012010} (\bibinfo {year} {2021})},\ \Eprint
  {http://arxiv.org/abs/1911.06092} {arXiv:1911.06092 [cond-mat.stat-mech]}
  \BibitemShut {NoStop}%
\bibitem [{\citenamefont {Hudomal}\ \emph {et~al.}(2022)\citenamefont
  {Hudomal}, \citenamefont {Desaules}, \citenamefont {Mukherjee}, \citenamefont
  {Su}, \citenamefont {Halimeh},\ and\ \citenamefont
  {Papi\'c}}]{Hudomal:2022vtv}%
  \BibitemOpen
  \bibfield  {author} {\bibinfo {author} {\bibfnamefont {Ana}\ \bibnamefont
  {Hudomal}}, \bibinfo {author} {\bibfnamefont {Jean-Yves}\ \bibnamefont
  {Desaules}}, \bibinfo {author} {\bibfnamefont {Bhaskar}\ \bibnamefont
  {Mukherjee}}, \bibinfo {author} {\bibfnamefont {Guo-Xian}\ \bibnamefont
  {Su}}, \bibinfo {author} {\bibfnamefont {Jad~C.}\ \bibnamefont {Halimeh}}, \
  and\ \bibinfo {author} {\bibfnamefont {Zlatko}\ \bibnamefont {Papi\'c}},\
  }\bibfield  {title} {\enquote {\bibinfo {title} {Driving quantum many-body
  scars in the pxp model},}\ }\href {\doibase 10.1103/PhysRevB.106.104302}
  {\bibfield  {journal} {\bibinfo  {journal} {Phys. Rev. B}\ }\textbf {\bibinfo
  {volume} {106}},\ \bibinfo {pages} {104302} (\bibinfo {year} {2022})},\
  \Eprint {http://arxiv.org/abs/2204.13718} {arXiv:2204.13718
  [cond-mat.quant-gas]} \BibitemShut {NoStop}%
\bibitem [{\citenamefont {Desaules}\ \emph
  {et~al.}(2022{\natexlab{d}})\citenamefont {Desaules}, \citenamefont {Bull},
  \citenamefont {Daniel},\ and\ \citenamefont {Papi\ifmmode~\acute{c}\else
  \'{c}\fi{}}}]{PhysRevB.105.245137}%
  \BibitemOpen
  \bibfield  {author} {\bibinfo {author} {\bibfnamefont {Jean-Yves}\
  \bibnamefont {Desaules}}, \bibinfo {author} {\bibfnamefont {Kieran}\
  \bibnamefont {Bull}}, \bibinfo {author} {\bibfnamefont {Aiden}\ \bibnamefont
  {Daniel}}, \ and\ \bibinfo {author} {\bibfnamefont {Zlatko}\ \bibnamefont
  {Papi\ifmmode~\acute{c}\else \'{c}\fi{}}},\ }\bibfield  {title} {\enquote
  {\bibinfo {title} {Hypergrid subgraphs and the origin of scarred quantum
  walks in many-body hilbert space},}\ }\href {\doibase
  10.1103/PhysRevB.105.245137} {\bibfield  {journal} {\bibinfo  {journal}
  {Phys. Rev. B}\ }\textbf {\bibinfo {volume} {105}},\ \bibinfo {pages}
  {245137} (\bibinfo {year} {2022}{\natexlab{d}})},\ \Eprint
  {http://arxiv.org/abs/2112.06885} {arXiv:2112.06885 [quant-ph]} \BibitemShut
  {NoStop}%
\bibitem [{\citenamefont {Weinberg}\ and\ \citenamefont
  {Bukov}(2017)}]{weinberg2017quspin}%
  \BibitemOpen
  \bibfield  {author} {\bibinfo {author} {\bibfnamefont {Phillip}\ \bibnamefont
  {Weinberg}}\ and\ \bibinfo {author} {\bibfnamefont {Marin}\ \bibnamefont
  {Bukov}},\ }\bibfield  {title} {\enquote {\bibinfo {title} {Quspin: a python
  package for dynamics and exact diagonalisation of quantum many body systems
  part i: spin chains},}\ }\href@noop {} {\bibfield  {journal} {\bibinfo
  {journal} {SciPost Physics}\ }\textbf {\bibinfo {volume} {2}},\ \bibinfo
  {pages} {003} (\bibinfo {year} {2017})},\ \Eprint
  {http://arxiv.org/abs/1610.03042} {arXiv:1610.03042 [physics.comp-ph]}
  \BibitemShut {NoStop}%
\bibitem [{\citenamefont {Weinberg}\ and\ \citenamefont
  {Bukov}(2019)}]{weinberg2019quspin}%
  \BibitemOpen
  \bibfield  {author} {\bibinfo {author} {\bibfnamefont {Phillip}\ \bibnamefont
  {Weinberg}}\ and\ \bibinfo {author} {\bibfnamefont {Marin}\ \bibnamefont
  {Bukov}},\ }\bibfield  {title} {\enquote {\bibinfo {title} {{QuSpin: a Python
  package for dynamics and exact diagonalisation of quantum many body systems.
  Part II: bosons, fermions and higher spins}},}\ }\href {\doibase
  10.21468/SciPostPhys.7.2.020} {\bibfield  {journal} {\bibinfo  {journal}
  {SciPost Phys.}\ }\textbf {\bibinfo {volume} {7}},\ \bibinfo {pages} {020}
  (\bibinfo {year} {2019})},\ \Eprint {http://arxiv.org/abs/1804.06782}
  {arXiv:1804.06782 [physics.comp-ph]} \BibitemShut {NoStop}%
\bibitem [{\citenamefont {Martínez-Tibaduiza}\ \emph
  {et~al.}(2020)\citenamefont {Martínez-Tibaduiza}, \citenamefont {Aragão},
  \citenamefont {Farina},\ and\ \citenamefont {Zarro}}]{MARTINEZ}%
  \BibitemOpen
  \bibfield  {author} {\bibinfo {author} {\bibfnamefont {D.}~\bibnamefont
  {Martínez-Tibaduiza}}, \bibinfo {author} {\bibfnamefont {A.H.}\ \bibnamefont
  {Aragão}}, \bibinfo {author} {\bibfnamefont {C.}~\bibnamefont {Farina}}, \
  and\ \bibinfo {author} {\bibfnamefont {C.A.D.}\ \bibnamefont {Zarro}},\
  }\bibfield  {title} {\enquote {\bibinfo {title} {New bch-like relations of
  the su(1,1), su(2) and so(2,1) lie algebras},}\ }\href {\doibase
  https://doi.org/10.1016/j.physleta.2020.126937} {\bibfield  {journal}
  {\bibinfo  {journal} {Physics Letters A}\ }\textbf {\bibinfo {volume}
  {384}},\ \bibinfo {pages} {126937} (\bibinfo {year} {2020})},\ \Eprint
  {http://arxiv.org/abs/2005.09500} {arXiv:2005.09500 [math-ph]} \BibitemShut
  {NoStop}%
\bibitem [{\citenamefont {Abramowitz}\ \emph {et~al.}(1988)\citenamefont
  {Abramowitz}, \citenamefont {Stegun},\ and\ \citenamefont
  {Romer}}]{abramowitz1988handbook}%
  \BibitemOpen
  \bibfield  {author} {\bibinfo {author} {\bibfnamefont {Milton}\ \bibnamefont
  {Abramowitz}}, \bibinfo {author} {\bibfnamefont {Irene~A}\ \bibnamefont
  {Stegun}}, \ and\ \bibinfo {author} {\bibfnamefont {Robert~H}\ \bibnamefont
  {Romer}},\ }\href@noop {} {\enquote {\bibinfo {title} {Handbook of
  mathematical functions with formulas, graphs, and mathematical tables},}\ }
  (\bibinfo {year} {1988})\BibitemShut {NoStop}%
\end{thebibliography}%

\section*{Appendix A: Complexity for $\mathrm{SU}(2)$ symmetry}\label{appa}
Here we briefly discuss the symmetry Krylov basis construction, if the Hamiltonian has any particular symmetry \cite{Balasubramanian:2022tpr}. Here we primarily focus on the SU(2) algebra for spin $j$, which is given by  $[J_{0}, J_{\pm}] = \pm J_{\pm}$ and $[J_{+}, J_{-}] = 2 J_{0}$. The corresponding Hamiltonian can be explicitly expressed in terms of the SU(2) generators as
\begin{align}
    H = \alpha (J_{+} + J_{-}) + \eta_{0} J_{0} + \delta I\,. \label{symH}
\end{align}
The associated Krylov basis vectors are finite-dimensional and given by $\ket{\mathcal{K}_{n}} = \ket{j, -j + n}$, where $n = 0, \cdots 2j$. The lowest-weight state corresponds to $n=0$, and we take it as the initial state. It is time evolved according to
\begin{align}
    \ket{\Psi(t)} = e^{- i H t} \ket{K_{0}} = \sum_{m=0}^{2j} \psi_{n}(t) \ket{\mathcal{K}_{n}}\,, \label{ev}
\end{align}
where $\psi_{n} (t)$'s are the Krylov basis coefficients. Note that the initial state is not an eigenstate of the Hamiltonian \eqref{symH}. The symmetry allows us to directly extract the associated Lanczos coefficients as \cite{Balasubramanian:2022tpr}
\begin{align}
    a_{n} = \eta_{0} (-j + n) + \delta \,,~~~ b_{n} = \alpha \sqrt{n (2j - n + 1)}\,. \label{ll}
\end{align}

\begin{figure}[h]
\subfigure[]{\includegraphics[height=5.7cm,width=\linewidth]{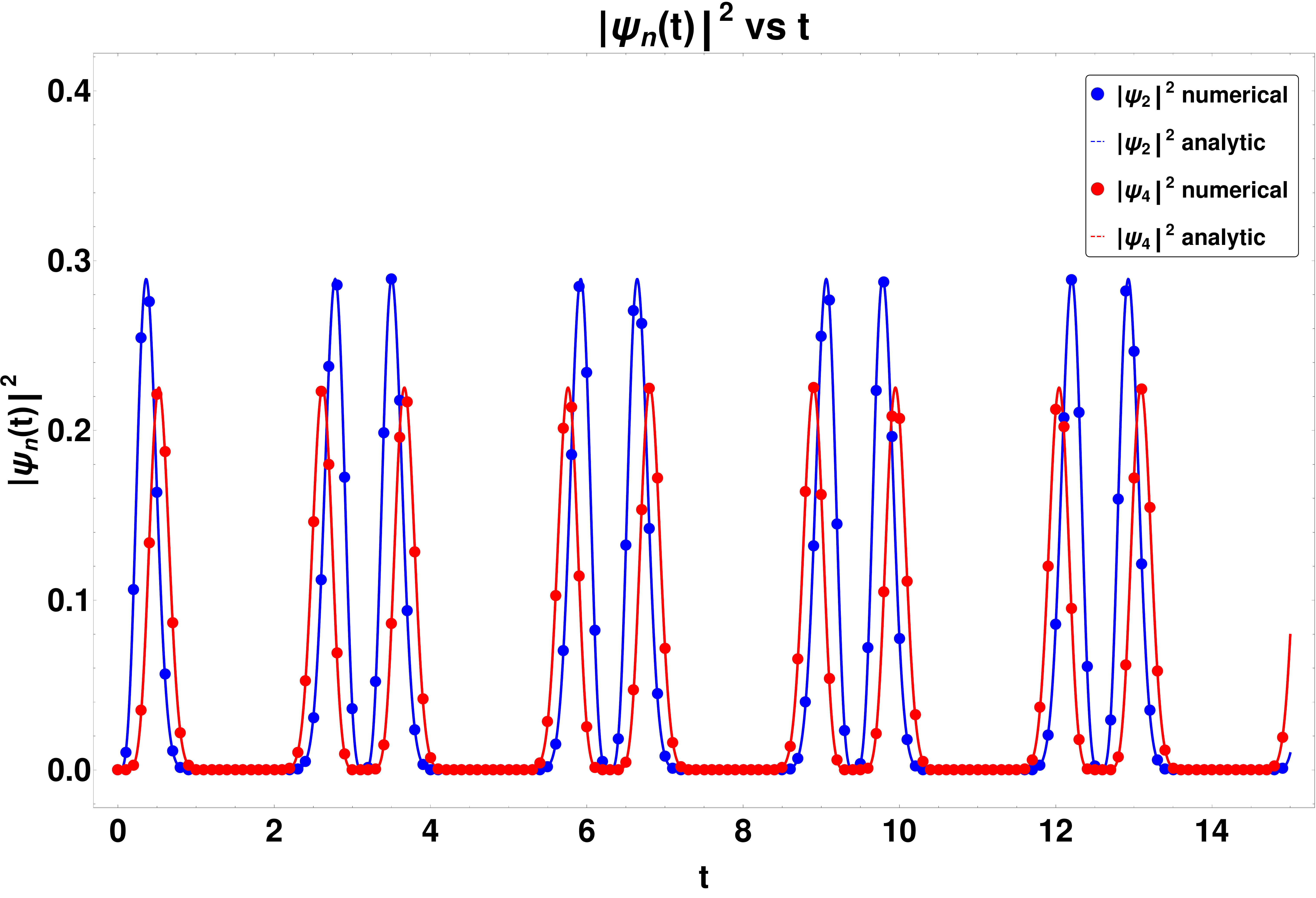}}
\hfill
\subfigure[]{\includegraphics[height=5.7cm,width=\linewidth]{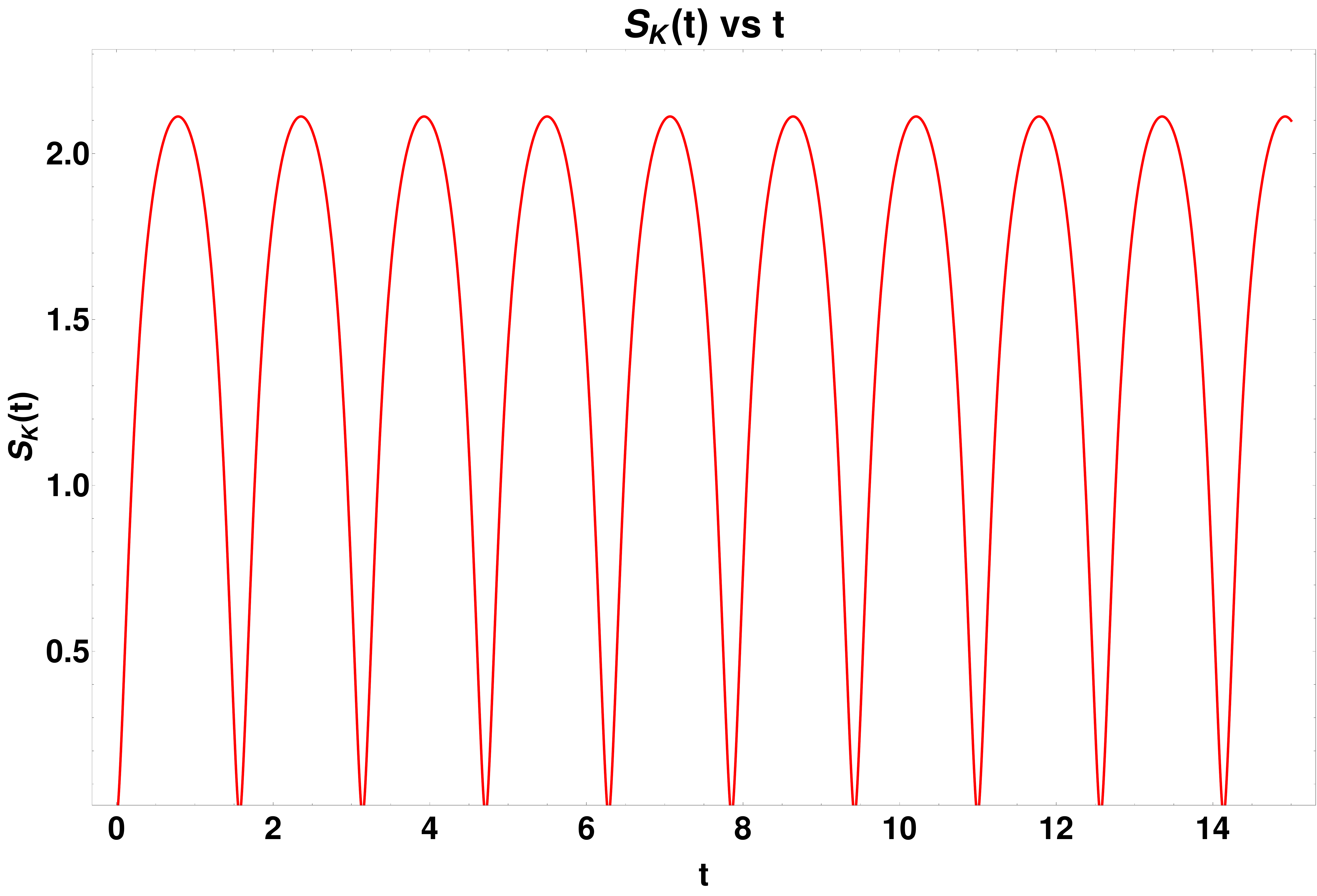}}
\caption{(a) Evolution of $\psi_{n}(t)$'s and the (b) evolution of entropy $S_K(t)$ for the paramagnetic model. We choose lattice size $N=16$.} \label{fig:dev}
\end{figure}
Let us now note the following relations \cite{MARTINEZ}. For an operator given by
\begin{align}
    \hat{G} = e^{\lambda_{+} T_{+} + \lambda_{-} T_{-} + \lambda_{0}T_{0} + \omega \id}\,,
\end{align}
where $T_{0}, T_{\pm}$ satisfy the commutation relations
\begin{align}
    [T_{+}, T_{-}] = 2 T_{0}\,, ~~~~
    [T_{0}, T_{\pm}] = \pm  T_{\pm}\,.
\end{align}
We have the following expression for $\hat{G}$
\begin{align}
    \hat{G} = e^{\omega}e^{\Lambda_{+}T_{+}}e^{\log(\Lambda_{0})T_{0}}e^{\Lambda_{-}T_{-}}\,,
\end{align}
and we have the functions $\Lambda_{\pm}, \Lambda_{0}$ are given by
\begin{align}
    \Lambda_{0} &= \left( \cosh\nu - \frac{ \lambda_{0}}{2\nu}\sinh\nu \right)^{-2} \\
    \Lambda_{\pm} &= \frac{2 \lambda_{\pm} \sinh(\nu)}{2 \nu \cosh\nu - \lambda_{0} \sinh\nu} \\
    \nu^2 &= \left( \frac{\lambda_{0}}{2} \right)^2 + \lambda_{+}\lambda_{-}\,.
\end{align}
Therefore, the time evolution operator is given as
\begin{align}
    e^{- i H t} = e^{- i \omega t}e^{A T_{+}}e^{B T_{0}}e^{C T_{-}}\,.
\end{align}
We find the following expressions for $A$, $B$, and $C$ (by noting that $\lambda_{+} = \lambda_{-} = \alpha$) in the Hamiltonian.
\begin{align}
    H = \alpha(T_{+} + T_{-}) + \eta_{0} T_{0} + \omega \id \,. \label{ham2}
\end{align}
For the given Hamiltonian \eqref{ham2}, we have the following expressions for $\nu, \Lambda_{\pm}$.
\begin{align}
    \nu &= \frac{i t}{2} \sqrt{ 4\alpha^2 +\eta_{0}^{2}}\,, \\
    B &= -2\log\left(\cos\Big( \frac{t}{2} \sqrt{ 4  \alpha^2 +  \eta_{0}^2}\right)\,, \notag \\ &+ i\frac{\eta_{0} \sin\left( \frac{t}{2} \sqrt{4 \alpha^2 +  \eta_{0}^2 }\right)}{\sqrt{4 \alpha^2 + \eta_{0}^2}} \Big)\,, \label{B} \\
    A = C &= \frac{2 \alpha}{i \sqrt{4 \alpha^2 + \eta_{0}^2}\cot\left(\frac{t}{2} \sqrt{ 4 \alpha^2 + \eta_{0}^2}\right) - \eta_{0}}\,. \label{AC}
\end{align}
Therefore, the time evolved state is given by
\begin{align}
    &\ket{\psi(t)} = e^{-i H t}\ket{j, -j}\, \notag \\
    &= e^{-i\omega t}e^{-B j}\sum_{n = 0}^{2 j}A^{n}\sqrt{\frac{\Gamma(2 j + 1)}{n! \Gamma(2 j - n + 1)}}\ket{j, -j + n}\,. \label{psit}
\end{align}
From here, the wave functions can be easily read off
\begin{align}
    \psi_{n} (t) = e^{-i\omega t}e^{-B j} A^{n}\sqrt{\frac{\Gamma(2 j + 1)}{n! \Gamma(2 j - n + 1)}} \,. \label{psin}
\end{align}
As one can see here, the expression for $\psi_{n} (t)$ for $\eta_{0} = 0$ and $\omega = 0$ depends solely on  $\alpha$, as is reflected in our numerical results.

From Eq.\eqref{psin}, one can see that the non-zero basis functions can be obtained until $n = 2j+1$. This is due to the existence of the reciprocal Gamma function in \eqref{psin} and it is well-known that $\lim_{z \rightarrow m} 1/\Gamma(-m) = 0$ for $m = 0, 1, 2, \ldots$ \cite{abramowitz1988handbook}. Thus, we get a simpler expression for the complexity as
\begin{align}
    C(t) = \frac{2 j}{1 + \frac{\eta^{2}_{0}}{4 \alpha^2}}\sin^{2}\left( \frac{t}{2} \sqrt{ 4 \alpha^2 + \eta_{0}^2}\right)\,. \label{Ct}
\end{align}
Note that the evolution of complexity depends on $\alpha$ and $\eta_{0}$, whereas the spin $j$ only sets its amplitude of it. Also, note the periodic nature of the $C(t)$. We can also compute the entropy
\begin{align}
    S_K(t) = -\sum_{n=0}^{N}  |\psi_{n} (t)|^2 \log |\psi_{n} (t)|^2\,.
\end{align}
The plot for entropy is shown in Fig.\,\ref{fig:dev}(b).
%\vspace{-.9cm}
%Corresponding to various choices of $\delta, \epsilon$, it is possible to study various algebras. For example, $SL(2,R)$  (or equivalently $SU(1,1)$) corresponds to $\delta = - 1,\; \epsilon = 1$. Similarly, the values $\delta = -i/2,\; \epsilon = i$ corresponds to $SO(2,1)$. In the case where either $\delta$ or $\epsilon$ are complex (but not both), the expression for Krylov complexity \eqref{Ct} will have to be modified accordingly. The expression derived in \eqref{Ct} holds only if $\alpha^2\delta\epsilon$ and $\delta^2 \eta^2_{0}$ are real.
\newline
\subsection*{Appendix B: Some commutation relations}
During our discussions on the algebra satisfied by the PXP Hamiltonian with first-order perturbation, we omit the full expressions for the commutators of the ladder operators. In this appendix, we explicitly list the same. In \eqref{HpmCommZZ}, we denote the $\mathcal{O}(\lambda^2)$ terms collectively as $Y^{(1)}$. The explicit expression for the same is as follows (we again resort to the notation $\tilde{\sigma}^{(*)}_{m} \equiv P_{m - 1} \sigma^{(*)}_{m} P_{m + 1}$):
\begin{widetext}
\begin{align*}
    Y^{(1)} &= \sum_{n} P_{2n-1}\tilde{\sigma}^{z}_{2n + 1} - P_{2n - 2}\tilde{\sigma}^{z}_{2n}
    + \sum_{n} \tilde{\sigma}^{z}_{2n}P_{2n + 2} - \tilde{\sigma}^{z}_{2n+1}P_{2n + 3} 
    + \frac{1}{2}\sum_{n} \tilde{\sigma}^{+}_{2n}\sigma^{-}_{2n+2}P_{2n + 3} - \tilde{\sigma}^{+}_{2n+1}\sigma^{-}_{2n+3}P_{2n + 4}\notag \\
    &+ \frac{1}{2}\sum_{n} \tilde{\sigma}^{-}_{2n}\sigma^{+}_{2n+2}P_{2n + 3} - \tilde{\sigma}^{-}_{2n+1}\sigma^{+}_{2n+3}P_{2n + 4} 
    + 2\sum_{n} P_{2n-1}\tilde{\sigma}^{z}_{2n + 1}P_{2n + 3} - P_{2n-2}\tilde{\sigma}^{z}_{2n}P_{2n + 2}\notag \\
    &- \frac{1}{2}\sum_{n} P_{2n-2}\tilde{\sigma}^{+}_{2n}\sigma^{-}_{2n+2}P_{2n + 3} - P_{2n-1}\tilde{\sigma}^{+}_{2n+1}\sigma^{-}_{2n+3}P_{2n + 4}
    - \frac{1}{2}\sum_{n} P_{2n-2}\tilde{\sigma}^{-}_{2n}\sigma^{+}_{2n+2}P_{2n + 3} - P_{2n-1}\tilde{\sigma}^{-}_{2n+1}\sigma^{+}_{2n+3}P_{2n + 4} 
    \notag \\
    &+ \frac{1}{2}\sum_{n}\tilde{\sigma}^{+}_{2n}\sigma^{-}_{2n+2}P_{2n + 3}P_{2n + 4} - \tilde{\sigma}^{+}_{2n+1}\sigma^{-}_{2n+3}P_{2n + 4}P_{2n + 5} 
    + \frac{1}{2}\sum_{n} \tilde{\sigma}^{-}_{2n}\sigma^{+}_{2n+2}P_{2n + 3}P_{2n + 4} - \tilde{\sigma}^{-}_{2n+1}\sigma^{+}_{2n+3}P_{2n + 4}P_{2n + 5}\,.
\end{align*}
\end{widetext}

\noindent
Similarly, during our discussion of the commutator \eqref{JpmComm2}, we do not write the explicit expression for $\tilde{X}^{(1)}_{\pm}$. To write a little more, it is prudent to first consider the commutator $[H^{(1)}_{0}, H^{(1)}_{\pm}] = \alpha^3 [J^{(1)}_{0}, J^{(1)}_{\pm}]$. We obtain the following result
\begin{align}
    [H^{(1)}_{0}, H^{(1)}_{\pm}] &= \pm H^{(1)}_{\pm} \pm \sum_{n = 1}^{16}f_{n}(\lambda)\tilde{X}^{(1)}_{n, \pm}\,,
\end{align}
where $\tilde{X}_{n, \pm}$ are terms containing multiple $P$ and $\sigma$ operators \cite{Bull}. There are $16$ such terms in this expression, and they are weighted by the functions $f_{n}(\lambda)$. These functions are polynomials of $\lambda$, of order up to $\lambda^3$. Thus, we rewrite the above expression in terms of $J^{(1)}_{0}$ and $J^{(1)}_{\pm}$ as \eqref{JpmComm2}, with $\tilde{X}^{(1)}_{\pm} =\frac{1-\alpha^{2}}{\alpha^{2}}J^{(1)}_{\pm} + \frac{1}{\alpha^{3}} \sum_{n = 1}^{16}f_{n}(\lambda)\tilde{X}^{(1)}_{n, \pm}$.

\end{document}